\RequirePackage{fix-cm}
\documentclass[twocolumn,epjc3]{svjour3}  
\smartqed  
\RequirePackage{graphicx}
\RequirePackage{bm}
\RequirePackage{amsmath}
\RequirePackage{amssymb}
\RequirePackage{color}
\RequirePackage{ulem}

\newcommand{\new}[1]{#1}

\journalname{Eur. Phys. J. A}
\begin{document}
\title{Transport coefficients of magnetized neutron star cores}

\author{Peter Shternin\thanksref{e1,addr1}
and
Dmitry Ofengeim\thanksref{addr1}
}

\thankstext{e1}{e-mail: pshternin@gmail.com}


\institute{Ioffe Institute,
Politekhnicheskaya 26, St. Petersburg, 194021, Russia \label{addr1}
}

\date{Received: date / Accepted: date}

\maketitle

\begin{abstract}
We review the calculations of the kinetic coefficients (thermal conductivity, shear viscosity, momentum transfer rates) of the neutron star core matter within the 
framework of the  Landau Fermi-liquid theory.
We restrict ourselves to the case of normal (i.e. non-superfluid) matter.  As an example we consider simplest $npe\mu$ composition of neutron star core matter.
Utilizing the CompOSE database of dense matter equations of state and several microscopic interactions we analyze the 
uncertainties in calculations of the kinetic coefficients that result from the insufficient knowledge of the properties of the dense nuclear matter and suggest possible approximate treatment. In our study we also take into account non-quantizing magnetic field. 
The presence of magnetic field makes transport anisotropic leading to the tensor structure of  kinetic coefficients.
We find that the moderate ($B\lesssim 10^{12}$~G) magnetic field do not affect considerably thermal conductivity of neutron star core matter, since the latter is mainly governed by the electrically neutral neutrons. In contrast, shear viscosity is affected even by the moderate $B\sim 10^8 - 10^{10}$~G.  Based on the in-vacuum nucleon interactions we provide practical expressions for calculation of transport coefficients for any equation of state of dense matter.

\end{abstract}

\section{Introduction}
\label{S:intro}

Neutron stars (NSs) are the most compact astrophysical objects known that are still stable against the gravitational collapse. This is possible because NSs are largely composed of strongly interacting degenerate baryons (although quark cores, other hadronic, or mixed models  are also discussed), which pressure is strong enough to counterbalance the gravity forces. This is strongly supported by the inferences on the masses ($\sim 1-2\,M_\odot$) and radii ($R\sim10-20$~km) obtained for these objects by astrophysical methods, which \new{correspond} to the mean densities  $\mathrm{few}\times\rho_0$ with $\rho_0=2.8\times 10^{14}$~g~cm$^{-3}$ being the nuclear saturation density. 

Understanding  properties of \new{matter} under such extreme conditions is a subject of fundamental importance for  modern physics. Such studies allow one to test the predictions of the nuclear matter theories for the conditions unreachable in the terrestrial settings over the astrophysical observations. 

\new{Various physical input is required to model processes and dynamical phenomena that can occur during the NS life. Among it there are the transport properties of the NS matter, see, e.g. \cite{Schmitt2018} for review.}

In the present study we revisit the calculation of the transport coefficients of NS core matter based on the Landau Fermi-liquid theory. We give relatively detailed description of the technique used to calculate these coefficients and to identify how they enter the evolution equations that are then used in the modelling of various physical processes. It is not possible to give a detailed account of many aspects of neutron star transport theory in one article, 
so we restrict ourselves here to the transport coefficients mediated by collisions between the particles composing baryonic NS cores. We left aside more exotic compositions such as meson condensates, or quark cores. We ignore processes related to the reactions. In this sense we do not consider the bulk viscosity, since the bulk viscosity  mediated by collisions is negligible.  
\new{It should be calculated using a different technique in comparison to other transport coefficients (see., e.g., \cite{Schmitt2018}).}
We also do not consider the possibility of nucleon pairing which can result \new{in} the superfluidity/superconductivity phenomena in NS cores. The transport coefficients of the superfluid/superconducting NS cores are much less explored and the consistent picture is not drawn yet.

We, however, include the magnetic field effects into consideration. \new{Magnetic field plays an important role  in NS physics.}
\new{Indeed,} the most common astrophysical manifestation of the NSs are the radio pulsars 
\new{the operation of which}
is driven by the magnetic field. The studies of magnetars show that the surface field can reach values up to $10^{14}$--$10^{15}$~G \cite{Kaspi2017ARA&A}.
\new{It is natural to assume that the NS core matter can be under the influence of a strong magnetic field}. 
The effect of the magnetic field on the transport properties of the NS crust is studied in great detail, see, e.g.~\cite{Potekhin1999AA,Potekhin2015SSRv,Ofengeim2015EL,Baiko2016MNRAS,Harutyunyan2016PhRvC}, \new{however} the transport coefficients of the magnetized NS cores received less attention. 

Magnetic field makes the motion of charged particles curvilinear, this affects their response to the external perturbations. Neutral particles feel the effects of magnetic field indirectly, through the interaction with charged species (the neutral particles can also have the magnetic momenta). The influence of the magnetic field on the charge particles in degenerate matter can be described by the 
cyclotron frequency on the Fermi surface of particle species $a$
\begin{equation}\label{eq:omega_BF}
    \omega_{\mathrm{BF}a}=\frac{q_a B}{m_a^*},
\end{equation}
where $q_a$ 
\new{is the electric charge  of the particle species $a$} 
and $m_a^*$ is its effective mass at the Fermi surface. In this paper we assume that the magnetic field is non-quantizing, i.e. $\omega_{\mathrm{BF}a}\lesssim T$. For typical NS core conditions, this inequality is most easily violated for lightest particles, i.e., electrons at $B\gtrsim 10^{14}\ T/(10^8~\mathrm{K})~\mathrm{G}$ 
\cite{Iakovlev1991Ap&SS,Yakovlev1991Ap&SS}. When $T\ll \omega_{\mathrm{BF}a}$, but $\omega_{\mathrm{BF}a}\ll T_{\mathrm{F}a}$, where $T_{\mathrm{F}a}$ is the degeneracy temperature of particle species $a$, the field is weakly quantizing. In this case particles populate many Landau levels. Transport coefficients in this case demonstrate oscillating behavior around the  classical (i.e. non-quatizing) values, and are well-described by the expressions which neglect quantization after the oscillations are smeared out. Therefore the results discussed here will be relevant for a moderately weak quantizing field as well.

At very large fields, known as the strongly quantizing regime, $\omega_{\mathrm{BF}a}\gtrsim T_{\mathrm{F}a}$, particles populate mainly the lowest Landau levels. For electrons, this corresponds to unrealistically large $B\gtrsim 10^{18}$~G \cite{Yakovlev1991Ap&SS}
and can barely happen in NS cores  (but not in the crust e.g., \cite{Potekhin1999AA}).

The calculations of the transport coefficients of NS core matter are hampered by the poorly known properties of the baryon interactions at supranuclear densities. The equation of state (EOS) of such matter is not known and many models are available on the market. Recent progress in theory and observations shrinks the range of available models, but the robust picture is yet to be established. 

From the practical point of view it is desirable to have the publically available repository \new{containing} relevant properties of dense matter for variety of proposed models in coherent manner, so they can be readily used for astrophysical implications. For the EOSs, this route is taken, e.g., by the CompOSE database\footnote{{https://compose.obspm.fr/}}. It contains a data on a large number of EOSs relevant for a NS and supernovae simulations. 

In principle, it would be convenient for such database to contain as well the set of the transport coefficients relevant for each EOS. However, no simple solution for this task is seen. Transport coefficients are not universal and  for each EOS should be calculated under the same underlying microscopic model. At the moment this does not look feasible. 

In the present study we identify what information is needed from the microscopic theory for calculating transport coefficients of magnetized NS cores.
 \new{Specifically, here we consider the beta-stable nucleonic matter with baryon number density $\lesssim 1\,$fm$^{-3}$, temperature $T\leqslant 10^{10}\,$K and magnetic field $\lesssim 10^{14}\,$G. }%
Utilizing a range of the nucleonic EOSs from the CompOSE database and a few microscopic interactions we illustrate the potential scatter that can emerge in calculations. We then elaborate on the `poor man' solution for nucleonic NS cores based on the in-vacuum nucleon interaction. This approach allows one to calculate transport coefficients for any EOS and we provide practical approximate expressions allowing to do this. 

The paper is organized as follows. 
In sec.~\ref{sec:Hydro} we review the first-order relativistic hydrodynamics equations in order to identify the occurrence of the transport coefficients studied here. 
We do not consider effects of the General Relativity since we deal with the microscopic calculations of the transport coefficients which are performed in the local Lorentz frame. The typical mean free path scale is much smaller than the macroscopic scale where the curvature of space-time manifests itself. 
In sec.~\ref{sec:LandauFL} we consider the transport theory of Fermi liquids. In particular, in sec.~\ref{sec:tensor} we briefly introduce the irreducible spherical tensor formalism convenient for studying the problem in magnetic field. In sec.~\ref{sec:collisions} we outline the general expressions for 
\new{calculation of transport coefficients} 
at lowest  variational order. In sec.~\ref{sec:interactions} the quasiparticle collisions governed by  electromagnetic (sec.~\ref{sec:em}) and strong  (sec.~\ref{sec:strong}) interactions are considered. 
In sec.~\ref{sec:discuss} we apply the described formalism to the transport coefficients in nucleonic NS cores. In sec.~\ref{sec:lambda_discuss} we consider effective scattering cross-sections. In sec.~\ref{sec:kin0_discuss} we describe partial contributions to transport coefficients  in non-magnetized matter and  sec.~\ref{sec:kinB_discuss} we discuss transport coefficients in presence of the magnetic field. We conclude in sec.~\ref{sec:conclude}.

We give the practical expressions for the 
\new{calculation of transport coefficients}
for nucleonic NS core matter in \ref{sec:app:pract}.

Throughout the
paper we set $\hbar=c=k_B=1$. The metric \new{tensor} convention is $g_{\mu\nu}=\mathrm{diag}(1,-1,-1,-1)$, and the Greek indices are used for the components of four-vectors while the Roman ones for components of three-vectors. Bold font is used for three-vectors. The Levi-Civita antisymmetric tensor $\epsilon^{\mu\nu\alpha\beta}$ is normalized   as $\epsilon^{0123}=1$. 
Dirac matrices obey $\gamma^5=-i\gamma^0\gamma^1\gamma^2\gamma^3$.

\section{Hydrodynamic equations}\label{sec:Hydro}
Let us start from formulating the relativistic hydrodynamic equations, e.g.  \cite{Rezzolla2013book,RomatschkeRomatschke2017book},  for the normal (i.e. non-superfluid) NS core matter  which is a mixture of $r$ species (e.g. neutrons, protons, electrons, etc.). We assume that collision timescales  between the various species are compatible with the equilibration timescale for the single species. Therefore it is appropriate to describe the NS core mixture as a single fluid. If this condition is not met (as can generally happen in terrestrial or astrophysical plasmas) or if the  superfluidity is taken into account, then more complicated multifluid hydrodynamics should be constructed, see, e.g.,
\cite{Gusakov2016PhRvD,Rau2020PhRvD,AnderssonComer2021LRR,Dommes2021arXiv}.

Hydrodynamic equations consist of the conservation 
equation for the  energy-momentum tensor ${\cal T}^{\mu\nu}$  and the 
conservation
equations for the particle currents $j^\mu_{(a)}$, $a=1,\dots, r$. The latter equations read
\begin{equation}\label{eq:j_cons}
    \partial_\mu j^\mu_{(a)}=0.
\end{equation}
In principle, the particle currents are not conserved since the weak reactions can operate in the NS core. In this case the right-hand side of eq.~(\ref{eq:j_cons}) should contain reaction terms. In case of reaction mixtures it is, in principle, more natural to consider the exactly conserved currents (i.e. baryon number current) instead of particle currents. However, the timescales of the weak reactions in the NS cores are much larger than the collision timescales we will deal below. Therefore we omit these terms for brevity\footnote{For the hyperonic cores there can exist also strong inelastic processes of a form $A+B\leftrightarrow C+D$. For simplicity, we assume the equilibrium state with respect to such reactions, so the particle currents are still conserved. }. 

We assume that the large-scale electromagnetic field can be present in the system. 
The equation for the energy-momentum tensor of the fluid is then
\begin{equation}\label{eq:Tmunu_cons}
    \partial_\mu {\cal T}^{\mu\nu} =  F^{\nu\lambda}J_\lambda,
\end{equation}
where $F^{\mu\nu}$ is the electromagnetic field tensor 
and $J^\mu$ is the electromagnetic current 
\begin{equation}\label{eq:electric_current_def}
    J^\mu=\sum_{a=1}^r q_a j^\mu_{(a)}.
\end{equation}
We assume for simplicity that the magnetic fields are not overwhelmingly  large so that the matter is unpolarized and the magnetic pressure can be neglected.
More general discussion can be found in \cite{Huang2011AnPhy,Finazzo2016PhRvD,Hernandez2017JHE}.

One defines the so-called local rest frame (LRF) of the fluid which for the ideal fluid
is defined as a frame where the  energy flow ${\cal T}^{0i}$, momentum components densities ${\cal T}^{i0}$, and particle flows $j_{(a)}^i$, $i=1,2,3$, vanish. For the fluids outside  equilibrium, the definition of the LRF becomes ambiguous  \cite{Rezzolla2013book,Landau1987Fluid}.
In  any case, the rest frame of the fluid is described by the hydrodynamic four-velocity $U^\mu$ normalized as
$U^\mu U_\mu=1$. In the  LRF, $U^\mu=(1,0,0,0)$, and in the laboratory frame $U^\mu=\gamma(1,\bm{V})$ where $\bm{V}$ is three-velocity and $\gamma=(1-V^2)^{-1/2}$ is the corresponding  Lorentz factor. We define the  the orthogonal projector 
\new{
\begin{equation}\label{eq:orth_proj}
    \Delta^{\mu\nu}=g^{\mu\nu}-U^\mu U^\nu
\end{equation}
}
and decompose the four-gradient operator as
\begin{equation}\label{eq:grad_decomp}
    \partial^\mu=U^\mu U^\nu\partial_\nu + \Delta^{\mu\nu} \partial_\nu\equiv U^\mu D + \nabla^\mu.
\end{equation}
In the LRF, $D\to \partial_t$ and $\nabla^\mu\to (0,-\bm{\nabla})$, where $\bm{\nabla}=\partial_{\bm{r}}$ is the spatial gradient operator.

Using the hydrodynamic velocity, the particle currents can be written as
\begin{equation}\label{eq:j_decomp}
    j^\mu_{(a)}=n_a U^\mu+\Delta j^\mu_{(a)},
\end{equation}
where $\Delta j^\mu_{(a)}$ is the dissipative correction, which vanishes in equilibrium, and $n_a$ is the number density of particle species $a$. Total particle number density is
\begin{equation}\label{eq:ntot_def}
    n=\sum_{a=1}^r n_a
\end{equation}
and the decomposition of the total particle current $j^\mu$ is
\begin{equation}\label{eq:j_tot_decomp}
    j^\mu\equiv \sum_{a=1}^r j^\mu_{(a)}\equiv n U^\mu+\Delta j^\mu,
\end{equation}
where $\Delta j^\mu$ is the total dissipative particle flux. 
Similarly, the decomposition of the energy-momentum tensor into the ideal and non-ideal part is
\begin{equation}\label{eq:Tmunu_decomp}
    {\cal T}^{\mu\nu}=(\mathcal{E}+P) U^\mu U^\nu - Pg^{\mu\nu} +\Delta {\cal T}^{\mu\nu},
\end{equation}
where $\mathcal{E}$ is the internal energy density of the fluid, $P$ is the \new{thermodynamic} pressure, and $\Delta {\cal T}^{\mu\nu}$ is the dissipative correction.
Local equilibrium thermodynamic quantities $n_a$, $\mathcal{E}$, and $P$ are assumed to be related by the thermodynamic laws to the local temperature $T(x)$ and chemical potentials $\mu_a(x)$ via the standard relations provided the equation of state (e.g., in the form $P=P(\{\mu_a\},T)$) is given. The thermodynamic laws are
\begin{subequations}\label{eq:thermodynamics}
\allowdisplaybreaks
\begin{eqnarray}
    d{\cal E} &=& TdS +\sum_a \mu_a d n_a, \label{eq:SecondLaw}\\
    H&\equiv& {\cal E}+P=TS + \sum_a \mu_a n_a, \label{eq:Enthalpy}\\
    dP &=& SdT + \sum_a n_a d\mu_a,\label{eq:GibbsDuhem}
\end{eqnarray}
\end{subequations}
where $H$ is the total enthalpy density 
and $S$ is the equilibrium entropy density. 

The decompositions (\ref{eq:j_decomp})--(\ref{eq:Tmunu_decomp}) describe fluids close to the local equilibrium state so that the dissipative corrections are small and can be systematically expanded in derivatives of the thermodynamic field variables (e.g., \cite{Kovtun2019JHEP})
\begin{eqnarray}
    \Delta {\cal T}^{\mu\nu}&=&{\cal O}(\partial)+{\cal O}(\partial^2)+\dots,\\
    \Delta j^\mu &=&{\cal O}(\partial)+{\cal O}(\partial^2)+\dots.
\end{eqnarray}
Truncating these expansions leads subsequently to the first-order hydrodynamics, second-order hydrodynamics, etc.

There is a principle ambiguity in how to define the fields $T(x)$, $\{\mu_a(x)\}$, and $U^\mu(x)$ in order to write the decompositions
(\ref{eq:j_decomp})--(\ref{eq:Tmunu_decomp}) in non-equilibrium case. The different possibilities are commonly referred as a `choice of the frame' (e.g., \cite{Kovtun2019JHEP}).

The standard choice assumes that the dissipative corrections are transverse to $U^\mu$:
\begin{subequations}\label{eq:matching}
\allowdisplaybreaks
\begin{eqnarray}
U_\mu\Delta j^\mu_{(a)} &=&0,\\
U_\mu U_\nu \Delta {\cal T}^{\mu\nu}&=&0,
\end{eqnarray}
\end{subequations}
and additional condition is required to fix $U^\mu$. Two traditional options are due to Eckart \cite{EckartPhysRev58} and Landau and Lifshitz \cite{Landau1987Fluid}
\begin{subequations}
\allowdisplaybreaks
\begin{eqnarray}
\Delta j^\mu &=&0,\quad \mathrm{Eckart}, \label{eq:EckartFrame}\\
U_\mu\Delta {\cal T}^{\mu\nu}&=&0,\quad \mathrm{Landau-Lifshitz}.\label{eq:LandauFrame}
\end{eqnarray}
\end{subequations}

It is well-known, however, that the first-order hydrodynamics equations in these frames suffer from acasual and unstable behavior \cite{HiscockPhysRevD85}, see detailed discussion in, e.g., \cite{Rezzolla2013book,AnderssonComer2021LRR,RomatschkeRomatschke2017book}.
In principle, that these theories allow for unstable modes and acasual heat propagation does not mean that the actual studies will necessary encounter these problems.  \new{In most} of studies concerned with the NS core transport, up to our knowledge, the first-order theory was enough. This is not the case for the heavy ion collisions and may change with the progress of the simulations of NS mergers. 

The solutions to these problems were proposed, for instance, on the basis of second-order theories of extended irreversible thermodynamics
\cite{Mueller1967ZPhy,Israel1976AnPhy,Israel1979AnPhy} or Carter's variational formalism (e.g. \cite{AnderssonComer2021LRR}), see \cite{Gavassino2021FrASS} for a recent review. 
Alternatively it was recently proposed that the first-order hydrodynamics equations can be made  stable if the frames beyond traditional Eckart or Landau-Lifshitz one are considered \cite{Tsumura2007PhLB,Tsumura2008PhLB,Van2012PhLB,Freistuehler2014RSPSA,Freistuehler2017RSPSA,Freistuehler2018JMP,Bemfica2018PhRvD,Bemfica2019arXiv,Bemfica2019PhRvD,Bemfica020arXiv,Kovtun2019JHEP,Hoult2020JHEP}. The latter formalism was recently explored numerically \cite{Pandya2021PhRvD} and equations were extended to the second order in general frame \cite{Noronha2021arXiv}.
Notice that in fact the certain frames can be preferred over others on the physical basis 
(i.e the so-called thermodynamics frame 
\cite{Van2012PhLB,Becattini2015EPJC}, see also \cite{Zubarev1974}).

The discussion of the validity and sufficiency of the  first order  or  second order description is beyond the scope of the present paper
\cite{AnderssonComer2021LRR,Gavassino2020PhRvD,Gavassino2021FrASS,Bemfica020arXiv}. We restrict ourselves to much less ambitious task. We are interested in the microscopic calculations of the transport coefficients (thermal conductivity, shear viscosity, diffusion coefficients) appearing already in the first-order theory. These coefficients are governed by quasiparticle collisions and  are invariant in the first order under the choice of the frame \cite{DeGroot1980Book,Bemfica2018PhRvD,Kovtun2019JHEP}. 

In order to identify these coefficients let us formulate the entropy production law.
The expression for the canonical entropy current is based on the first law of thermodynamics \cite{Rezzolla2013book}
\begin{subequations}
\allowdisplaybreaks
\begin{eqnarray}
    TS^\mu&=&PU^\mu+U_\nu {\cal T}^{\mu\nu} -\sum_a \mu_a j^\mu_{(a)} \label{eq:S_canon}\\
    &=& TSU^\mu +U_\nu\Delta {\cal T}^{\mu\nu}-\sum_a \mu_a \Delta j^\mu_{(a)}.\label{eq:S_canon_diss}
\end{eqnarray}
\end{subequations}
In principle, in a non-equilibirum state the true entropy current, which 
strictly obeys the second law of thermodynamics $\partial_\mu S^\mu \geqslant 0$, is not necessarily given by eq.~(\ref{eq:S_canon}), but it can also contain correction terms
\cite{Bhattacharyy2014JHEP,Bhattacharyy2014JHEPa,Becattini2019PhRvD,Dowling2020PhRvD}. However, in the domain of validity of the first-order theory, one can remain with eq.~(\ref{eq:S_canon}). Then the inequality $\partial_\mu S^\mu \geqslant 0$,  in principle, becomes approximate \cite{Kovtun2019JHEP,Gavassino2020PhRvD,Bemfica020arXiv}.

\new{Using eqs.~(\ref{eq:j_cons}), (\ref{eq:Tmunu_cons}), (\ref{eq:j_decomp}), (\ref{eq:Tmunu_decomp}), and (\ref{eq:thermodynamics}), the}  divergence of the canonical entropy current \new{can be written as} 
\begin{equation}\label{eq:zeta_gen}
    \varsigma\equiv \partial_\mu S^\mu = \Delta {\cal T}^{\mu\nu}\partial_\mu\frac{U_\nu}{T}-\sum_a\Delta j_{(a)}^\mu \partial_\mu\frac{\mu_a}{T}-\frac{1}{T}E^\lambda J_\lambda,
\end{equation}
which is valid in any frame \cite{Bemfica020arXiv}. Taking into account the discussion above, we further assume that the matching conditions (\ref{eq:matching}) hold, but do not yet fix $U^\mu$.

The term $E^\lambda\equiv U_\nu F^{\lambda\nu}$ in eq.~(\ref{eq:zeta_gen}) is the electric field four-vector resulting from the decomposition of the  electromagnetic field tensor 
\begin{equation}
    F^{\mu\nu}=E^\mu U^\nu - E^\nu U^\mu +\epsilon^{\mu\nu\alpha\beta} U_\alpha B_\beta,\label{eq:Fmunu}
\end{equation}
where $B^\mu\equiv \frac{1}{2}\epsilon^{\mu\nu\alpha\beta} U_\nu F_{\alpha\beta}$ is the  magnetic field four-vector. We assume that the magnetic field is large, i.e. $B={\cal O}(1)$, while the electric field is induced by the dissipative processes, so $E={\cal O}(\partial)$.

The equations of motion are obtained by employing eqs.~(\ref{eq:grad_decomp}) and (\ref{eq:j_decomp}) in  eq.~(\ref{eq:j_cons})  and by contracting eq.~(\ref{eq:Tmunu_cons}) with $U_\nu$ and $\Delta^\lambda_\nu$  \cite{Rezzolla2013book}. One obtains
\begin{subequations}\label{eq:laws_of_motion}
\allowdisplaybreaks
\begin{eqnarray}
Dn_a&+&n_a \nabla_\mu U^\mu = - \partial_\mu \Delta j_{(a)}^\mu, \label{eq:particle_continuity}\\
    D{\cal E} &+& H\nabla_\mu U^\mu = - U_\nu\partial_\mu\Delta {\cal T}^{\mu\nu}+E^\sigma J_\sigma,\label{eq:energy_cons}\\
    H DU^\lambda &-&\nabla^\lambda P-\Delta^{\lambda}_\nu F^{\nu\sigma} J_\sigma = -
    \Delta^{\lambda}_{\nu}\partial_\mu\Delta {\cal T}^{\mu\nu}.\label{eq:NavierStokes}
\end{eqnarray}
\end{subequations}
At the first order in gradients, right-hand sides of eqs.~(\ref{eq:laws_of_motion}) vanish. 

To proceed further, let us define particle fractions as $y_a\equiv n_a/n$ and introduce diffusion currents $i^\mu_{(a)}$ via 
\begin{equation}\label{eq:diff_current_def}
    j_{(a)}^\mu=y_a j^\mu + (\Delta j^\mu_{(a)} - y_a\Delta j^\mu) \equiv y_a j^\mu + i^\mu_{(a)}.
\end{equation}
Notice that at first order $i^\mu_{(a)}$ are independent of the choice of $U^\mu$ (and of the frame selection in general).
The diffusion currents sum to zero:
\begin{equation}\label{eq:diff_curr_sum}
    \sum_{a=1}^r i^\mu_{(a)} = 0.
\end{equation}
We also assume that the system is electrically neutral in the LRF, $\sum_a q_a n_a=0$. The electromagnetic current is then 
\begin{equation}\label{eq:el_curr_diff_curr}
    J^\mu=\sum_a q_a \Delta j_{(a)}^\mu = \sum_a q_a i^\mu_{(a)}.
\end{equation}
\new{Due to charge neutrality, the electromagnetic current is orthogonal to the fluid velocity, i.e. $U_\mu J^\mu = 0$. This retains only the magnetic part of the Lorentz force, $\Delta^\lambda_\nu F^{\nu\sigma} J_\sigma = U_\alpha \epsilon^{\alpha\lambda\mu\nu}J_\mu B_\nu$, in the left-hand side of eq.~(\ref{eq:NavierStokes}).}
Notice that in multifluid hydrodynamics, different `fluids' might have different LRFs, and the definition of local charge neutrality becomes more subtle, see, e.g., \cite{Metens1990PhFlB}.

 The dissipative correction to the energy-momentum tensor $\Delta {\cal T}^{\mu\nu}$ can be further decomposed as
\begin{equation}\label{eq:Tmunu_diss_decomp}
    \Delta {\cal T}^{\mu\nu}=U^\mu {\cal Q}^\nu +U^\nu {\cal Q}^\mu + \Pi^{\mu\nu},
\end{equation}
where ${\cal Q}^\mu$ is the dissipative part of the energy flux
\begin{equation}\label{eq:en_flux}
    {\cal Q}^\mu=U_\nu \Delta {\cal T}^{\nu\lambda}\Delta^\mu_\lambda\equiv I^\mu_{(q)}+h\Delta j^\mu=J^\mu_{(q)}+\sum_a h_a \Delta j^\mu_{(a)},
\end{equation}
$I^\mu_{(q)}$ is the heat flux (which is frame-independent in first order), $h\equiv H/n$ is the specific enthalpy per particle, $J^\mu_{(q)}$ is the reduced heat flux, $J^\mu_{(q)}=I^\mu_{(q)}-\sum_a h_a i^\mu_{(a)}$, and $h_a$ is the specific enthalpy of the particle species $a$, so that $h=\sum_a y_a h_a$. Notice that the specific enthalpies cannot be defined in the phenomenological single-fluid theory. They, however, will arise naturally from the Fermi-liquid kinetic theory \new{(discussed below)}.

The term $\Pi^{\mu\nu}$ in eq.~(\ref{eq:Tmunu_diss_decomp}) is the non-equilibrium part of the stress tensor which in view of eq.~(\ref{eq:matching}) is 
\begin{equation}\label{eq:Pimunu_def}
  \Pi^{\mu\nu}=\Delta^\mu_{\lambda} {\cal T}^{\lambda\delta} \Delta^{\nu}_\delta+  P\Delta^{\mu\nu}=\Delta^\mu_{\lambda}\ \Delta{\cal T}^{\lambda\delta} \Delta^{\nu}_\delta.
\end{equation}

Using the decompositions (\ref{eq:Tmunu_diss_decomp})--(\ref{eq:en_flux}), the entropy current in (\ref{eq:S_canon_diss}) can be rewritten in the following forms 
\begin{subequations}\label{eq:S_canon_firstorder}
\allowdisplaybreaks
\begin{eqnarray}
S^\mu&=&\frac{S}{n}j^\mu +\frac{1}{T}\left(I^\mu_q-\sum_a\mu_a i^\mu_{(a)}\right)\\
&=&\sum_a s_a j^\mu_{(a)}+\frac{J^\mu_{(q)}}{T}
\end{eqnarray}
\end{subequations}
which do not depend on $U^\mu$ \cite{DeGroot1980Book}. The quantities $s_a$ in eq.~(\ref{eq:S_canon_firstorder}) are partial specific entropies to which all said above about $h_a$ applies as well. 

Let us now rewrite the entropy production rate (\ref{eq:zeta_gen}) with the help of eqs.~(\ref{eq:Tmunu_diss_decomp}), (\ref{eq:en_flux}), (\ref{eq:NavierStokes}),  and (\ref{eq:thermodynamics}) in several equivalent forms \cite{vanErkelens1977PhyA}
\begin{subequations}\label{eq:zetas}
\allowdisplaybreaks
\begin{eqnarray}
    T\varsigma 
    &=&\Pi^{\mu\nu} \nabla_\mu U_\nu -I_{(q)}^\mu\left( \frac{\nabla_\mu T}{T} - DU_\mu\right)\nonumber\\
    &&-
   T \sum_a i^\mu_{(a)}\nabla_\mu\frac{\mu_a}{T}+\frac{1}{n}j_\mu F^{\mu\lambda}J_\lambda \label{eq:zeta1}\\
    &=&\Pi^{\mu\nu} \nabla_\mu U_\nu -J_{(q)}^\mu\left( \frac{\nabla_\mu T}{T} - DU_\mu\right)\nonumber\\
    &&-
    \sum_a i^\mu_{(a)}\left(d_{(a)\mu}^{\mathrm{E}}-\frac{h_a}{hn}\Delta_{\mu\nu}F^{\nu\lambda} J_\lambda\right)\label{eq:zeta2}\\
    &=&\Pi^{\mu\nu} \nabla_\mu U_\nu -J_{(q)}^\mu\left( \frac{\nabla_\mu T}{T} - DU_\mu\right)\nonumber\\
    &&-
    \sum_a \Delta j^\mu_{(a)}\left(d_{(a)\mu}-\frac{h_a}{hn}\Delta_{\mu\nu}F^{\nu\lambda} J_\lambda\right). \label{eq:zeta3}
\end{eqnarray}
\end{subequations}
In order to derive these equations, we expressed the acceleration $DU_\mu$ from (\ref{eq:NavierStokes}) taken at ${\cal O}(\partial)$ (i.e. neglecting its right-hand side). However, we traditionally kept \new{the acceleration term} in the combination $T^{-1}\nabla_\mu T-DU_\mu$ which multiplies the heat fluxes in eqs.~(\ref{eq:zetas}).

In eqs.~(\ref{eq:zetas})  we introduced four-vectors
\begin{subequations}
\begin{equation}\label{eq:dEmu}
    d_{(a)}^{\mathrm{E}\mu}=\widetilde{d}_{(a)}^\mu-\frac{q_a}{n}j_\nu {F^{\nu\mu}},
\end{equation}
\begin{equation}\label{eq:dmu}
    d_{(a)}^\mu = \widetilde{d}_{(a)}^\mu+ q_aE^\mu,
\end{equation}
\end{subequations}
where
\begin{equation}\label{eq:dtilde}
    \widetilde{d}_{(a)}^\mu = T\nabla^\mu\frac{\mu_a}{T}
    -T\frac{h_a}{h}\sum_b y_b\nabla^\mu \frac{\mu_b}{T}.
\end{equation}
All these vectors are linearly dependent since
\begin{equation}\label{eq:dtrue_lindep}
    \sum_{a=1}^{r}y_a \widetilde{d}_{(a)}^{\mu} =0,
\end{equation}
and, owing to charge neutrality,
\begin{equation}
    \sum_{a=1}^r  y_ad^{\mathrm{E}\mu}_{(a)}=\sum_{a=1}^r  y_ad_{(a)}^{\mu}=0.
\end{equation}

The first eq.~(\ref{eq:zeta1}) is manifestly frame independent, since it contains frame-independent fluxes. 
The second eq.~(\ref{eq:zeta2}) is rewritten using the reduced heat flux, since it naturally emerges from the kinetic theory and is also frame-independent. 

Notice that the thermodynamic forces coupled to the diffusion currents contain [the last term of eq.~(\ref{eq:zeta2})] the electric field in the specific frame [namely, particle, or the Eckart one, cf. eqs.~(\ref{eq:dEmu}) and (\ref{eq:dmu})]. 
Notice also that the $\Delta_{\mu\nu}$ projector in the magnetic term can be dropped since $i^\mu_{(a)}$ is orthogonal to $U^\mu$.

Consider, finally, the third form of the entropy production equation (\ref{eq:zeta3}). Here dissipative particle  current and the vector $d^\mu_{(a)}$ depend on the choice of frame (but not the entropy production itself). However, eq.~(\ref{eq:zeta3})
 has the structure that is  readily obtained from the kinetic theory by integrating the corresponding single-particle transport equations, see below. In the Eckart frame [eq.~(\ref{eq:EckartFrame})] eqs.~(\ref{eq:zeta2}) and (\ref{eq:zeta3}) coincide. This suggest that it is convenient to work in the Eckart frame for the calculation of the transport coefficients.

Equation (\ref{eq:zeta2}) has a form of bilinear combinations of thermodynamic forces and fluxes 
\begin{equation}\label{eq:entropy_bilinear}
    T\varsigma = -\sum_{k=\zeta,\eta,q,Da} {\cal Y}_k\cdot {\cal X}_k, 
\end{equation}
where ${\cal X}_k$ are thermodynamic forces, corresponding to different transport phenomena. Namely,
\begin{subequations}
\begin{equation}
    {\cal X}_\zeta=\nabla^\mu U_\mu \label{eq:Xzeta}
\end{equation}
corresponds to the bulk viscosity ($k=\zeta$), 
\begin{equation}
    {\cal X}^{\mu\nu}_\eta=\frac{1}{2}\left(\Delta^\mu_\sigma\Delta^\nu_\tau+\Delta^\nu_\sigma\Delta^\mu_\tau-\frac{2}{3}\Delta^{\mu\nu}
    \Delta_{\sigma\tau}\right)\nabla^\sigma U^\tau\label{eq:Xeta}
\end{equation}
corresponds to the shear viscosity ($k=\eta$),
\begin{equation}
    {\cal X}^\mu_{q}= \frac{\nabla^\mu T}{T} - DU^\mu\label{eq:Xq}
\end{equation}
corresponds to thermal conductivity ($k=\kappa$), and \new{$r$ thermodynamic forces ($k=Da$ with $a=1\dots r$)}
\begin{equation}
    {\cal X}^\mu_{Da}=d^{\mathrm{E}\mu}_{(a)}-\frac{h_a}{hn}\Delta^{\mu\nu}F_{\nu\lambda} J^\lambda\label{eq:Xa}
\end{equation}
drive the diffusion \new{processes}.
\end{subequations}
In eq.~(\ref{eq:entropy_bilinear}), ${\cal Y}_k$ are the corresponding thermodynamic fluxes 
\begin{subequations}
\begin{eqnarray}
    {\cal Y}_\zeta&=&-\frac{1}{3}\Delta_{\mu\nu}\Pi^{\mu\nu}\equiv \Pi ,\label{eq:Yzeta}\\
    {\cal Y}^{\mu\nu}_\eta&=&-\left(\Delta^\mu_\sigma\Delta^\nu_\tau-\frac{1}{3}\Delta^{\mu\nu}
    \Delta_{\sigma\tau}\right)\Pi^{\sigma\tau}\equiv -\Pi^{\langle\mu\nu\rangle},\label{eq:Yeta}\\
    {\cal Y}^\mu_{q}&=&J^\mu_{(q)},\label{eq:Yq}\\
    {\cal Y}^\mu_{Da}&=&i^\mu_{(a)}.\label{eq:Ya}
\end{eqnarray}
\end{subequations}
In eqs.~(\ref{eq:Yzeta})--(\ref{eq:Yeta}) we decomposed the stress tensor in the isotropic part (described by the viscous pressure $\Pi$) and the traceless symmetric part (shear stress tensor), for which we use the short-hand notation $\Pi^{\langle \mu\nu\rangle}$.

The irreversible thermodynamics states that the fluxes are the linear combinations of forces (and vice versa), namely
\begin{equation}\label{eq:transp_onsager}
    {\cal Y}_k=-\sum_{k'}\hat{L}_{kk'} {\cal X}_{k'},
\end{equation}
where $\hat{L}_{kk'}$ is the matrix of \new{the transport coefficients.}
Here and up to the end of this section we omit the tensor component indices for brevity, and use hats to stress tensor character of corresponding  quantities (in order to eliminate the explication of tensor component indices).
Entropy production now is given by the quadratic form on the thermodynamic forces ${\cal X}_k$. The transport coefficients matrix $\hat{L}_{kk'}$ needs to be semi-positive definite for the second law of thermodynamics to be valid. 
In addition, the matrix $\hat{L}_{kk'}$ obeys Onsager reciprocal relations \cite{Landau5eng}
\begin{equation}\label{eq:Onsager_reci}
    \hat{L}_{kk'}(\bm{B})=\pm \hat{L}^\mathrm{T}_{k'k}(-\bm{B}),
\end{equation}
where superscript T means transposition with respect to the tensor multiindex, and  one needs to use the plus sign if the thermodynamic forces $k$ and $k'$ have the same time-reversal symmetry, and minus in other case (i.e. for the cross-coefficients between viscosity and diffusion; however these coefficients are zero \new{due to inversion symmetry}).

\new{
The Curie principle states that in the isotropic media the \new{thermodynamic} fluxes \new{and forces} of different tensor dimensions do not mix \cite{deGrootMazur1984}. In the presence of the magnetic field, the system possesses lower-degree axial symmetry and one can write 
\begin{eqnarray}
    {\cal Y}_\zeta&=&-\zeta {\cal X}_\zeta-\hat{\zeta_1}\hat{{\cal X}}_\eta\label{eq:bulk_def}\\
    \hat{{\cal Y}}_{\eta}&=&2\hat{\eta} \hat{{\cal X}}_\eta-\hat{\zeta_1}{\cal X}_\zeta\label{eq:shear_def},
\end{eqnarray}
where $\zeta$ is the bulk viscosity coefficient, $\hat{\eta}$ is the shear viscosity which, in general, is a four-rank tensor, and $\hat{\zeta}_1$ is the cross-term viscosity coefficient which is a traceless symmetric second-rank tensor. Due to inversion symmetry there is no cross terms between the vector fluxes and the viscous forces and vice versa. Below (sec.~\ref{sec:tensor}) we will see that for a wide class of systems including NS cores one may consider $\hat{\zeta}_1=0$ due to a particular form of the Lorentz force \cite{Landau10eng}.} \new{If more general interaction with magnetic field is considered (e.g. the particles' magnetic moments are taken into account), one may have $\hat{\zeta}_1 \ne 0$.}

For the vector fluxes the general situation is more cumbersome. One can expand eq.~(\ref{eq:transp_onsager}) in the explicit form
\begin{subequations}\label{eq:transp_vector}
\begin{eqnarray}
    J_q &=&-\hat{L}_{qq} {\cal X}_q - \sum_{a} \hat{L}_{qa} {\cal X}_{Da},\label{eq:Jq_lin}\\
    i_{(a)}&=&-\hat{L}_{aq} {\cal X}_q - \sum_{b} \hat{L}_{ab} {\cal X}_{Db},\label{eq:ia_lin}
\end{eqnarray}
\end{subequations}
which contains thermal conductivity, diffusion, and thermodiffusion processes. At the level of first-order irreversible thermodynamics there is a freedom to choose the thermodynamic forces and fluxes in various ways by changing a basis of the expansion of the entropy production as a quadratic form. For instance, it is possible to use thermodynamic fluxes ${\cal Y}_k$ in place of forces, and take thermodynamic forces ${\cal X}_k$ as the fluxes. Notice that the thermodynamic force ${\cal X}_{Da}$ as defined in eq.~(\ref{eq:Xa}) contains the term proportional to $J$, which in turn is the linear combination of the diffusion currents, \new{i.e.} ${\cal Y}_{Da}$. This means that eq.~(\ref{eq:ia_lin}) can be viewed as a system of linear equations for $i_{(a)}$. Solving this system, one expresses the diffusion currents $i_{(a)}$  via the linear combination of the thermal force ${\cal X}_q$ and the true diffusion forces $\widetilde{{\cal X}}_{Da}$, which do not contain the magnetic field term, i.e.
\begin{equation}\label{eq:Xa_red}
    \widetilde{{\cal X}}_{Da}=
    d^{\mathrm{E}}_{(a)}.
\end{equation}
The entropy production is still given by the quadratic form with some different transport coefficient matrix related to the $\hat{L}_{kk'}$. We do not write the explicit transformation between these formulations here. Instead we rewrite the diffusion law in the so-called Stephan-Maxwell form. To do this we first rewrite the linear transport laws in the form  where  the set of thermodynamic forces contains ${\cal X}_q$ and the diffusion currents  $i_{(a)}$, while the thermodynamic fluxes are ${\cal X}_{Da}$ and still $J_q$: 
\begin{subequations}\label{eq:kin_vec_SM1}
\begin{eqnarray}
    J_q &=&-T \hat{\kappa} {\cal X}_q - \sum_{a} \hat{R}_{qa} i_{(a)},\label{eq:Jq_SM1}\\
    {{\cal X}}_{Da}&=&-\hat{R}_{aq} {\cal X}_q - \sum_{b} \hat{R}_{ab} i_{(b)},\label{eq:Xa_SM1}
\end{eqnarray}
\end{subequations}
where the thermal conductivity tensor $\hat{\kappa}$ is introduced, $\hat{R}_{qa}(\bm{B})=\hat{R}^{\mathrm{T}}_{aq}(-\bm{B})$, and $\hat{R}_{ab}(\bm{B})=\hat{R}^{\mathrm{T}}_{ba}(-\bm{B})$ is the conjugate set of the tensor transport coefficients.
Multiplying eq.~(\ref{eq:Xa_SM1}) by $n_a$, summing over $a$, and employing eqs.~(\ref{eq:diff_curr_sum}--\ref{eq:el_curr_diff_curr}) and eq.~(\ref{eq:dtrue_lindep}), one observes 
\begin{equation}
    \sum_b q_b\hat{\Delta}\hat{F} i_{(b)} = \sum_{a} \hat{R}_{aq} {\cal X}_q + \sum_{ab} n_a \hat{R}_{ab} i_{(b)}.
\end{equation}
Since this equality should be valid for any ${\cal X}_q$ and any $i_{(b)}$ satisfying eq.~(\ref{eq:diff_curr_sum}),  the coefficients $\hat{R}_{ab}$, $\hat{R}_{aq}$ should satisfy
\cite{zhdanov2002transport} 
\begin{subequations}
\begin{equation}\label{eq:Raq_rank}
    \sum_a n_a\hat{R}_{aq}=0,
\end{equation}
\begin{equation}\label{eq:R_rank}
    \sum_a n_a\hat{R}_{ab}-q_b\hat{\Delta}\hat{F}=n\hat{R},
\end{equation}
\end{subequations}
where the tensor $\hat{R}$ does not depend on $b$.
Now eq.~(\ref{eq:Xa_SM1}) can be rewritten in the Stephan-Maxwell form
\begin{eqnarray}
    n_a{{\cal X}}_{Da}&=&-n_a\hat{R}_{aq} {\cal X}_q + \sum_{b} \hat{J}_{ab}\left(\frac{ i_{(b)}}{n_b}-\frac{ i_{(a)}}{n_a}\right)\nonumber\\
    &-&q_a\hat{\Delta}\hat{F}i_{(a)},\label{eq:Xa_SM}
\end{eqnarray}
where
\begin{equation}
    \hat{J}_{ab}=-n_an_b(\hat{R}_{ab}-\hat{R})
\end{equation}
is the set of the $r(r-1)/2$ independent friction tensor coefficients also known as the momentum \new{transfer} rates. 
Since according to eq.~(\ref{eq:Raq_rank}) there is $r-1$ independent thermal diffusion coefficient $\hat{R}_{aq}$, there are $r(r+1)/2$ tensors in total which describe the heat and particle diffusion. Notice that the magnetic term in the second line in eq.~(\ref{eq:Xa_SM}) does not contribute to the entropy production rate. Indeed, according to eq.~(\ref{eq:zeta2}) entropy production contains the product $i_{(a)}{\cal X}_{Da}$. The magnetic term contribution from (\ref{eq:Xa_SM}) is then \new{$q_ai_{(a)}\hat{\Delta}\hat{F}i_{(a)}=q_ai_{(a)}\hat{F}i_{(a)}$} which vanishes due to asymmetry of the electromagnetic field tensor. 

In the isotropic case (in our context -- in the absence of magnetic field) all transport coefficients described above are scalars. In presence of the magnetic field, they become tensors with a certain symmetry with respect to the magnetic field direction. We discuss this structure in detail in sec.~\ref{sec:tensor} based on the irreducible spherical tensor formalism.

In the terrestrial settings transports coefficients defined in the effective hydrodynamics theories can be (at least in principle) obtained from experiment. In the astrophysical settings this is more complicated. Therefore the reliable values for transport coefficients should be derived on the basic of some microscopic theory. 

Within the linear response regime, the expressions for the transport coefficients can be given by the Kubo-type formulae (e.g., 
\cite{Mahan93book}). Appropriate expressions for the relativistic hydrodynamics (including magnetized case) can be found, e.g., in \cite{Huang2011AnPhy,Finazzo2016PhRvD,Hernandez2017JHE,HarutyunyanParticles2018}  and references therein. The practical analytical calculations of the transport coefficients in this approach can be complicated since they require resummation of infinite number of diagrams. 

Another alternative that works in the weak-coupling limit is the kinetic theory framework. Here we assume that the low-temperature conditions in NS cores \new{allow} to represent matter as a mixture of weakly interacting quasiparticles and describe it within the Landau Fermi-liquid theory \cite{BaymPethick}. Then the transport coefficients can be calculated from the kinetic theory for Fermi-liquids, which we describe below.

\section{Transport theory of Landau Fermi-liquids}\label{sec:LandauFL}
\subsection{General setup and definitions}\label{sec:LandauFL_gen}

Landau Fermi-liquid theory does not consider the ground state  of the system. In contrast it deals with the slightly excited states and assumes that these states of the condensed system are described in terms of weekly interacting quasiparticles which have a one-to-one correspondence to the actual particle states of the system. 
The Landau Fermi-liquid theory was initially formulated in the non-relativistic setup. The relativistic generalization closely following the original consideration was constructed for the single-component fluid by Baym and Chin \cite{BaymChin1976NuPhA} (see also the generalization for mixtures \cite{Gusakov2009PhRvC}). The manifestly covariant generalization exists \cite{vanWeert1984PhLA,vanWeert1985PhyA,vanWeert1986PhyA} 
based on the expansion of the pressure variation instead of the energy density variation.

For the purpose of the transport coefficients calculation, it is easiest to work in the rest frame of the fluid. Since the gradients of the hydrodynamic velocity enter the equations for the thermodynamic forces, one also considers the laboratory frames that are close to the LRF, i.e. which have non-relativistic velocities $V\ll 1$. In this case the formulation by Baym and Chin \cite{BaymChin1976NuPhA} is natural. After necessary velocity gradients are identified in equations, one can set $V=0$. The resulting equations are similar to their non-relativistic counterparts, having in difference mainly the relativistic quasiparticle dispersion law \cite{BaymChin1976NuPhA}, see also \cite{DeGroot1980Book}.

The quasiparticle states are characterized by the quasiclassical distribution functions $f_a(\bm{p},\bm{r},t)$. Below we will omit the coordinate dependence of the distribution functions for brevity. Distribution functions also depend on the spin quantum numbers (and other quantum numbers, if present). We do not consider here interesting spin-dependent effects, and restrict ourselves to the spin-unpolarized state. Let us abbreviate 
\begin{equation}
\int_{\bm p}\equiv\sum_{\sigma}\int \frac{d^3{\bm p}}{(2\pi)^3} \, ,
\end{equation}
where $\sigma$ is a spin state index.
Notice that since we work in the LRF, we do not introduce the Loretz-invariant volume element here. This allows to consider relativistic and non-relativistic cases on the same footing. 

The particle densities (zero components of the particle currents) are assumed to be given by the integration of the quasiparticles distribution functions 
\begin{equation}\label{eq:part_dens}
    j^0_{(a)}=\int_{\bm p} f_a(\bm{p}).
\end{equation}
In principle, one usually defines the family of the space-like hyperplanes orthogonal to some time-like vector $\pi^\mu$, so that the densities of hydrodynamic variables are defined as, e.g. $\pi_\mu j^\mu_{(a)}$ \cite{Zubarev1974,Hayata2015PhRvD}. We leave this generalization aside and take $\pi^\mu=(1,0,0,0)$.

The momentum density ${\cal T}^{i0}$ is also defined to be a known functional of  $f_a(\bm{p})$
\begin{equation}\label{eq:mom_dens}
    {\cal T}^{i0} =\sum_a\int_{\bm{p}} p^i f_a(\bm{p}).
\end{equation}
In contrast, the energy density ${\cal T}^{00}$ is considered as an unknown functional of the set of distribution functions $\{f_a(\bm{p})\}$, $a=1,\dots, r$. \new{However} for the small departures from the equilibrium ground state, the 
\new{variations in} energy-momentum density can be written as
\begin{equation}\label{eq:Tmu0_FL}
    \delta {{\cal T}^{\mu 0}} =\sum_a\delta {\cal T}^{\mu 0}_{(a)}= \sum_{a}
    \int_{\bm{p}} p^\mu_{(a)}(\bm{p}) \delta f_a(\bm{p}), 
\end{equation}
where
\begin{equation}\label{eq:qp_fourmomentum}
    p^\mu_{(a)}\equiv(\varepsilon_a(\bm{p}),\bm{p})
\end{equation}
and $\varepsilon_a(\bm{p})$ is the quasiparticle energy, which itself is the functional of the distribution functions $\varepsilon_a(\bm{p})=\varepsilon_a(\bm{p})[\{f_b(\bm{p})\}]$. The variational derivative of the quasiparticle energies with respect to the distribution functions
\begin{equation}\label{eq:Landau_parameters}
    \delta\varepsilon_a(\bm{p})=\sum_{b}\int_{\bm{p}'} f_{ab}(\bm{p},\bm{p}') \delta f_b(\bm{p}')
\end{equation}
defines the Landau Fermi-liquid interaction $f_{ab}(\bm{p},\bm{p}')$.  

Entropy density of (quasi)particle species $a$ has a purely combinatorial nature and is given by the same expression as for the non-interacting gas
\begin{equation}\label{eq:entropy_kin}
    S_a=-\int_{\bm{p}} \Bigl[ f_a(\bm{p})\log f_a(\bm{p}) + (1-f_a(\bm{p}))\log(1-f_a(\bm{p})) \Bigr].
\end{equation}

The local equilibrium state descibed by a set of the local equilibrium distribution functions $\{f_a^{\mathrm{l.e.}}\}$ is obtained by maximizing total entropy density subject to constrains $j_{(a)}^{0} [f^{\mathrm{l.e}}]= j_{(a)}^{0}$ and ${\cal T}^{\mu 0}[f^{\mathrm{l.e}}]= {\cal T}^{\mu 0}$, where $j^0_{a}$ and ${\cal T}^{\mu 0}$ are the actual local values of particle density and energy-momentum density, respectively, which are well-defined as expectation values of the corresponding quantum-mechanical operators for a given state of the system. This conditional extremum problem amounts to maximization of  the functional 
\new{\begin{eqnarray}
    \sum_a &&\left[S_a[f^{\mathrm{l.e}}]- \alpha_a \left( j_{(a)}^{0}[f^{\mathrm{l.e}}]- j_{(a)}^{0}\right)\right.\nonumber\\
    &&\left.-\beta_\mu\left( {\cal T}_{(a)}^{\mu 0}[f^{\mathrm{l.e}}]- {\cal T}_{(a)}^{\mu 0}\right) \right],\label{eq:entropy_le}
\end{eqnarray}
where $\alpha_a$ and $\beta^\mu$ are Lagrange multipliers, which are identified as \cite{Zubarev1974} 
\begin{equation}\label{eq:LagrangeMultipliers}
     \alpha_a=-\frac{\mu_a}{T},\quad
    \beta^\mu=\frac{U^\mu}{T}.
\end{equation}
}
Equating to zero the variation of eq.~(\ref{eq:entropy_le}) over distribution functions $f_a$ results in the local equilibrium distribution functions
\begin{equation}\label{eq:f_l.e.}
    f_a^{\mathrm{l.e.}}=f_F\left(\frac{p^\mu_{(a),\mathrm{l.e.}} U_\mu(\bm{r},t)-\mu_a(\bm{r},t)}{T(\bm{r},t)}\right),
\end{equation}
\new{where 
\begin{equation}\label{eq:FFermi}
    f_{\mathrm{F}}(x)=\left[\exp\left(x\right)+1\right]^{-1}
\end{equation}
is the Fermi function.}
Notice that here the local equilibrium dispersion law appears in  $p^{\mu}_{(a),\mathrm{l.e.}}=(\varepsilon_a^{\mathrm{l.e.}}(\bm{p}),\bm{p})$, which itself is the functional of eq.~(\ref{eq:f_l.e.}).

Using equation (\ref{eq:LagrangeMultipliers}) in variation of eq.~(\ref{eq:entropy_le}) results in the thermodynamic relation
\begin{equation}
    U_\mu \mathrm{d} {\cal T}^{\mu 0}=T\mathrm{d}S+\sum_a \mu_a \mathrm{d} j^0_{(a)},
\end{equation}
\new{where $S=\sum_a S_a$,}
which when written in the LRF reduces to  eq.~(\ref{eq:SecondLaw}).

Fermi-liquid theory describes low-temperature systems close to the $T=0$ ground state. In this case eq.~(\ref{eq:f_l.e.}) reduces to the Heaviside step function 
\begin{equation}
f_a^{\mathrm{l.e.}}(\bm{p})=\mathrm{\Theta}(p_{{F}a} - p),    
\end{equation}
where 
\begin{equation}\label{eq:pFa_def}
    p_{\mathrm{F}a}=(3\pi^2 n_a)^{1/3}
\end{equation}
is \new{the} quasiparticle species $a$ Fermi momentum.
\new{This means that all states with $p<p_{\mathrm{F}a}$} are occupied and those with $p>p_{\mathrm{F}a}$ are vacant. For small perturbations from equilibrium, the distribution function varies only in the  vicinity of the Fermi surface. One defines the quasiparticle Fermi velocity
\begin{equation}\label{eq:vFa_def}
    v_{\mathrm{F}a}=\left(\frac{\partial\varepsilon_a^{\mathrm{l.e.}}(p)}{\partial p}\right)_{p=p_{\mathrm{F}a}}
\end{equation}
and (Landau) effective mass on the Fermi surface
\begin{equation}\label{eq:ma_eff}
    m_a^*=\frac{p_{\mathrm{F}a}}{v_{\mathrm{F}a}}.
\end{equation}

The evolution of the distribution function is described by the Landau-Boltzmann transport equation \cite{BaymChin1976NuPhA,BaymPethick}
\begin{eqnarray}\label{eq:kin_eq}
    \frac{\partial f_a }{\partial t} +\frac{\partial \varepsilon_a}{\partial{\bm{p}}}\nabla f_a &-& 
    \left(\frac{\partial \varepsilon_a}{\partial{\mathbf{r}}}-\bm{F}_a\right)
    \nabla_{\bm{p}} f_a
    = I_a[\left\{f_b\right\}],
    \end{eqnarray}
where the important difference from the Boltzman equation for a gas \cite{DeGroot1980Book,CercignaniKremer2002book} is contained in the appearance of the ${\partial \varepsilon_a}/\partial{\mathbf{r}}$ term. In eq.~(\ref{eq:kin_eq}), $\bm{F}_a$ is the external force (not included in the miscroscale mean field) which we here take as the Lorentz force
\begin{equation}\label{eq:Lorentz}
    \bm{F}_a=q_a\left(\bm{E}+\left[\bm{v}_a\times\bm{B}\right]\right),
\end{equation}
where $\bm{v}_a=\partial\varepsilon_a/\partial\bm{p}$. Finally, the term $I_a[\left\{f_b\right\}]$ is the collision integral for the quasiparticle species $a$ which describes the change of the distribution function due to collisions and depends in principle on the full set of the distribution functions, $b=1,\dots, r$.

Transport eq.~(\ref{eq:kin_eq}) allows to derive a general equation of transfer for any state variable $\psi$.
Introducing
\begin{equation}\label{eq:kin_average}
    \langle \psi \rangle = \int_{\bm{p}} \psi f_a,
\end{equation}
and integrating (\ref{eq:kin_eq}) multiplied by $\psi$ one obtains
\begin{eqnarray}\label{eq:gen_trans}
    \partial_\mu \langle v^\mu_a \psi\rangle &=& \frac{\partial\langle\psi\rangle}{\partial t} + 
    \nabla \langle \bm{v}_a\psi\rangle\nonumber\\
    &=&\left\langle\frac{\partial\psi}{\partial t}+\bm{v}_a\nabla\psi-\left(\nabla\varepsilon_a-\bm{F}_a\right)\frac{\partial \psi}{\partial \bm{p}}\right\rangle\nonumber\\
    &&+\int_{\bm{p}}\psi I_a,
\end{eqnarray}
where we assumed that $\nabla_{\bm{p}}\cdot\bm{F}_a=0$ like in the case of Lorentz force. The term $\langle v^\mu_a \psi \rangle$, where 
\begin{equation}\label{eq:v_qp}
    v^\mu_a=\frac{\partial\varepsilon_a}{\partial p^\mu_{(a)}}=(1,\bm{v}_a),
\end{equation}
gives the flux of the variable $\psi$ and the right-hand side of eq.~(\ref{eq:gen_trans}) is the quantity $\psi$ production (or source) term.

Setting $\psi=1$ in eq.~(\ref{eq:gen_trans}) results in the particle current conservation laws eqs.~(\ref{eq:j_cons}) with $j^\mu_{(a)}=\langle v^\mu_{a} \rangle$.
 Notice, that in general, $\bm{v}_a\neq \bm{p}_a/\varepsilon_a$ as holds in the free space. Here it is assumed that the collision integrals conserve particle numbers\new{, i.e.
 $
     \int_{\bm{p}} I_a =0
 $,
 } since we do not consider reactions. 
 
 Similarly the transfer equations for four-momenta $\psi=p^{\mu}_{(a)}$ summed over the particle species lead, with the help of eq.~(\ref{eq:Tmu0_FL}), to the energy-momentum tensor conservation law eq.~(\ref{eq:Tmunu_cons}) with the definition \cite{BaymPethick,BaymChin1976NuPhA}
\begin{equation}\label{eq:Tmunu_FL}
    {\cal T}^{\mu\nu}_{(a)}=\int_{\bm{p}} p^\mu v_a^\nu f_a(\bm{p})-g^{\mu\nu} 
    \left(\int_{\bm{p}} \varepsilon_a(\bm{p}) f_a(\bm{p})-{\cal T}^{00}_{(a)} \right).
\end{equation}
Notice that for the ideal realtivistic gas \cite{DeGroot1980Book,CercignaniKremer2002book} the last term is exactly zero. In deriving eq.~(\ref{eq:Tmunu_FL}), we assumed that the collisions conserve energy and momentum
\begin{equation}
    \sum_a  p^{\mu}_{(a)}I_a=0,
\end{equation}
i.e. there are no external scattering mechanisms and the energy-momentum  leakage due to emission processes is neglected. 
Substituting of the local equilibrium function eq.~(\ref{eq:f_l.e.}) into eq.~(\ref{eq:entropy_kin}), integrating by parts,  summing over particles, and using the definition (\ref{eq:Tmu0_FL}) one obtains in LRF the eq.~(\ref{eq:Enthalpy}) and, hence, the Gibbs-Duhem relation eq.~(\ref{eq:GibbsDuhem}).

Comparing eqs.~(\ref{eq:entropy_kin}) and (\ref{eq:kin_average}) one observes that the entropy transfer equation can be derived by setting  $\psi=\psi_S^{(a)}=-\log f_a +(1-f_a^{-1})\log(1-f_a)$. Current for $\psi_S^{(a)}$ is identified with the partial entropy current of the particle species $a$ and eq.~(\ref{eq:gen_trans}) results in 
\begin{equation}\label{eq:Sprod_full}
    \partial_\mu \langle v^\mu_{a} \psi_{S}^{(a)}\rangle \equiv \varsigma_a= \int_{\bm{p}}\log\left(\frac{1-f_a}{f_a}\right)\, I_a [\{f_b\}].
\end{equation}
The right-hand side, summed over particle species, gives the total entropy production, i.e. $\varsigma=\sum_a\varsigma_a$. It vanishes for the local equilibrium functions eq.~(\ref{eq:f_l.e.}) if the collision probabilities which enter the  collision integral also correspond to the local equilibrium state. 

The collision integral in the right-hand side of eq.~(\ref{eq:kin_eq}) contains contribution from  the binary quasiparticle collisions between all species and has the Uehling-Uhlenbeck form
\begin{equation}
    I_a[\{f_b\}]=\sum_b I_{ab},
\end{equation}
where
\begin{eqnarray}
I_{ab}[f] &=& -\frac{1}{1+\delta_{ab}}\int_{{\bm p}_{1'}}\int_{{\bm p}_{2}}\int_{{\bm p}_{2'}} w_{ab}({\bm p},{\bm p}_{2};{\bm p}_{1'},{\bm p}_{2'})\nonumber\\&\times&
\left[f_{1}f_{2}(1-f_{1'})(1-f_{2'})\right.\nonumber
\\
&&\left.
-(1-f_{1})(1-f_2)f_{1'}f_{2'}\right] \, , \label{boltz_collint}
\end{eqnarray}
where we abbreviated $f_1=f_a(\bm{p})$, $f_{1'}=f_a(\bm{p}_{1'})$, $f_2=f_b(\bm{p}_2)$, $f_{2'}=f_b(\bm{p}_{2'})$, and
the kernel $w_{ab}({\bm p},{\bm p}_{2};{\bm p}_{1'},{\bm p}_{2'})$ is the differential probability of the quasiparticle collisions.\footnote{In case of inelastic collisions (reactions), the final quasiparticle states can correspond to different particle species, i.e. $f_{1'}=f_c(\bm{p}_{1'})$ and $f_{2'}=f_d(\bm{p}_{2'})$ for a binary reaction $a+b\leftrightarrow c+d$. }
It is seen that the collision integral in this form vanishes for the local distribution functions eq.~(\ref{eq:f_l.e.}) again if the collision probabilities are calculated for the local equilibrium quasiparticles. However, in general, the local equilibrium functions do not solve the transport eq.~(\ref{eq:kin_eq}), since they do not  give zero in the driving term. 
Both sides of kinetic equation vanish in the global equilibrium state (for the global equilibrium distribution functions). Deviation of the local equilibrium state from the global equilibrium results in the dissipative processes that tend to eliminate these differences. 

 Unitary of the scattering \new{matrix} for binary collisions, \new{which enters $w_{ab}$}, leads to the Boltzmann $H$-theorem \cite{DeGroot1980Book,CercignaniKremer2002book}, i.e. to the entropy increase law $\varsigma\geqslant 0$ in eq.~(\ref{eq:Sprod_full})
for the collision integral eq.~(\ref{boltz_collint}).

The laws of dissipative hydrodynamics are derived from the kinetic theory by considering small deviations around the local equilibrium state. There exist two general methods that perform this expansion. One is the Grad's moments method \cite{Grad1949}  and other is the Chapman-Enskog expansion method \cite{ChapmanCowling1999}. 

The moments method is based on the expansion of the non-equilibrium distribution function in some orthogonal set constructed from $p^\mu$; the lowest order moments are the physical flows. The expansion is then truncated at a certain finite number of expansion coefficients (moments). 

The Chapman-Enskog expansion employs the small parameter, namely the Knudsen number $\mathrm{Kn}=\lambda/L$, where $\lambda$ is the typical mean free path or the microscopic scale, and $L$ is the typical scale of the state variables gradients, or the macroscopic scale. The distribution functions are then progressively expanded in  orders in $\mathrm{Kn}$.

Both methods \new{were derived for non-relativistic systems and} are extended to the relativistic sector e.g., \cite{DeGroot1980Book,CercignaniKremer2002book}. \new{Both approaches} suffer from certain limitations, see \cite{Denicol2014,Gabbana2020PhR} for a detailed discussion. In the non-relativistic case, at lowest order Grad's method and Chapman-Enskog methods give equivalent formulations, while this is not so in the relativistic case. The relativistic generalization of the Grad's moments method allowed \cite{Israel1979AnPhy} to construct the casual second-order hydrodynamic equations. On the other hand, the Chapman-Enskog formulation is asymptotically  correct (in small $\mathrm{Kn}$ limit). There are evidences from the numerical analysis of the solution of the relativistic Boltzmann equations, that the Chapman-Enskog procedure is favored, e.g.,  \cite{Gabbana2020PhR} and references therein, see,
however, \cite{GarciaPerciante2020JSP}.
The methods combining advantages of the both procedures are proposed in the non-relativistic (e.g., \cite{zhdanov2002transport}) and relativistic \cite{Denicol2014,Denicol2016arXiv} setup.

Since the Chapman-Enskog method provides asymptotically correct limit for  transport equations, the first-order transport coefficients can be reliably calculated in this approach. Below we employ the Chapman-Enskog method at the lowest (linear) order in $\mathrm{Kn}$.
To a certain extent, the Champan-Enskog procedure in a Fermi-liquid turns out to be similar to those for the relativistic gas kinetic theory, described in detail in, e.g.,  \cite{DeGroot1980Book,CercignaniKremer2002book,Denicol2014}.

\subsection{Chapman-Enskog procedure}\label{sec:Chapman}
In order to use the Chapman-Enskog procedure one needs to linearize the transport equation (\ref{eq:kin_eq}) around the equilibrium distribution function in terms which are progressively larger in powers of the Knudsen number 
\begin{equation}\label{eq:Chapman-Enskog}
    f_a=f_a^{\mathrm{eq}}+\delta f_a^{(0)}+\delta f_a^{(1)}+\dots
\end{equation}
where $f_a^{\mathrm{eq}}$ is the distribution function in global equilibrium and 
\begin{equation}\label{eq:deltaf0}
    \delta f_a^{(0)}=f_a^{\mathrm{l.e.}}-f^{\mathrm{eq}}_a
\end{equation}
is the difference between the local and global equilibrium distribution functions.
The quasiparticle energies are functionals of the distribution function and are subject to the similar expansion
\begin{equation}\label{eq:Chapman_energy}
    \varepsilon_a=\varepsilon_a^{\mathrm{eq}}+\delta\varepsilon_a^{(0)}+\delta\varepsilon_a^{(1)}+\dots
\end{equation}

In the first order one retains zero-order terms in the driving term [left-hand side of eq.~(\ref{eq:kin_eq})] with the exception of the magnetic part of the Lorentz force and first-order terms in the collision integral. Linearization of the collision integral requires a certain care. Here it is necessary to bear in mind that the conservation laws in the collision probabilities $w_{ab}$ contain true quasiparticle energies. Therefore the collision integral will vanish exactly for any  distribution functions of the local equilibrium form   if one uses there the true quasiparticle spectrum instead of the local equilibrium one, i.e., if one substitutes $p^\mu_{(a),\mathrm{l.e.}}\to p^\mu_{(a)}$ \new{in eq.~(\ref{eq:f_l.e.})} \cite{BaymPethick,Landau10eng}.
It is instructive to introduce the deviation denoted with bar via
\begin{equation}\label{eq:deltafbar_def}
    f_a(\bm{p})=f_a^{\mathrm{l.e.}}(p^\mu_{(a),\mathrm{l.e.}}) +\delta f_a(\bm{p})=
    f_a^{\mathrm{l.e.}}(p^\mu_{(a)})+\overline{\delta f_a(\bm{p})}.
\end{equation}
Importantly, the thermodynamic fluxes 
in the first order 
are expressed via the functions $\overline{\delta f_a}$ \cite{Landau10eng}: 
\begin{subequations}\label{eq:fluxes_fbar}
\begin{eqnarray}
    \Delta\bm{j}_{(a)}&=&\int_{\bm{p}} \bm{v}_a\, \overline{\delta f}_a,\label{eq:deltaj_kin}\\
    \bm{J}_{(q)}&=&\sum_a\int_{\bm{p}} \bm{v}_a\,(\varepsilon_a-h_a)\, \overline{\delta f}_a,\label{eq:Jq_kin}\\
    \Pi^{\langle ij\rangle}&=&\sum_a \int_{\bm{p}} p^{\langle i} v_a^{j\rangle} \, \overline{\delta f}_a.\label{eq:Pi_kin}
\end{eqnarray}
\end{subequations}
\new{To obtain eqs.~(\ref{eq:Jq_kin}) and~(\ref{eq:Pi_kin}) we used eq.~(\ref{eq:Tmunu_FL}) and the definitions   (\ref{eq:en_flux}) and~(\ref{eq:Pimunu_def}) in the LRF [i.e., in the limit $U^\mu \to (1,0,0,0)$)]. Here and below
\begin{equation}
A^{\langle ij \rangle} = \frac{1}{2}\left( A^{ij} + A^{ji} - \frac{2}{3} \delta^{ij} A^{kk} \right)
\end{equation}
is the short-hand notation for the traceless symmetric part of a 3-dimensional tensor $A^{ij}$.\footnote{
\new{The tensor $\Pi^{\langle ij \rangle}$ coincides with the spatial components of the tensor $\Pi^{\langle \mu\nu \rangle}$ in eq.~(\ref{eq:Yeta}) at $U^\mu \to (1,0,0,0)$.}
}} 
All variables in eqs.~(\ref{eq:fluxes_fbar}), i.e. $\bm{v}_a$, $\varepsilon_a$, and $h_a$, now correspond to the global equilibrium. 

The magnetic part of the Lorentz force also vanishes exactly for the distribution function $f_a^{\mathrm{l.e.}}(p^\mu_{(a)})$. We assume that the magnetic field is large, therefore it should be kept in the driving term at the first order. This term thus also contains $\overline{\delta f_a}$ \cite{Landau10eng}. 

The linearized transport equation (\ref{eq:kin_eq}) is then given by
\begin{eqnarray}
    \frac{\partial \delta f_a^{(0)}}{\partial t}&+&\bm{v}_a\nabla \delta f_a^{(0)}-
\left(\frac{\partial\delta \varepsilon^{(0)}_a}{\partial{\bm{r}}}-q_a\bm{E}\right)
    \nabla_{\bm{p}} f^{\mathrm{eq}}_a\nonumber\\
    &=&- q_a\left[\bm{v}_a\times\bm{B}\right]\nabla_{\bm{p}} \overline{\delta f}^{(1)}_a + I_a\left[\left\{\overline{\delta f}^{(1)}_r\right\}\right],\label{eq:Boltz_lin0}
\end{eqnarray}
where $\bm{v}_a$, quasiparticle energies and collision probabilities in $I_a$ are calculated for the global equilibrium state. Derivatives in the left-hand side are due to gradients of the macroscopic fields $T(x)$, $\mu_a(x)$, and $\bm{V}(x)$.

In order to identify the terms containing velocity gradients, it is necessary to consider the equations in the fixed inertial laboratory frame that moves with the instant  velocity $-\bm{V}$, $V\ll 1$, relatively to the LRF. Let us indicate state variables in LRF with a bar for a moment, i.e. $\overline{\bm{p}}$, $\overline{\varepsilon}_a$, and $\overline{f}_a$. General principles of Lorentz invariance require that the transformation laws for quasiparticle energy and momentum are the same as for the free particles \cite{BaymChin1976NuPhA}, therefore 
\begin{equation}\label{eq:dirac_mass_general}
    \overline{\varepsilon}^2_a-\overline{\bm{p}}^2=\varepsilon^2_a-\bm{p}^2\equiv M^{2}_{Da}(\bm{p},\bm{r}),
\end{equation}
where the (Dirac) mass $M_{Da}$ here in principle depends on $\bm{p}$. Notice, that if there is no dependence of $M_{Da}$ on $\bm{p}$, then the standard relation $\bm{v}_a=\bm{p}/\varepsilon_a$ holds.

When $V\ll 1$,  transformation laws for the quasiparticle energy and momenta are \cite{BaymChin1976NuPhA}
\begin{subequations}
\begin{eqnarray}
    \overline{\varepsilon}_a&=&\varepsilon_a - \bm{p}\bm{V},\label{eq:e_transform}\\
    \overline{\bm{p}}&=&\bm{p}-\varepsilon_a \bm{V}.\label{p_transform}
\end{eqnarray}
\end{subequations}
and the distribution function in the moving frame is related to the LRF distribution function $\overline{f}_a(\overline{\bm{p}})$ as
\begin{equation}\label{eq:fp_transf}
    f_a(\bm{p})=\overline{f}_a(\overline{\bm{p}}).
\end{equation}
The variation of the distribution function at fixed $\bm{p}$ can be written according to the chain differentiation rule
\begin{equation}\label{eq:f_var_V}
    \delta f_a = \delta \overline{f}_a(\overline{\bm{p}}) +\delta\overline{\bm{p}} \nabla_{\overline{\bm{p}}} \overline{f}_a =\delta \overline{f}_a(\overline{\bm{p}})-\varepsilon \frac{\partial\overline{\varepsilon}}{\partial\overline{\bm{p}}}\delta \bm{V} \frac{\partial \overline{f}_a}{\partial\overline{\varepsilon}},
\end{equation}
where the variation in the first term is taken at fixed $\overline{\bm{p}}$.
Similarly, the quasiparticle energy variation is
\begin{equation}\label{eq:energy_var_v}
    \delta\varepsilon_a = \delta \overline{\varepsilon}_a- \varepsilon_a \frac{\partial\overline{\varepsilon}_a}{\partial\overline{\bm{p}}}\delta \bm{V}+ \bm{p}\delta\bm{V}.
\end{equation}
After substitution of eqs.~(\ref{eq:f_var_V}) and (\ref{eq:energy_var_v}) to eq.~(\ref{eq:Boltz_lin0}) one can set $\bm{V}=0$, $\overline{\bm{p}}=\bm{p}$, and $\overline{\varepsilon}_a=\varepsilon_a$.

The proper (i.e. independent of $\delta \bm{V}$) zero-order variation of the distribution function in the LRF can be written as
\begin{equation}\label{eq:deltaf0}
    \delta f_a^{(0)} = f_F'(x_a)\left(\frac{\delta \varepsilon_a^{(0)}-\delta\mu_a}{T} +(\varepsilon_a-\mu_a)\delta\frac{1}{T}\right),
\end{equation}
where the dimensionless quantity 
\begin{equation}\label{eq:x_def}
    x_a=\frac{\varepsilon_a-\mu_a}{T}
\end{equation}
is introduced.

Collecting all terms, one obtains for the left-hand side of eq.~(\ref{eq:Boltz_lin0})
\begin{eqnarray}
 \mathrm{l.h.s.}= &&- f_F'(x_a) \left\{\frac{\varepsilon_a-h_a}{T}\bm{v}_a\left(\frac{\nabla T}{T}+\dot{\bm{V}}
 \right)\right.\nonumber\\
 &+&\frac{1}{T}\left.\bm{v}_a\left(\bm{d}_{(a)} +\frac{h_a}{ h n}[\bm{J}\times\bm{B}] \right)\right.
 \nonumber\\ 
 &+& \left. \frac{p_i v_{aj}}{T} V_{ij}\right.
 \nonumber\\ 
 &+&\frac{1}{3}(\bm{p}\bm{v}_a)\mathrm{div}\bm{V} -\left.\frac{\partial}{\partial t} \left(\frac{\varepsilon_a^{\mathrm{l.e.}}-\mu_a
 }{T}\right)\right\},\label{eq:lhs_fin}
\end{eqnarray}
where (\ref{eq:NavierStokes}) in the limit $V\ll 1$ is used in the second line to eliminate the acceleration $\dot{\bm{V}}$. In eq.~(\ref{eq:lhs_fin}),  
 $\bm{d}_{(a)}$ is the spatial part of the four-vector $d^\mu_{(a)}$, eq.~(\ref{eq:dmu}). Namely in the LRF, $d_{(a)}^\mu=(0,-\bm{d}_{(a)})$,  where
\begin{equation}\label{eq:da_three}
    \bm{d}_{(a)}=T\nabla\frac{\mu_a}{T}
    -T\frac{h_a}{h}\sum_b y_b\nabla \frac{\mu_b}{T}- q_a\bm{E}.
\end{equation}
The tensor  $V_{ij}$ in eq.~(\ref{eq:lhs_fin}) is 
\begin{equation}
V_{ij}=\new{\partial_{\langle i} V_{j\rangle}=}\frac{1}{2}\left(\frac{\partial V_i}{\partial x_j}+\frac{\partial V_i}{\partial x_j}-\frac{2}{3}\delta_{ij}\mathrm{div}\bm{V}\right)
\end{equation}
and is the spatial part of eq.~(\ref{eq:Xeta}) in the limit $V\ll 1$.

Different lines in eq.~(\ref{eq:lhs_fin}) correspond to different transport processes. Namely, the first line corresponds to the heat conduction while the second line to diffusion processes. These two processes are driven by the vector thermodynamic forces, and in principle they mix (sec.~\ref{sec:Hydro}). 
The third line of eq.~(\ref{eq:lhs_fin}) corresponds to the shear  viscosity, whose driving force is the second-order traceless tensor $V_{ij}$.
The fourth line of eq.~(\ref{eq:lhs_fin}) corresponds to the scalar bulk viscosity processes \cite{BaymPethick}.
As stated above, we do not consider bulk viscosity here, since it is mainly governed not by the quasiparticles collisions but by the reactions, and in this case in the kinetic approach a non-stationary problem should be solved. As such, we further assume that $\mathrm{div}\bm{V}=0$.

It is convenient to further transform the  right-hand side of eq.~(\ref{eq:Boltz_lin0}), defining \cite{BaymPethick}
\begin{equation}\label{eq:deltaf1bar}
    \overline{\delta f_a(\bm{p})}^{(1)} \equiv - \frac{1}{T} f_F'(x_a) \Phi_a(\bm{p}),
\end{equation}
where $\Phi_a(\bm{p})$ with $a=1,\dots,r$ are unknown functions to be found. Notice that eqs.~(\ref{eq:deltaf1bar}) and (\ref{eq:lhs_fin}) ensure formally that the deviations from the local equilibrium distribution functions are localized in the vicinity of the Fermi surface due to appearance of the term $f'_{F}(x)$ in these equations.

With this substitution, the magnetic operator in eq.~(\ref{eq:Boltz_lin0}) becomes
\begin{equation}\label{eq:magop_lin}
    \frac{q_a}{T}f_F'(x_a)\left[\bm{v}_a\times\bm{B}\right]\nabla_{\bm{p}} \Phi_a(\bm{p}),
\end{equation} 
while the  collision integrals  eq.~(\ref{boltz_collint}) take the linearized form
\begin{eqnarray}
I_{ab\ \rm lin}[\Phi]&=&-\frac{1}{T(1+\delta_{ab})}\int_{{\bm p}_{1'}}\int_{{\bm p}_{2}}\int_{{\bm p}_{2'}} w^{\mathrm{eq}}_{ab}({\bm p},{\bm p}_{2};{\bm p}_{1'},{\bm p}_{2'})\nonumber\\
&&\times{\cal F}_{ab} (\Phi_1+\Phi_{2}-\Phi_{1'}-\Phi_{2'}) \, ,\label{eq:Boltz_lin}
\end{eqnarray}
where we introduced the Pauli blocking factor
\begin{equation}\label{eq:Pauli}
    {\cal F}_{ab} = f_1^{\mathrm{eq}}f_2^{\mathrm{eq}}(1-f_{1'}^{\mathrm{eq}})(1-f_{2'}^{\mathrm{eq}}),
\end{equation}
$\Phi_1=\Phi_a(\bm{p})$, $\Phi_{1'}=\Phi_a(\bm{p}_{1'})$, $\Phi_2=\Phi_b(\bm{p}_2)$, and $\Phi_{2'}=\Phi_b(\bm{p}_{2'})$. We will further drop the superscript \new{`eq'}.

The problem thus reduces to finding the functions $\Phi_a(\bm{p})$ from the system of the linearized transport equations
\begin{eqnarray}
  &-& \frac{1}{T}f_F'(x_a) \left\{(\varepsilon_a-h_a)\bm{v}_a\left(\frac{\nabla T}{T}+\dot{\bm{V}}
 \right)\right.\nonumber\\
 &&+\left.\bm{v}_a\left(\bm{d}_{(a)} +\frac{h_a}{ h n}[\bm{J}\times\bm{B}] \right)+ {p_i v_{aj}} V_{ij}\right\}\nonumber\\
 &=&\sum_b I_{ab\,\mathrm{lin}}[\Phi]+\frac{q_a}{T}f_F'(x_a)\left[\bm{v}_a\times\bm{B}\right]\nabla_{\bm{p}} \Phi_a(\bm{p}).
 \label{eq:kin_lin_fin}
\end{eqnarray}

Consider now the entropy production rate given by eq.~(\ref{eq:Sprod_full}). Linearization of the first term under the integral gives
\begin{equation}
    \log\frac{1-f_a}{f_a}\approx \frac{\varepsilon_a-\mu_a}{T} - \frac{1}{T}\Phi_a.
\end{equation}
The first term correspond to the entropy exchange between the different species.  When summed over species, this part of eq.~(\ref{eq:Sprod_full})  vanishes (for exact collision integral, as discussed above) and one is left with
\begin{equation}\label{eq:entrprod_lin}
    T\varsigma = -\sum_a\int_{\bm{p}} \Phi_a(\bm{p}) I_a\left[\left\{\Phi_b\right\}\right].
\end{equation}
It is evident from this form of entropy production rate, that $\varsigma$ is indeed second order in gradients. The linearized collision operator should be seminegative-definite in order to ensure the second law. 
Multiplying the left-hand side of the linearized Boltzmann eq.~(\ref{eq:kin_lin_fin}) by $\Phi_a(\bm{p}s)$, integrating over $\bm{p}$ and summing over species we obtain eq.~(\ref{eq:zeta3}) for the entropy generation (formally written in the LRF). Notice that the magnetic term in the right-hand side of (\ref{eq:kin_lin_fin}) does not contribute to entropy production.

Equations~(\ref{eq:kin_lin_fin}) do not determine the functions $\Phi_a(\bm{p})$ completely, since the collision integrals are all zero for any set of the functions \new{$\Phi^{\mathrm{ker}}_a=\mathfrak{a}_a+\mathfrak{b}^\mu p_{(a)}^\mu$, where $\mathfrak{a}_a$ ($a=1\,\dots, r$) and $\mathfrak{b}^\mu$ are arbitrary constants.}
These constants, which fix the solution of the homogeneous equation, should be determined by the additional conditions known as the conditions of fit \cite{DeGroot1980Book}. 
Different conditions of fit can be traced to the different choices of the hydrodynamic frames \cite{Bemfica2018PhRvD}, which is important for stability of the hydrodynamical equations, as discussed in sec.~\ref{sec:Hydro}. According to the general discussion, in order to find the frame-invariant collision transport coefficients at first order, it is actually possible to impose any conditions of fit. According to eqs.~(\ref{eq:fluxes_fbar}), the uniform solutions given by a set of the constants \new{$\{\frak{a}_a\}$} do not contribute to the dissipation fluxes we consider (notice, that this would not be true for  the bulk viscosity). The same is true for the `energy' kernel solutions in the form \new{$\frak{b}^0 \varepsilon_{(a)}^0$}. The conditions of fit that fix \new{$\{\frak{a}_a\}$} and \new{$\frak{b}^0$} are already contained in our definition of the local equilibrium state; they also correspond to the conditions in eqs.~(\ref{eq:matching}). The remaining kernel solutions of the form \new{$\bm{\frak{b}}\bm{p}$} do not modify the heat current eq.~(\ref{eq:Jq_kin}) or the shear stress tensor eq.~(\ref{eq:Pi_kin}), but affect the  dissipative particle currents eq.~(\ref{eq:deltaj_kin}). Clearly, the conditions of fit for fixing the constant vector \new{$\bm{\frak{b}}$} are nothing more that the conditions that fix the hydrodynamic frame. At the same time, we are interested in  diffusion fluxes $\bm{i}_{(a)}$ that do not depend on \new{$\bm{\frak{b}}$}. Therefore we suggest that the vector \new{$\bm{\frak{b}}$} is fixed by imposing the Eckart condition of fit (i.e. \new{$\bm{\frak{b}}=0$}), when the diffusion transport coefficients can be calculated directly from (\ref{eq:deltaj_kin}) in accordance with the discussion around eq.~(\ref{eq:zetas}). This considerations allow us to forget about the kernel solutions and assume that the  deviation functions $\Phi_a$ are linear combinations of thermodynamic forces ${\cal X}_k$
\begin{equation}\label{eq:Phia_Xk}
    \Phi_a(\bm{p})=\sum_k\hat{{\cal G}}^{a}_k(\bm{p})\cdot {\cal X}_k,
\end{equation}
where $\hat{{\cal G}}^{a}_k(\bm{p})$ are some, in general tensor, functions.

\subsection{Tensor relations}\label{sec:tensor}
The Curie principle states that the responses to the thermodynamic forces of different tensor ranks do not mix \cite{deGrootMazur1984}. The microscopic basis of this statement is the scalar character of the collision integral in the isotropic medium. In anisotropic medium (i.e. anisotropic crystal) this principle can be violated. In the neutron star context this is the case when the magnetic field is very strong and the quasiparticle collisions are affected by magnetic field effects. Another example is the generic anisotropic structure, i.e. anisotropic pasta phase at the bottom of the crust, or the anisotropic pairing texture possible for the triplet neutron pairing. Both these effects are almost unexplored in the regard to NSs. Below we assume that magnetic field does not affect the quasiparticle scattering probabilities and the collision integral is isotropic. In degenerate matter the latter is a good approximation when $\omega_{\mathrm{BF}a} \ll q v_{\mathrm{F}a}$, where $q$ is the typical momentum transfer in collisions and is valid for non-quantizing fields considered in this paper.

In order to fully take into account the symmetry of the problem, 
it is instructive to use the irreducible tensors technique e.g., \cite{DeGroot1980Book,Denicol2014}.
To this end one performs expansion over the irreducible representations of a little group corresponding to timelike vector $U^\mu$. In the LRF this means the expansion over the irreducible representations of the group of 3D rotations. Usually in the kinetic theory the Cartesian irreducible tensor formalism is used. For our purposes, however, it is instructive to use the formalism of the irreducible spherical tensors (e.g., \cite{Varshalovich}) which allows for a simple account of the axial symmetry of the problem in the presence of the magnetic field. 

Irreducible tensor set (under rotations) is a set of quantities which transform under an irreducible representation of the rotation group. Irreducible spherical tensors, in particular, transform in the same way as the eigenfunctions of the angular momentum operator, i.e. the spherical harmonics. Each vector $\bm{T}$ can be expressed as the rank-1 irreducible spherical tensor $T_\mu$, $\mu=-1,0,1$, employing  spherical basis \cite{Varshalovich}. The transformation law between Cartesian and spherical basis is described by the unitary matrix $U_{\mu i}$, i.e. $T_\mu=U_{\mu i} T_i$. Explicitly,
\begin{equation}
    \left(\begin{array}{c}
        T_{-1}   \\
         T_0\\
         T_{+1}
    \end{array}
    \right) = 
    \left(
    \begin{array}{ccc}
        \frac{1}{\sqrt{2}} & -\frac{i}{\sqrt{2}}  &0\\
         0&0   &1\\
         -\frac{1}{\sqrt{2}}& -\frac{i}{\sqrt{2}} & 0 
    \end{array}
    \right)\left(
    \begin{array}{c}
         T_x  \\
         T_y  \\
         T_z
    \end{array}
    \right).
\end{equation}

Similarly, any second-rank tensor can be transformed to the spherical basis as 
\begin{equation}
    T_{\mu\mu'} = U_{\mu i} U_{\mu'j} T_{ij}.
\end{equation}
Then the tensor $T_{\mu\mu'}$ can be cast over the irreducible components $T_{Kq}$ via the unitary Clebsch-Gordan transformation
\begin{equation}
    T_{\mu\mu'}=\sum_{Kq} i^K C^{Kq}_{1\mu 1\mu'} T_{Lq},\quad T_{Kq}=\sum_{\mu\mu'} (-i)^K C^{Kq}_{1\mu 1\mu'} T_{\mu\mu'},
\end{equation}
where
$C^{Kq}_{1\mu 1\mu'}$ is the  Clebsch-Gordan coefficient \cite{Varshalovich}.
The phase factor $i^K$ is introduced here to ensure that the resulting tensors behave similarly to spherical functions under complex conjugation, i.e. $\left(T_{Kq}\right)^*=(-1)^q T_{K-q}$. 
Then
\begin{subequations}
\begin{eqnarray}
    T_{Kq}&=&\sum_{\mu\mu'} (-i)^K C^{Kq}_{1\mu 1\mu'} U_{\mu i} U_{\mu'j }T_{ij}, \\
    T_{ij}&=&\sum_{Kq} i^K C^{Kq}_{1\mu 1\mu'} U^\dag_{\mu i} U^\dag_{\mu'j } T_{Kq}.
\end{eqnarray}
\end{subequations}

Explicitly, for the traceless symmetric tensor of the second rank $T_{ij}$
\begin{subequations}
\begin{equation}
    T_{20}=-\sqrt{\frac{3}{2}}T_{zz},
\end{equation}
\begin{equation}
    T_{2\pm 1}=\pm T_{xz}+iT_{yz},
\end{equation}
\begin{equation}
    T_{2\pm 2} = \frac{1}{2}\left(T_{yy}-T_{xx}\right)\mp i T_{xy}.
\end{equation}
\end{subequations}

We now rewrite the linearized transport eq.~(\ref{eq:kin_lin_fin}) in the irreducible spherical tensor formalism. The left-hand side of  eq.~(\ref{eq:kin_lin_fin}) can be written in the following general form (cf. e.g., \cite{Anderson1987}) 
\begin{equation}\label{eq:lhs_irred}
    -\frac{1}{T}f_F'(x)\sum_k\sqrt{\frac{4\pi}{2\ell_k+1}} D^a_k(p) \left(Y_{\ell_k}(\hat{\bm{p}})\cdot ({\cal X}_{k})_{\ell_k}\right),
\end{equation}
where $\ell_k$ is the tensor dimension of the thermodynamic force ${\cal X}_k$, dot defines the scalar product
\begin{equation}
    \left(Y_{\ell_k}(\hat{\bm{p}})\cdot ({\cal X}_{k})_{\ell_k}\right)=\sum_{m=-\ell_k}^{\ell_k} Y^*_{\ell_k m}(\hat{\bm{p}}) ({\cal X}_{k})_{\ell_k m},
\end{equation}
$Y_{\ell_k m}(\hat{\bm{p}})$ \new{are} the spherical harmonics,  $\hat{\bm{p}}$ denotes the direction of $\bm{p}$, asterisk means complex conjugate, and $({\cal X}_{k})_{\ell_k m}$ is the $m$'th component of the thermodynamic force ${\cal X}_{k}$ taken in the irreducible spherical tensor form. The functions $D^a_k(p)$ in eq.~(\ref{eq:lhs_irred}) contain the remaining scalar terms in eq.~(\ref{eq:kin_lin_fin}) which depend on $p$ but not on the direction $\hat{\bm{p}}$. They are given in tab.~\ref{tab:transp_params}.

\begin{table*}
\caption{}
\label{tab:transp_params}
\begin{tabular*}{\textwidth}{@{\extracolsep{\fill}}lccccc@{}}
\hline
Force ($k$) & \multicolumn{1}{c}{$\ell$} & \multicolumn{1}{c}{$D_k^a$} & \multicolumn{1}{c}{$\widetilde{D}_k$} & \multicolumn{1}{c}{$\widetilde{X}_k$} &
\multicolumn{1}{c}{$\xi_k$}
\\
\hline
$\kappa$ & 1 & $v_a (\varepsilon_a-h_a)$ & $T$ & $x$ & $-1$ \\
$\eta$ & 2 & $\sqrt{2/3}\,(pv_a)$ & $\sqrt{2/3}\,p_{\mathrm{F}a}$ & $1$ & $+1$ \\
$Da$ & 1 & $v_a$ & $1$ & $1$ & $+1$\\
\hline
\end{tabular*}
\end{table*}

In order to transform the right-hand side of eq.~(\ref{eq:kin_lin_fin}) in a similar way, the deviation function $\Phi_a(\bm{p})$  can be expanded in the spherical harmonics as
\begin{equation}\label{eq:phi_spherical}
    \Phi_a(\bm{p})=\sum_{\ell m} \left(\Phi_a\right)_{\ell m} (p) Y^*_{\ell m }(\hat{\bm{p}}).
\end{equation}
The collision integral in most general form in this basis is the matrix in $\ell m$ indices
\begin{equation}\label{eq:collint_spherical_generat}
    I_a[\{\Phi_b\}]=\sum_{\ell m \ell' m'} I_a[\{(\Phi_b)_{\ell' m'}\}]^{\ell m}_{\ell' m'} Y^*_{\ell m}(\hat{\bm{p}}).
\end{equation}
However, when the collision probability is isotropic, the linearized collision integral is diagonal in $\ell m$ and, moreover, does not depend on $m$
\begin{equation}\label{eq:collint_spherical}
     I_a^\ell[\{(\Phi_b)_{\ell' m'}\}]^{\ell m}_{\ell' m'} = I_a^\ell[\{(\Phi_b)_{\ell m}\}] \delta_{\ell\ell'}\delta_{mm'}.
\end{equation}

The magnetization term eq.~(\ref{eq:magop_lin}) can be written in the irreducible form by 
introducing the $\bm{p}$-space angular momentum operator
\begin{equation}\label{eq:angmom_p}
    \hat{\bm{\ell}}_{\bm{p}}=-i\left[\bm{p}\times\nabla_{\bm{p}}\right].
\end{equation}
Then eq.~(\ref{eq:magop_lin}) becomes
\begin{equation}\label{eq:mag_angmom}
    q_a\left[\bm{v}_a\times\bm{B}\right]\nabla_{\bm{p}} \overline{\delta f}^{(1)}_a = 
    i \frac{q_a v_a}{p T}f_F'(x_a) (\bm{B}\cdot\hat{\bm{\ell}}_{\bm{p}})  \Phi_a(\bm{p}).
\end{equation}
This operator is diagonal in the spherical basis eq.~(\ref{eq:phi_spherical}) if the direction of the magnetic field is taken as the $Z$ axis. The equations can be solved in this frame and then rotated into the general frame, as shown below.

Thus for the isotropic collision operator, the equations for different tensor components of $\Phi_a$ decouple and for each component $\left({\Phi_a}\right)_{\ell m}$ we obtain \begin{eqnarray}\label{eq:kineq_irred}
    &&-\frac{1}{T}f_F'(x)\sum_{k|\ell_k=\ell}\sqrt{\frac{4\pi}{2\ell+1}} D^a_k(p) 
    ({{\cal X}_k})_{\ell m} = \nonumber\\
    && I^{\ell}_a[\{(\Phi_b)_{\ell m}\}] + i\frac{1}{T}f_F'(x) m \omega_{\mathrm{B}a}(p)(\Phi_a)_{\ell m},
\end{eqnarray}
where the summation is carried over those thermodynamic forces, that have $\ell_k=\ell$.  The forces of the same rank, i.e. $\ell_k=\ell_{k'}$ are present in the left-hand side of eq.~(\ref{eq:kineq_irred}) and do mix. Namely, this is the case for the  thermal conductivity and diffusion processes which have $\ell_\kappa=\ell_{Da}=1$, see eq.~(\ref{eq:transp_vector}). As such we `proved' the Curie principle for the linearized kinetic equation.
\new{Notice that the conservation of $\ell_k$ by the magnetic operator (\ref{eq:mag_angmom}) ensures absence of the cross-term viscosity $\hat{\zeta}_1$ introduced in eqs.~(\ref{eq:bulk_def})--(\ref{eq:shear_def}).}
The term $\omega_{\mathrm{B}a}$ in eq.~(\ref{eq:kineq_irred}) is the (momentum-dependent) cyclotron frequency of the particle species $a$
\begin{equation}\label{eq:omegaB}
    \omega_{\mathrm{B}a}=\frac{q_a B v_a}{p}.
\end{equation}

Therefore using the irreducible tensor formalism we have splitted the initial system of transport equations  into the set  of $2\ell+1$ independent eqs.~(\ref{eq:kineq_irred}) for each $\ell$. Owing to the isotropic character of the collision integral and a simple form of the magnetic term in eqs.~(\ref{eq:kineq_irred}), one observes that
\begin{equation}\label{eq:Phi_resc}
    (\Phi_a(B))_{\ell m} = (\Phi_a(mB))_{\ell 1},
\end{equation}
thus  eq.~(\ref{eq:kineq_irred}) needs to be solved only for $m=1$. The components of $\Phi_a$ for other $m$'s are obtained by a simple magnetic field rescaling $B\to mB$ \cite{Viehland1974JChPh,vanErkelens1978PhyA}.

The general solution of the irreducible linear eq.~(\ref{eq:kineq_irred}) is a linear combination [cf. eq.~(\ref{eq:Phia_Xk})]\footnote{Notice that $\left({\cal G}^a_{k}(p) \right)_{\ell m}$ are coefficients in the linear combination eq.~(\ref{eq:Phia_Xk_irred}) written in the specific frame, where $\bm{B}$ is the polar axis,  and are not the spherical components of the tensor $\hat{G}^a_k$ in eq.~(\ref{eq:Phia_Xk}). They can be related by the transformation similar as performed in eq.~(\ref{eq:Yk_LK}).}
\begin{equation}\label{eq:Phia_Xk_irred}
    \left(\Phi_a\right)_{\ell m}(p)=
    \sum_k \left({\cal G}^a_{k}(p) \right)_{\ell m} \left({{\cal X}_k}\right)_{\ell m}.
\end{equation}
Substituting eq.~(\ref{eq:kineq_irred}) to 
eqs.~(\ref{eq:fluxes_fbar}) using  eqs.~(\ref{eq:deltaf1bar}) and (\ref{eq:phi_spherical}) thermodynamic fluxes in spherical tensor notations reduce to 
\begin{eqnarray}\label{eq:Lkk}
    ({\cal Y}_k)_{\ell m} &=& -\sum_a\int_{\bm p} \frac{1}{T}f'_F(x) \sqrt{\frac{4\pi}{2\ell+1}} D^a_k(p)Y_{\ell m} (\hat{\bm{p}}) \Phi_a(\bm{p})\nonumber\\
    &=& - \sum_{k'} L^{kk'}_{\ell m} ({\cal X}_{k'})_{\ell m}, 
\end{eqnarray}
with the matrix of the kinetic coefficients $L^{kk'}_{\ell m}$ which is diagonal in $\ell m$. Explicitly, the matrix $L^{kk'}_{\ell m}$ is related to functions  $\left({\cal G}^a_{k}(p)\right)_{\ell m}$ in eq.~(\ref{eq:Phia_Xk_irred}) as
\begin{equation}\label{eq:Lkk_Gk}
    L^{kk'}_{\ell m} =\sum_a \frac{g_{sa}}{2\pi^2}\int p^2 dp\, \frac{f_F'(x)}{T} \sqrt{\frac{4\pi}{2\ell+1}} D^a_k(p)\left({\cal G}^a_{k'}(p)\right)_{\ell m},
\end{equation}
where $g_{sa}=2$ (for fermions) is the spin statistical weight. 

It is instructive to rotate the eq.~(\ref{eq:Lkk}) to a general coordinate system in the LRF using the rotation matrices (Wigner ${\cal D}$-function \cite{Varshalovich}) \begin{eqnarray}
    \left({{\cal Y}_k}\right)_{\ell m} &=& \sum_{ k'm' m''} \left({\cal D}^\ell_{mm'}(\hat{\bm{b}},0)\right)^* L^{kk'}_{\ell m'} \left({\cal X}_{k'}\right)_{\ell m''} {\cal D}^\ell_{m''m'}(\hat{\bm{b}},0) \nonumber\\
    &=& \sqrt{\frac{4\pi}{2\ell+1}}\sum_{k'} \sum_{K=0}^{2\ell} 
    L^{kk'}_K\left[\left({\cal X}_{k'}\right)_{\ell}\otimes Y_K(\hat{\bm{b}})\right]_{\ell m},\label{eq:Yk_LK}
\end{eqnarray}
where $\hat{\bm{b}}\equiv \bm{B}/B$, square brackets denote irreducible spherical tensor product
\begin{equation}\label{eq:irred_product}
    \left[\left({\cal X}_{k'}\right)_\ell\otimes Y_K(\hat{\bm{b}})\right]_{\ell m}=
    \sum_{m''Q} C^{\ell m}_{\ell m'' KQ} \left({\cal X}_{k'}\right)_{\ell m''} Y_{KQ}(\hat{\bm{b}}),  
\end{equation}
and $L^{kk'}_K$ is a different set of $2\ell+1$ kinetic coefficients linearly related to $2\ell+1$ coefficients $L^{kk'}_{\ell m}$ as
\begin{equation}\label{eq:LK_def}
    L^{kk'}_K=\sum_{m} (-1)^{\ell-m} C^{K0}_{\ell m \ell -m} L^{kk'}_{\ell m}.
\end{equation}

As eq.~(\ref{eq:Yk_LK}) is written in a form of an expansion over the spherical harmonics in the direction of magnetic field, it can be convenient in practice. Notice, that the component index $m$ can now be dropped in eq.~(\ref{eq:Yk_LK}) since this is now the relation between the tensors for the general orientation of the LRF coordinate system with respect to $\bm{B}$. The relation (\ref{eq:Yk_LK}) can be transformed from the LRF to the general frame by applying the appropriate Lorentz boosts to the tensor quantities in this relation. In this way the generalized transport coefficients (i.e. as specified in \cite{Dommes2020PhRvD}) can be determined. At the end of the day they are expressed via the $(2\ell+1)$ coefficients $L^{kk'}_{\ell m}$ or $L^{kk'}_K$ for each $kk'$ pair.
According to discussion around eq.~(\ref{eq:Phi_resc}), it is enough to calculate the function 
\begin{equation}\label{eq:Ltilde}
    \widetilde{L}_{kk'}(B)\equiv L^{kk'}_{\ell_k 1}(B),
\end{equation}
since $L^{kk'}_{\ell m}(B)=L^{kk'}_{\ell 1}(mB)$.

The spherical tensor formalism above can be easily connected to the Cartesian one. Let us start from the vector ($\ell=1$) thermodynamic forces. 
 For example, for the thermal conductivity problem (ignoring the thermal diffusion term) one gets 
\begin{subequations}\label{eq:kappa_Cartesian}
\begin{eqnarray}
{\cal Y}_{qz}&=&-T \widetilde{\kappa}(B=0) {\cal X}_{qz},\label{eq:kappa_parallel}\\
(\mp {\cal Y}_{qx} - i {\cal Y}_{qy})&=&-T\widetilde{\kappa}(\pm B) (\mp {\cal X}_{qx}- i{\cal X}_{qy})\label{eq:kappa_transv}.
\end{eqnarray}
\end{subequations}
Comparing with the standard definitions \cite{Landau10eng}, one identifies $\kappa_\parallel\equiv \widetilde{\kappa}(B=0)$ as the parallel component of heat conductivity, $\kappa_\perp\equiv \mathrm{Re}\,\widetilde{\kappa}(B)$ as the transverse heat conductivity, and $\kappa_{\wedge}\equiv - \mathrm{Im}\, \widetilde{\kappa}(B)$ as the Hall thermal conductivity component. Similar expressions hold for other transport coefficients related to the vector thermodynamic forces. The conduction parallel to magnetic field does not depend on $B$, while the conduction across the magnetic field depends on it.

The relations between the Cartesian components of the stress-energy tensor and the tensor $V_{ij}$ ($\ell=2$) are 
\begin{subequations}
\begin{eqnarray}
\Pi_{zz}&=&2\widetilde{\eta}(B=0) V_{zz},\\
(\Pi_{zx}\pm i\Pi_{zy})&=&2\widetilde{\eta}(\pm B)(V_{zx}\pm i V_{zy}),\\
(\Pi_{xx}-\Pi_{yy}\pm 2i\Pi_{xy})&=&2\widetilde{\eta}(\pm2B)\nonumber\\
&&\times(V_{xx}-V_{yy}\pm 2i V_{xy}).
\end{eqnarray}
\end{subequations}
Let us compare this equations with traditional formulation containing five shear viscosity coefficients $\eta_0\dots\eta_4$ \cite{Landau10eng}
\begin{subequations}
\begin{eqnarray}
\Pi_{zz}&=&2\eta_0 V_{zz},\\
\Pi_{xx}&=&-\eta_0 V_{zz}+\eta_1(V_{xx}-V_{yy})+2\eta_3 V_{xy},\\
\Pi_{yy}&=&-\eta_0 V_{zz}-\eta_1(V_{xx}-V_{yy})-2\eta_3 V_{xy},\\
\Pi_{xy}&=&2\eta_1 V_{xy}-\eta_3 (V_{xx}-V_{yy}),\\
\Pi_{xz}&=&2\eta_2V_{xz}+2\eta_4 V_{yz},\\
\Pi_{yz}&=&2\eta_2 V_{yz}-2\eta_4 V_{xz}.
\end{eqnarray}
\end{subequations}
Then one identifies $\eta_0=\widetilde{\eta}(B=0)$, $\eta_1=\mathrm{Re}\,\widetilde{\eta}(2B)$, $\eta_3=-\mathrm{Im}\,\widetilde{\eta}(2B)$,
$\eta_2=\mathrm{Re}\,\widetilde{\eta}(B)$, 
$\eta_4=-\mathrm{Im}\,\widetilde{\eta}(B)$.

\subsection{Relaxation time approximation}\label{sec:rta}
The above considerations are general, but they can be analyzed in the most transparent form if the 
relaxation time approximation for the collision integral is used, which reads
\begin{equation}\label{eq:rta}
    I_a^\ell[\Phi_a]= \frac{1}{T}f_F'(x_a)\frac{(\Phi_a)_{\ell m}}{\tau^{(\ell)}_{a}(\varepsilon_a)}.
\end{equation}
The covariant form of the relaxation time approximation is the Anderson-Witting model \cite{Anderson1974Phy}  where in the general frame eq.~(\ref{eq:rta}) is multiplied by a factor $(p^\mu U_\mu)/p^0$ (equal to 1 in the LRF). This form violates the current conservation as well as the energy and momentum conservation, respected by the collision integrals eq.~(\ref{boltz_collint}). In this sense it is not correct to employ the relaxation time approximation for description of transport of a closed (self-contained) plasma. In the non-relativistic case, the generalization of the relaxation time approximation was given by \cite{Bhatnagar1954PhRv} and the current-conserving relaxation time approximations for relativistic plasmas were recently formulated by 
\cite{Formanek2021arXiv,Rocha2021PhRvL}.
Here these peculiarities do not matter since we investigate the general tensor structure of the transport coefficients. One may view the relaxation time in eq.~(\ref{eq:rta}) as a manifestation of the external scattering mechanisms which do not respect the conservation laws (an example is the scattering of electron off the ionic lattice in the NS crust). Within the same reasoning, we consider the single component case for simplicity, but retain the electric charge of this component. Then the electric conductivity and thermal diffusion effects are present, governed by the external scattering mechanism.

In the relaxation time approximation, the solution of the linearized eq.~(\ref{eq:kineq_irred}) is given in the  form eq.~(\ref{eq:Phia_Xk_irred}) where simply
\begin{equation}\label{eq:Philm_rta}
    \left({\cal G}^a_k(p)\right)_{\ell m}=- \sqrt{\frac{4\pi}{2\ell+1}} D^a_k(p)  \frac{\tau^{(\ell)}_a}{1+im\omega_{\mathrm{B}a}\tau^{(\ell)}_a},
\end{equation}
and according to eq.~(\ref{eq:Lkk_Gk})

\begin{eqnarray}
    L^{kk'}_{\ell m} &=&- \frac{g_s}{2\pi^2(2\ell+1)}\int p^2 dp \frac{f_F'(x)}{T} D^a_k(p)D^a_{k'}(p) \nonumber \\
    &\times& 
    \frac{\tau^{(\ell)}_a}{1+im\omega_{\mathrm{B}a}\tau^{(\ell)}_a}.\label{eq:Lkk_rta}
\end{eqnarray}

Three transport coefficients in the vector sector (with $k,\,k'=q,\, Da$) are
\begin{eqnarray}\label{eq:Ltilde_vec}
    \left\{\begin{array}{l}
    \widetilde{L}_{qq}(B)\\
    \widetilde{L}_{qa}(B)\\
    \widetilde{L}_{aa}(B)\\
    \end{array}
    \right\}=&-&\frac{1}{3\pi^2} \int p^2 v_a^2 dp \frac{f'_F(x)}{T} \left\{\begin{array}{c}
    (\varepsilon_a-h_a)^2\\
    (\varepsilon_a-h_a)\\
    1
    \end{array}
    \right\} \nonumber\\
    &\times&\frac{\tau^{(1)}_a(\varepsilon_a)}{1+i \omega_{\mathrm{B}a} \tau^{(1)}_a(\varepsilon_a)}.
\end{eqnarray}
The coefficient $\widetilde{L}_{aa}$ when multiplied by $q_a^2$ is the charge conductivity, the coefficient $\widetilde{L}_{qa}$ describes termodiffusion effects, and $\widetilde{L}_{qq}$ is related to the heat conduction, although the traditional thermal conductivity coefficient in this case is given by 
\begin{equation}
    T \widetilde{\kappa}=\widetilde{L}_{qq}-\frac{\widetilde{L}_{qa}^{2}}{\widetilde{L}_{aa}},
\end{equation}
according to eqs.~(\ref{eq:kin_vec_SM1}). In the  degenerate matter, $h_a\approx \mu_a$ and the factor $f_F'(x)\approx -\delta(x)$, and  integration in eq.~(\ref{eq:Ltilde_vec}) results in
\begin{subequations}\label{eq:kappa_sigma_rta}
\begin{eqnarray}
    \widetilde{\kappa}&=&\frac{\pi^2 n_a T}{3m_a^*}\frac{\tau^{(1)}_a(\mu_a)}{1+i\omega_{\mathrm{BF}a}\tau^{(1)}_a(\mu_a)},\label{eq:kappa_rta}\\
        \widetilde{L}_{aa}&=&\frac{n_a }{m_a^*}\frac{\tau^{(1)}_a(\mu_a)}{1+i\omega_{\mathrm{BF}a}\tau^{(1)}_a(\mu_a)},\label{eq:sigma_rta}\\
\end{eqnarray}
\end{subequations}
where $\omega_{\mathrm{BF}a}$
is the cyclotron frequency at the Fermi surface defined in eq.~(\ref{eq:omega_BF}). 
Here we employed the fact that the thermal diffusion coefficient vanishes at the first approximation  $(\tau^{(1)}_a(\varepsilon_a)=\mathrm{const})$ (see the discussion below), thus $\widetilde{\kappa}\approx T^{-1}\widetilde{L}_{qq}$. Equation~(\ref{eq:sigma_rta}) is the Drudde formula and the Wiedemann-Franz rule $\widetilde{\kappa}=\pi^2 T\widetilde{L}_{aa}/3$ holds in the degenerate case in the relaxation time approximation \cite{ZimanBook}.

In the limit of large magnetization, $\omega_{\mathrm{BF}a}\tau^{(1)}(\mu)\gg 1$, one finds
\begin{subequations}\label{eq:kappa_rta_lim}
\begin{eqnarray}
    \kappa_\perp=\mathrm{Re}\,\widetilde{\kappa} &=& \frac{\pi^2 n_a T}{3m_a^*}\frac{1}{\omega_{\mathrm{BF}a}^2\tau^{(1)}_a(\mu_a)}\propto B^{-2}\label{eq:kappa_perp_rta_lim}\\
    \kappa_{\wedge}=-\mathrm{Im}\,\widetilde{\kappa} &=& \frac{\pi^2 n_a T}{3m_a^*}\frac{1}{\omega_{\mathrm{BF}a}}\propto B^{-1}.\label{eq:kappa_hall_rta_lim}
\end{eqnarray}
\end{subequations}
Notice that the limiting value of the Hall component of the thermal conductivity in eq.~(\ref{eq:kappa_hall_rta_lim}) does not depend on the collision mechanism and depends only on the quasiparticle number density $n_a$ and the magnetic field $B$.

Similarly, for the shear viscosity ($\ell=2$), we obtain from eqs.~(\ref{eq:shear_def}), (\ref{eq:Lkk_rta}), (\ref{eq:Ltilde}), and tab.~\ref{tab:transp_params}, the following expression
\begin{equation}\label{eq:eta_rta}
    \widetilde{\eta}=-\frac{1}{15\pi^2} \int p^4v_a^2 dp \frac{f'_F(x)}{T}  \frac{\tau^{(2)}_a(\varepsilon_a)}{1+i \omega_{\mathrm{B}a} \tau^{(2)}_a(\varepsilon_a)},
\end{equation}
which for the strongly degenerate matter integrates to \begin{equation}\label{eq:eta_rta_deg}
    \widetilde{\eta}=\frac{n_a p_{\mathrm{F}a}^2}{5m_a^*}  \frac{\tau^{(2)}_a(\mu_a)}{1+i \omega_{\mathrm{BF}a} \tau^{(2)}_a(\mu_a)}.
\end{equation}

In the limit of strong magnetization, \new{$\omega_{BFa}\tau^{(2)}_a(\mu_a)\gg 1$}, we obtain
\begin{subequations}
\begin{eqnarray}
\eta_{1}&=&\frac{\eta_2}{4}=\frac{n_a p_{\mathrm{F}a}^2}{5m_a^*}  \frac{1}{4 \omega_{\mathrm{BF}a}^2 \tau^{(2)}_a(\mu_a)}.\label{eq:eta1_strong}\\
\eta_{3}&=&\frac{\eta_4}{2}=\frac{n_a p_{\mathrm{F}a}^2}{5m_a^*}\frac{1}{2\omega_{\mathrm{BF}a}}.\label{eq:eta3_strong}
\end{eqnarray}
\end{subequations}
Like for the thermal conductivity, the `Hall' components of the shear viscosity, $\eta_3$ and $\eta_4$, do not depend on $\tau^{(2)}$ and depend only on the quasiparticle number density.

\begin{figure}
  \includegraphics[width=\columnwidth]{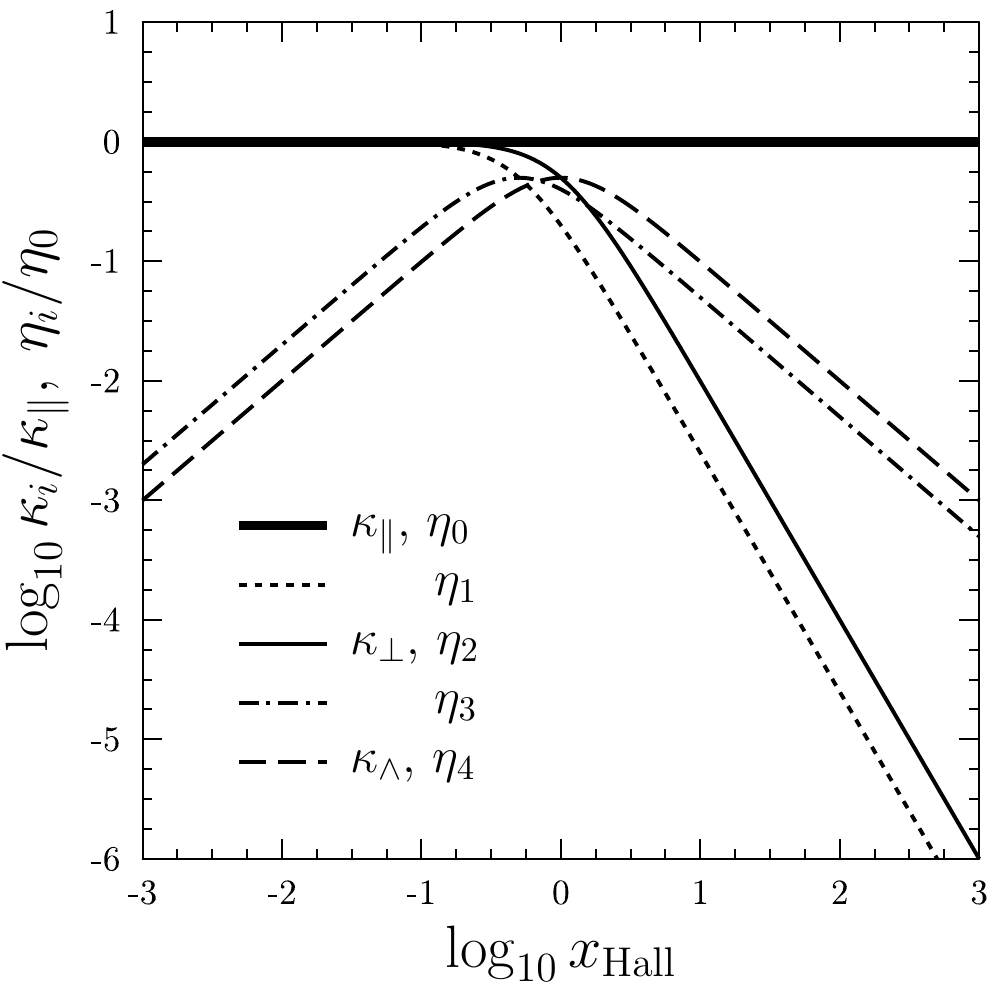}
\caption{Thermal conductivity components ($\kappa_\parallel$, $\kappa_\perp$, $\kappa_{\wedge}$) in units of $\kappa_\parallel$ and shear viscosity components ($\eta_0,\dots, \eta_4$) in units of $\eta_0$ in relaxation-time approximation as function of the Hall parameter $x_\mathrm{Hall}$, eq.~(\ref{eq:xHall}), see text for details. }
\label{fig:kinHall}       
\end{figure}

Although the relaxation time approximation is probably the simplest one, it catches the qualitative behavior of the tensor components of the transport coefficients with magnetic field. Let us introduce the Hall parameter
\begin{equation}\label{eq:xHall}
    x_{\mathrm{Hall}} =\omega_{\mathrm{BF}a}\tau_a(\mu_a),
\end{equation}
where $\tau_a(\mu_a)$ is either $\tau^{(1)}_a(\mu_a)$ or $\tau^{(2)}_a(\mu_a)$ depending on the tensor rank of the corresponding thermodynamic force. 
The particles are said to be magnetized when their $x_{\mathrm{Hall}}\gtrsim 1$.
In fig.~\ref{fig:kinHall} we plot the  dependence of various components of thermal conductivity and shear viscosity tensors on $x_{\mathrm{Hall}}$ following eqs.~(\ref{eq:kappa_sigma_rta}) and (\ref{eq:eta_rta_deg}).
All quantities  in fig.~\ref{fig:kinHall} are plotted relative to the longitudinal components ($\eta_0$ or $\kappa_\parallel$) of the transport coefficients in question. Until $x_{\mathrm{Hall}}\gtrsim 1$, transverse and longitudinal components of the transport coefficients are almost indistinguishable and the corresponding Hall components are small but gradually increase with $x_{\mathrm{Hall}}$. At $x_{\mathrm{Hall}}=1$, they reach maximum with $\kappa_\wedge/\kappa_\parallel=\eta_4/\eta_0=1/2$. At the same point $\kappa_\perp/\kappa_\parallel=\eta_2/\eta_0=1/2$. For larger $x_{\mathrm{Hall}}$, both transverse and Hall components of transport coefficient tensors decrease, albeit with different slopes. Namely, at large Hall parameters, one gets 
$\kappa_\perp/\kappa_\parallel=\eta_2/\eta_0=x_{\mathrm{Hall}}^{-2}$ and  $\kappa_\wedge/\kappa_\parallel=\eta_4/\eta_0=x_{\mathrm{Hall}}^{-1}$.

In general case, the relaxation time approximation does not hold. In particular this is the situation in the NS cores, where the relaxation time approximation is not applicable (sec.~\ref{sec:collisions}). 
In this case, one employs the general function $\Phi_{\ell m}$ in eqs.~(\ref{eq:Philm_rta})--(\ref{eq:Ltilde_vec}), (\ref{eq:eta_rta}) instead of the solution (\ref{eq:Philm_rta}).  Before turning to the calculation of transport coefficients in NS cores, let us briefly outline the variational method of the solution of the linearized transport equation.

\subsection{Variational principle}\label{sec:VarPrinciple}
The linearized transport eq.~(\ref{eq:kin_lin_fin}) is the linear integral equation for the set of functions $\Phi_a$, $a=1,\dots, r$, 
which in this section we denote collectively as $\Phi$, 
that can be symbolically written as
\begin{equation}\label{eq:int_eq_gen}
    X=(\hat{L}+\hat{M})\Phi,
\end{equation}
where $\hat{L}$ is the collision operator, which is symmetric and semi-negative definite in the Hilbert space of the functions $\Phi$, $\hat{M}$ is the magnetic operator which is anti-symmetric, and $X$ is the left-hand side. $\hat{L}$ and $\hat{M}$ here are understood as matrices in particle species state.

The general transport eq.~(\ref{eq:int_eq_gen}) can rarely be solved analytically. One of  the examples is the relaxation time approximation described in the previous section. For the traditional single-component Fermi-liquids, where the collision integral has the specific properties detailed in the next section, it is possible to construct the exact solutions of eq.~(\ref{eq:int_eq_gen}) \cite{JensenSmith1968PhLA,BrookerSykes1968PhRvL,SykesBrooker1970AnPhy}. This method was generalized to multicomponent problem by 
\cite{FlowersItoh1979ApJ,Anderson1987} and to the non-Hermitian case in \cite{PethickSchwenk2009PhRvC}.

In other cases other mathematical methods are necessary, for instance quadrature methods \cite{PolyaninBook}. Here we briefly outline the idea of the variational method of transport theory \cite{ZimanBook,DeGroot1980Book,Ichiyanagi1994PhR}.
Consider first the case $\bm{B}=0$, and, therefore, $\hat{M}=0$, and define the scalar product
\begin{equation}\label{eq:var_scalar}
    \langle \Psi|\Phi\rangle = \int_{\bm{p}} \Psi(\bm{p})^*\Phi(\bm{p})
\end{equation}
(in the multicomponent system the summation over species index is understood). 
Left multiplying eq.~(\ref{eq:int_eq_gen}) with $\langle \Phi|$, one obtains
\begin{equation}\label{eq:var_condition}
    \langle \Phi | X \rangle = \langle \Phi |\hat{L}|\Phi\rangle,
\end{equation}
where the right-hand side of this equation is clearly 
the entropy production rate according to eq.~(\ref{eq:entrprod_lin})
\begin{equation}\label{eq:varsigma_var}
    T \varsigma= - \langle \Phi | \hat{L} | \Phi\rangle.
\end{equation}
It can be proved \cite{ZimanBook} that the solution of eq.~(\ref{eq:int_eq_gen}) maximizes eq.~(\ref{eq:varsigma_var}) subject to condition eq.~(\ref{eq:var_condition}). This is one of the equivalent formulations of the variational principle of the transport theory \cite{ZimanBook}. One observes, that it is closely related to the second law of thermodynamics (see also \cite{Ichiyanagi1994PhR} for a review).
The variational principle can be reformulated to give the boundaries on the diagonal transport coefficients in the Onsager linear relations.

In practice, the variational principle is used via the expansion of the set of test functions over some convenient finite basis set $\{|\Phi_n\rangle\}_{n=1\dots N}$
\begin{equation}\label{eq:phi_var_expansion}
    \Phi=\sum_{n=1}^N c_n \Phi_n,
\end{equation}
with coefficients $c_n$ being the variational parameters. Then the variational equations take the form of the system of linear equations
\cite{ZimanBook,DeGroot1980Book}
\begin{equation}
    \langle \Phi_n |X \rangle= \sum_{m=1}^N \langle \Phi_n | \hat{L}|\Phi_m\rangle c_m
\end{equation}
which define the variational coefficients $c_n$. In a sense, this method approximates the integral kernel $\hat{L}$ with some degenerate  kernel \cite{PolyaninBook}.

Unfortunately, in the magnetized case the straightforward application of the variational principle is not possible. The physical reason is that the magnetic field term drops out of the entropy generation rate. One can formulate the similar principle for the eqs.~(\ref{eq:kineq_irred}) for functions $\Phi_{\ell m}$ (or using the function $\Phi(-\bm{B})$ in place of the complex conjugate function in eq.~(\ref{eq:var_scalar}) \cite{ZimanBook}). In this case the variational functional is only  stationary (have a saddle point) but not the extremal one, however the relevant equations for the variational basis will also take the form
\begin{equation}\label{eq:magvar}
    \langle \Phi_n |X \rangle= \sum_m \langle \Phi_n | \hat{L}+\hat{M}|\Phi_m\rangle c_m.
\end{equation}
This method for non-Hermitian operators is essentially the Bubnov-Galerkin method, see e.g., \cite{DeGroot1980Book,PolyaninBook}. We are not going in the discussion of the convergence of this procedure and assume that the sufficient conditions are fulfilled. In fact, we will use below the simplest variational solution based on the single variational function ($N=1$) which proved itself to be rather accurate under the NS conditions.

\subsection{Transport coefficients for the general Fermi-liquid collision integral}\label{sec:collisions}

For the multicomponent Fermi liquid inside NS cores the relaxation time approximations is not valid, therefore it is necessary to consider the generic linearized collision integral eq.~(\ref{eq:Boltz_lin}). Remember, that we assume that collision integral is scalar, i.e. the transition probability $w_{ab}$  depends only on the relative orientations of the momenta of the colliding particles. Following the formalism of sec.~\ref{sec:tensor}, the spherical components of the collision integral [see eq.~(\ref{eq:collint_spherical})] take the form 
\begin{equation}\label{eq:Ia_Iab_lm}
    I_{a}^{\ell}[\{\left(\Phi_b\right)_{\ell m}\}]=\sum_b  I_{ab}^{\ell}[\{\left(\Phi_b\right)_{\ell m}\}],
\end{equation}
where (see \ref{app} for details)
\begin{eqnarray}
    I_{ab}^{\ell}
    &=& -\frac{1}{T(1+\delta_{ab})} \int_{{\bm p}_{1'}}\int_{{\bm p}_{2}}\int_{{\bm p}_{2'}} w_{ab}({\bm p}_1,{\bm p}_{2},{\bm p}_{1'},{\bm p}_{2'})\nonumber \\
    &\times& {\cal F}_{ab}\left(
    ({\Phi_{a}}(p_1))_{\ell m}+({\Phi_{b}}(p_2))_{\ell m} {\cal P}_{\ell}(\hat{\bm{p}}_1\hat{\bm{p}}_2)\right.\nonumber \\
    &-&\left. ({\Phi_{a}}(p_{1'}))_{\ell m}{\cal P}_{\ell}(\hat{\bm{p}}_1\hat{\bm{p}}_{1'})\right.\nonumber \\
    &-&\left.({\Phi_{b}}(p_{2'}) )_{\ell m}{\cal P}_{\ell}(\hat{\bm{p}}_1\hat{\bm{p}}_{2'})\right),
    \label{eq:Iab_lm}
\end{eqnarray}
an ${\cal P}_\ell(y)$ are Legendre polynomials of the order $\ell$. Here we redefine $\bm{p}_1=\bm{p}$ in comparison with eq.~(\ref{eq:Boltz_lin}), that makes the description of the scattering process as $12\to 1'2'$ more symmetric (on the price of introducing the redundant index 1 in the left-hand side of the kinetic equation). Equation~(\ref{eq:Iab_lm}) is written for the elastic binary collisions, where the particle species in the input and output channels are the same. In case of the reactions (or inelastic collisions), in general deviation functions for four different species  (for the reaction in a form $a+b\leftrightarrow c+d$) enter 
eq.~(\ref{eq:Iab_lm}). This equation is still linear and the whole formalism below can be adapted to this case (assuming that the matter is in equilibrium with respect to such a reaction, i.e. $\mu_a+\mu_b=\mu_c+\mu_d$) with small modifications, although the kinematics of collisions become more involved.

The transition probability can be written as
\begin{equation}\label{eq:wab_def}
    w_{ab}=(2\pi)^4\delta^{(4)}(P'-P) |T_{ab}(12\to 1'2')|^2,
\end{equation}
where $\delta^{(4)}(P'-P)$ represent the four-momenta conservation with $P^\mu=p_1^\mu+p_2^\mu$ and ${P'}^{\mu}=p_{1'}^\mu+p_{2'}^\mu$ being the total quasiparticle pair four-momenta before and after the collision, respectively, and $T_{ab}(12\to 1'2')$ is the transition matrix element. Since we are considering the spin-unpolarised case, it is convenient to define the spin-averaged squared transition matrix element as
\begin{equation}\label{eq:Qab_def}
    {\cal Q}_{ab}=\frac{1}{4(1+\delta_{ab})} \sum_{\mathrm{spins}} |T_{ab}(12\to 1'2' )|^2.
\end{equation}

Owing to a strong degeneracy of Fermi liquids, it is possible to decompose angular and energy integration, putting quasiparticles on the Fermi surfaces whenever possible. Moreover, only in the vicinity the Fermi surface the well-defined quasiparticles exist. Then 
one can express (dropping species index for brevity)  $\mathrm{d} \bm{p}=m^* p \mathrm{d}\varepsilon \mathrm{d}\Omega_{\bm{p}}$. Moving from the energy integration to the integration over the dimensionless energy variable 
$x$, eq.~(\ref{eq:x_def}),
one can extend the  lower integration limit for $x$ from $-\mu/T$ to $-\infty$. This allows to introduce an additional symmetry variable, namely the parity with respect to the $x\leftrightarrow -x$ inversion, which we denote as $\xi=\pm 1$.
The collision integral conserves parity, therefore the perturbations that are odd and even  in $x$ decouple.
As stated already, at the same level of approximation one can take $h_a\approx \mu_a$ and neglect temperature corrections to the chemical potentials. Then the heat conductivity driving term becomes odd in $x$, while the shear viscosity and diffusion parts are even in $x$. 
This means, in particular, that  in the first approximation the thermal diffusion transport  coefficients vanish and therefore thermal conductivity and diffusion problems decouple. Based on this, we  can consider the shear viscosity, thermal conductivity, and diffusion problems separately \cite{BaymPethick,Anderson1987}.

\begin{figure}
  \includegraphics[width=\columnwidth]{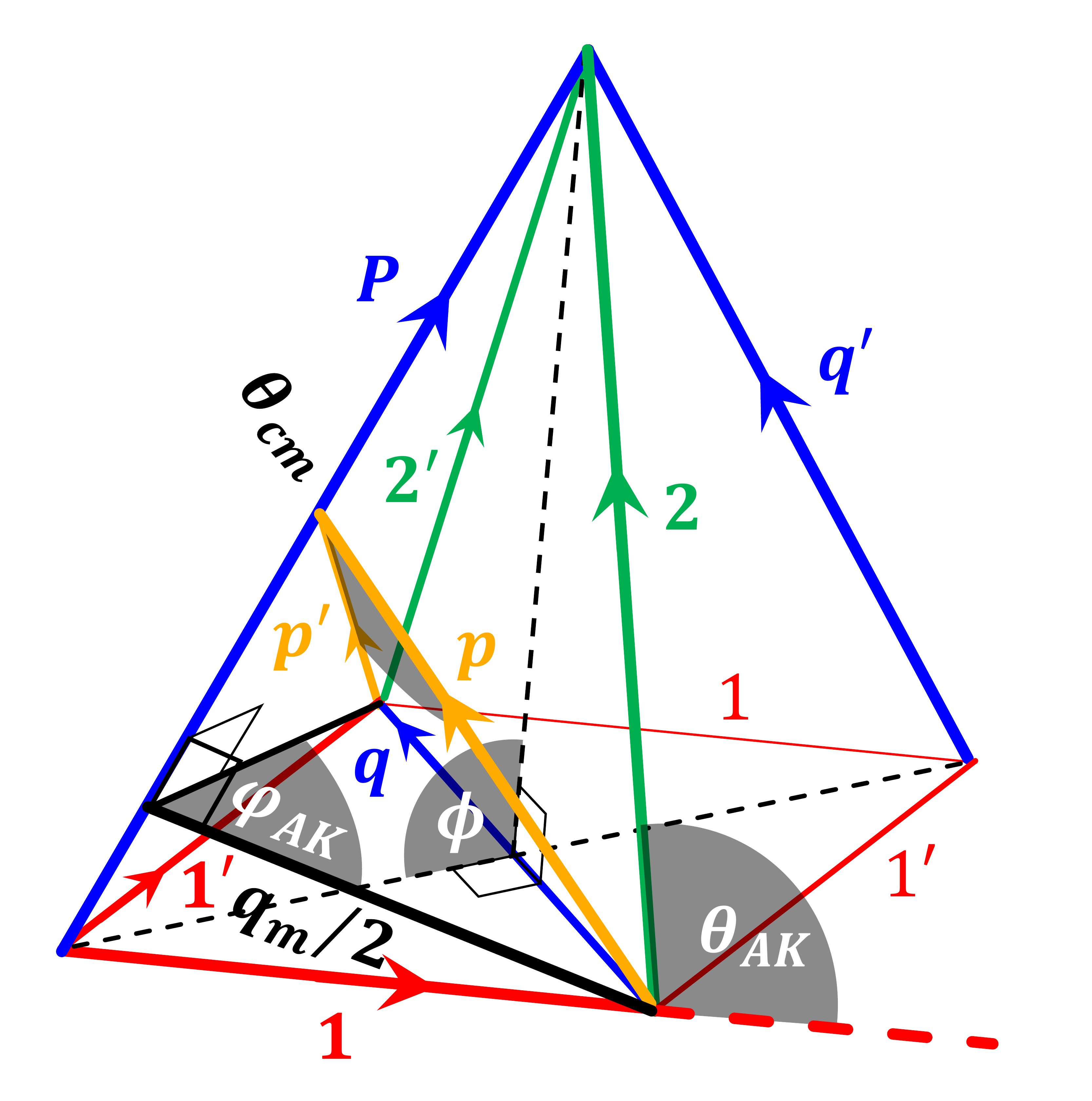}
\caption{Collision geometry for the process $12\to1'2'$ assuming $p_1=p_{1'}$ and $p_2=p_{2'}$. Various quasiparticle momenta and their combinations, as well as the relevant angles are identified, as discussed in the text.}
\label{fig:angles}       
\end{figure}

The relative orientation of four fixed-length vectors in space is fixed by two angular variables, see fig.~\ref{fig:angles}. Therefore of six angular integrations in eq.~(\ref{eq:Iab_lm}) only two remains (formally, three integrations are eliminated by the momentum-conserving delta-function and one integration over the third Euler angle of the body-frame coordinate system, say azimuthal angle of $\bm{p}_2$ around $\bm{p}_1$, is trivial). 
Different choices of these angular variables can be convenient depending on the properties of the collision matrix element. Figure~\ref{fig:angles} shows the geometry and various convenient angles and vectors.
The momenta of the $a$ and $b$ species particles before the collision are $\bm{p}_1$ and $\bm{p}_2$, respectively while their momenta after the collision are $\bm{p}_{1'}$ and $\bm{p}_{2'}$, respectively. Total colliding pair momentum $\bm{P}=\bm{p}_1+\bm{p}_2=\bm{p}_{1'}+\bm{p}_{2'}$ is conserved during collisions. The momentum transferred in collision from $b$ quasiparticle to $a$ quasiparticle is $\bm{q}=\bm{p}_{1'}-\bm{p}_1=\bm{p}_{2}-\bm{p}_{2'}$ and the transferred energy is $\omega=\varepsilon_{1'}-\varepsilon_{1}=\varepsilon_{2}-\varepsilon_{2'}$. In degenerate matter, the transferred energy is of the order of temperature (due to the Pauli blocking factor ${\cal F}_{ab}$ in eq.~(\ref{eq:Iab_lm})) and therefore is small. In addition, one defines the momentum $\bm{q}'=\bm{p}_{2'}-\bm{p}_1$ which describes the momentum transfer in exchange channel.

In what follows it is convenient to define the angular average of some function $A(\{\bm{p}_i\})$ as\footnote{Actually, this expression is not exactly an average, since for $A=1$ it gives $\pi(q_{\mathrm{max}}-q_{\mathrm{min}})$, see (\ref{eq:ang_int}) below. This definition, however, reduces the number of unnecessary factors in expressions.}
\begin{eqnarray}\label{eq:ang_av}
   \left\langle A(\{\bm{p}_i\})\right\rangle &=&\frac{p_1p_2p_{1'}p_{2'}}{16\pi^2} \int\mathrm{d}\hat{\bm{p}}_{1} \mathrm{d}\hat{\bm{p}}_{1'}\mathrm{d}\hat{\bm{p}}_{2}\mathrm{d}\hat{\bm{p}}_{2'}\ A(\{\bm{p}_i\})\nonumber\\
    &\times&
     \delta(\bm{p}_{1'}+\bm{p}_{2'}-\bm{p}_{1}-\bm{p}_2).
\end{eqnarray}
If the function $A(\{\bm{p}_i\})$ depends only on the relative orientations of the quasiparticle momenta, as in  eq.~(\ref{eq:Iab_lm}), the integration over $\hat{\bm{p}}_1$ gives just $4\pi$, and can be excluded from eq.~(\ref{eq:ang_av}).

Scattering matrices are frequently obtained from the microscopic theory as functions of the transferred momentum $\bm{q}$. Therefore, one of the convenient angular variables is $q=|\bm{q}|$ and for the second one one can use the angle $\phi$ between the planes ($\bm{p}_1\bm{p}_{1'}$) and ($\bm{p}_2\bm{p}_{2'}$), see fig.~\ref{fig:angles}. In this case the angular averages reduce to the integration over  $q$ and $\phi$:
\begin{eqnarray}\label{eq:ang_int}
  \left\langle A(\{\bm{p}_i\})\right\rangle=
    \int_{q_\mathrm{min}}^{q_\mathrm{max}} \mathrm{d} q \int_0^{\pi}\mathrm{d}\phi\ A(\{\bm{p}_i\}),
\end{eqnarray}
where $q_{\mathrm{max}}=\min(p_1+p_{1'},p_2+p_{2'})$ and $q_{\mathrm{min}}=\max(|p_1-p_{1'}|,|p_2-p_{2'}|)$.
In NS cores one has $|\omega| v_{\mathrm{F}a,b}\ll q$ and $q_\mathrm{min}\approx \max(|\omega|/v_{\mathrm{F}a},|\omega|/v_{\mathrm{F}b})$ \cite{Schmitt2018}. One can set $p_1\approx p_{1'}\approx p_{\mathrm{F}a}$, $p_2\approx p_{2'}\approx p_{\mathrm{F}b}$, $q_\mathrm{min}=0$ and $q_{\mathrm{max}}=\min(2{p_{\mathrm{F}a}},2p_{\mathrm{F}b})$. 

Alternatively, we notice that the cross-sections for the in-medium problems are frequently calculated from microscopic theory as a function of the total momentum $P$. In this case it is convenient to use $P$ and the Abrikosov-Khalatnikov angle $\phi_{\mathrm{AK}}$ \cite{Abrikosov1959RPPh} (see fig.~\ref{fig:angles}) instead of $q$  and $\phi$. The angle  $\phi_{\mathrm{AK}}$ is the angle between the $(\bm{p}_1\bm{p}_2)$ and $(\bm{p}_{1'}\bm{p}_{2'})$ planes. In this case, the angular average becomes
\begin{eqnarray}\label{eq:ang_int_nuc}
  \left\langle A(\{\bm{p}_i\})\right\rangle=
    \int_{P_\mathrm{min}}^{P_\mathrm{max}} \mathrm{d} P \int_0^{\pi}\mathrm{d}\phi_{\mathrm{AK}}\ A(\{\bm{p}_i\}),
\end{eqnarray}
where $P_\mathrm{min}=|p_{\mathrm{F}a}-p_{\mathrm{F}b}|$ and $P_\mathrm{max}=p_{\mathrm{F}a}+p_{\mathrm{F}b}$.
One can relate $P$ to the second Abrikosov-Khalatnikov angle, namely the angle $\theta_{\mathrm{AK}}$ between $\bm{p}_2$ and $\bm{p}_1$ via
\begin{equation}
    P^2=p_1^2+p_2^2+2p_1p_2\cos\theta_{\mathrm{AK}},
\end{equation}
which is convenient when $w_{ab}$ is known as a function of $\theta_{\mathrm{AK}}$ and $\phi_{\mathrm{AK}}$. In this case one obtains the standard definitions of angular averages of the Fermi-liquid theory \cite{BaymPethick,Anderson1987}.

Finally, it can be convenient to use $P$ and $q$ as the integration variables. This is achieved by utilizing the relation 
\begin{equation}
    q=q_m\sin\frac{\phi_{\mathrm{AK}}}{2},
\end{equation}
where $q_m(P)$ is the maximal value of  $q$ for a given $P$ and is equal to twice the height in the $(p_1,p_2,P)$ triangle (see fig.~\ref{fig:angles})
\begin{equation}\label{eq:qm}
    q_m=\frac{2p_1p_2 \sin\Theta_{\mathrm{AK}}}{P}=\frac{\sqrt{4p_{\rm Fa}^2 p_{\rm Fb}^2 - (p_{\rm Fa}^2+p_{\rm Fb}^2-P^2)^2}}{P}.
\end{equation}
Then
\begin{equation}\label{eq:ang_av_Pq}
    \left\langle A(\{\bm{p}_i\})\right\rangle= 
    2\int_{P_\mathrm{min}}^{P_\mathrm{max}} \mathrm{d} P \int_0^{q_m}\frac{\mathrm{d}q}{\sqrt{q_m^2-q^2}}\ \new{A(\{\bm{p}_i\})}. 
\end{equation}
The concise description of the different choices of the angular variables can be found in \cite{PethickSchwenk2009PhRvC}.

The remaining integration is performed over energy variables, or over $x_i$, $i=(1,\,2,\,1',\,2')$, in the  dimensionless form. One integration is eliminated by the energy conservation and generally only the two integrations remain. One of them can be carried out further if the transition probability can be placed on shell, i.e., if it does not depend on the transferred energy $\omega=\varepsilon_{1'}-\varepsilon_{1}$.
This is the standard approximation for the traditional Fermi-liquids \cite{BaymPethick}. However, in principle, transition probability can depend on the transferred energy. This is the typical situation encountered in the classical, non-degenerate plasmas. But, it turns out that the electromagnetic interaction  in the relativistic plasma (and also the QCD interactions for the quark matter) also possesses this property \cite{Heiselberg1992NuPhA}.

To proceed further, it is convenient to express
\begin{equation}\label{eq:Dka_deg}
D_k^a(p_1)\approx v_{\mathrm{F}a}\widetilde{D}_k(p_{\mathrm{F}a})\widetilde{X}_k(x_1),
\end{equation}
where $\widetilde{X}_k(x_1)$ is the $x$-dependent part of the driving term (namely, $x_1$ for the thermal conductivity and 1 for other processes considered), and $\widetilde{D}_k(p_{\mathrm{F}a})$ is the remainder, which depends on the on-shell momentum $p_{\mathrm{F}a}$. This decomposition is possible since we work in the vicinity of the Fermi surface which is guaranteed by the presence of the factor $f'_F(x)$ in all integral expressions for physical quantities. 
\new{The} functions entering  eq.~(\ref{eq:Phia_def}) are also given in tab.~\ref{tab:transp_params}.

Consider now the thermal conductivity or shear viscosity problems. As said above, these transport problems can be considered separately owing to the symmetry restrictions. The diffusion problem will be treated later. 
The irreducible sherical componsnts of the deviation function $\Phi^k_a$ can be represented ($k=\kappa,\eta$) as 
\begin{eqnarray}
    \left(\Phi^k_a(p_1)\right)_{\ell m}&=&-\sqrt{\frac{4\pi}{2\ell+1}} \widetilde{D}_k(p_{\mathrm{F}a}) \left ({\cal X}_k\right)_{\ell m}\nonumber\\
    &\times&\lambda^{(k,\mathrm{eff})}_{a}\left(\Psi^{k}_a(x_1)\right)_{\ell m},\label{eq:Phia_def}
\end{eqnarray}
where $\left(\Psi^{k}_a(x_1)\right)_{\ell m}$ are the dimensionless functions of the energy variable. Notice that there is no summation over $k$ in comparison to eq.~(\ref{eq:Philm_rta}) due to decoupling of the transport phenomena considered.

The quantities $\lambda^{(k,\mathrm{eff})}_{a}$ in eq.~(\ref{eq:Phia_def}) are the auxiliary effective mean free paths. Traditionally, in the Fermi-liquid transport theory one utilizes the concept of the effective relaxation times (e.g., \cite{BaymPethick}, see also sec.~\ref{sec:tensor}), which are related to the effective mean free paths as $\lambda_{\mathrm{eff}}=v_{\mathrm{F}}\tau_{\mathrm{eff}}$. Since the NS cores contain the relativistic leptons with Fermi velocities of the order of the speed of light and the heavy non-relativistic baryons (e.g., protons), the relaxation times for these quasiparticles are quite different, while the effective mean free paths are closer \cite{Shternin2020PhRvD}, and usage of $\lambda_{\mathrm{eff}}$ instead of  $\tau_{\mathrm{eff}}$ seems to be more convenient, although this is of course a matter of taste. In the standard approach, one takes the single quasiparticle excitation mean free path (relaxation time) for $\lambda^{(k,\mathrm{eff})}_{a}$ \cite{BaymPethick,Anderson1987}. However this approach is problematic in  the relativistic plasma in NS cores, since this quantity diverges at the Fermi surface due to the properties of the long-range electromagnetic interaction (e.g., \cite{Blaizot1996NuPhA}). This potentially could present a problem for the whole theory, however the transport is governed by the quasiparticles slightly distorted from the Fermi surface and the resulting transport  mean free path stays finite.
In principle, the auxiliary effective mean free paths   $\lambda^{(k,\mathrm{eff})}_{a}$ are  then selected based on the aesthetic arguments (e.g. in such a way that the system of transport equations is made maximally compact). 

Once the functions $\Psi^k_a(x)$ defined in eq.~(\ref{eq:Phia_def}) are found from the solution of the linearized transport equation, the complex thermal conductivity and shear viscosity coefficients can be found. The corresponding expressions are given by eqs.~(\ref{eq:kappa_psi}) and (\ref{eq:eta_psi}) in \ref{app}. The complete systems of linearized transport equation for thermal conductivity and shear viscosity problems for the case when the scattering probability can depend on the transferred energy $\omega$ are rather lengthy and are given in \ref{app}. Notice, that when $w_{ab}$ does not depend on $\omega$, the situation simplifies significantly and the exact solution of the system of the multicomponent transport equations can be constructed \cite{Anderson1987}. In general situation, the exact analytical solution is not known.

Here we employ the simplest variational solution  based on the one-parametric family of the test functions as described in sec.~\ref{sec:VarPrinciple}. In this respect, we consider the equation for $m=1$ (sec.~\ref{sec:tensor}) and  replace
\begin{equation}\label{eq:first_var_app}
    \lambda^{(k,\mathrm{eff})}_{a}(\Psi^{k}_a(x_1))_{\ell1}\to \widetilde{\lambda}^k_a \widetilde{X}_k(x_1)
\end{equation} 
in eq.~(\ref{eq:Phia_def}), where $\widetilde{\lambda}^k_a$ is now a (complex) variational parameter. The tilde above $\widetilde{\lambda}^k_a$ has the same meaning as in eq.~(\ref{eq:Ltilde}). Substitution eq.~(\ref{eq:first_var_app}) is a simplest one which respects the parity of the function $\widetilde{\Psi}(x_1)$ and clearly resembles the solution (\ref{eq:Philm_rta}) obtained in the relaxation time approximation.

Using eq.~(\ref{eq:magvar}) and eq.~(\ref{eq:phi_var_expansion}) with $N=1$  and evaluating the scalar products one obtains the  following system of equations for $k=\kappa,\eta$
\begin{equation}\label{eq:kinvar_system}
    1=\sum_{b}\left(\Lambda^k_{ab}\widetilde{\lambda}^k_a+{\Lambda'}^{k}_{ab}\widetilde{\lambda}^k_b\right) +i\frac{\omega_{\mathrm{BF}a}}{v_{\mathrm{F}a}}\widetilde{\lambda}^k_a
\end{equation}
at the lowest order variational solution. The matrices $\Lambda^k_{ab}$ and ${\Lambda'}^k_{ab}$
contain all information about the quasiparticle collisions and are related to the transport cross-section in the system. They differ for the different transport problems considered, see below, but the general structure is the same. We will call further $\Lambda^k_{ab}$ as transport matrix. Equation (\ref{eq:kinvar_system}) is written in the form that isolates different pair collision mechanisms. It is instructive to rewrite it in an explicit form of the $r\times r$ linear system for $\widetilde{\lambda}^k_a$
\begin{equation}\label{eq:kinvar_system1}
    1=\left(\Lambda^k_{a}+i\frac{\omega_{\mathrm{BF}a}}{v_{\mathrm{F}a}}\right)\widetilde{\lambda}^k_a+ \sum_{b\neq a}{\Lambda'}^{k}_{ab}\widetilde{\lambda}^k_b,\quad a=1\dots r,
\end{equation}
where the diagonal elements of the (transport) matrix of the system are
\begin{equation}\label{eq:Lambda_tot}
    \Lambda^k_a=\sum_b \Lambda^k_{ab} +{\Lambda'}^k_{aa}.
\end{equation}

Once the parameters $\widetilde{\lambda}^k_a$ are found from the system of equations (\ref{eq:kinvar_system1}), the \new{partial} thermal conductivity and shear viscosity  are given by the standard expressions (cf. eqs.~(\ref{eq:kappa_rta}) and (\ref{eq:eta_rta_deg})) 
\begin{eqnarray}
    \widetilde{\kappa}_a&=&\frac{\pi^2 n_a T}{3p_{\mathrm{F}a}}\widetilde{\lambda}_a^{\kappa},\label{eq:kappa_var}\\
    \widetilde{\eta}_a&=&\frac{n_ap_{\mathrm{F}a}}{5} \widetilde{\lambda}_a^{\eta}.\label{eq:eta_var}
\end{eqnarray}
\new{According to eq.~(\ref{eq:Lkk_Gk}), the total $\kappa$ and $\eta$ are given by a sum of the partial contributions over particle species.}

The transport matrices $\Lambda^k_{ab}$ in eq.~(\ref{eq:kinvar_system}) are given by the angular averages of the transition matrix element with certain angular factors depending on the transport coefficient considered. Specifically, for the thermal conductivity, $k=\kappa$ ($\ell=1$, $\xi=1$)
\begin{subequations}\label{eq:trmat_kappa_ab}
\begin{eqnarray}
    \Lambda^\kappa_{ab} &=& \frac{3 T^2 m_a^{*2}m_b^{*2}}{4\pi^4 p_{\mathrm{F}a}^2}
    \int_w \left\langle\left(\frac{w^2}{\pi^2} + \left[\frac{1}{3}-\frac{w^2}{6\pi^2}\right]\frac{q^2}{p_{\mathrm{F}a}^2}\right){\cal Q}_{ab}\right\rangle,\nonumber\\&& \label{eq:Lambda_ab_kappa}\\
    {\Lambda'}^{\kappa}_{ab} &=& -\frac{3 T^2 m_a^{*2}m_b^{*2}}{4\pi^4 p_{\mathrm{F}a}^2}
    \int_w \frac{w^2}{\pi^2} \left\langle\left(\frac{\bm{p}_1(\bm{p}_{2}+\bm{p}_{2'})}{2p_{\mathrm{F}a}p_{\mathrm{F}b}}\right){\cal Q}_{ab}\right\rangle,\nonumber\\
    &&\label{eq:Lambda1_ab_kappa}
\end{eqnarray}
\end{subequations}
where $w=\omega/T$ is the dimensionless transferred energy
and the abbreviation 
\begin{equation}\label{eq:intw_def}
    \int_w  = \int_0^\infty \mathrm{d}w \frac{(w/2)^2}{\sinh^2(w/2)} 
\end{equation}
is introduced. 

For the shear viscosity, $k=\eta$ ($\ell=2$, $\xi=+1$) the transport matrix elements are
\begin{subequations}\label{eq:trmat_eta_ab}
\begin{eqnarray}
    \Lambda^\eta_{ab} &=& \frac{3 T^2 m_a^{*2}m_b^{*2}}{4\pi^4 p_{\mathrm{F}a}^2}
    \int_w \left\langle\frac{q^2}{p_{\mathrm{F}a}^2}\left(1-\frac{q^2}{4p_{\mathrm{F}a}^2}\right){\cal Q}_{ab}\right\rangle,\nonumber\\
    &&\label{eq:Lambda_ab_eta}\\
    {\Lambda'}^{\eta}_{ab} &=& -\frac{3 T^2 m_a^{*2}m_b^{*2}}{4\pi^4 p_{\mathrm{F}a}^2}
    \int_w  \left\langle\frac{q^2}{p_{\mathrm{F}a}^2}\left(\frac{\bm{p}_1(\bm{p}_{2}+\bm{p}_{2'})}{2p_{\mathrm{F}a}p_{\mathrm{F}b}}\right){\cal Q}_{ab}\right\rangle.\nonumber\\
    &&\label{eq:Lambda1_ab_eta}
\end{eqnarray}
\end{subequations}

In principle, the determination of the diffusion coefficients should proceed in similar way. The certain care is needed since there are $r-1$ independent forces in the left-hand side of the Boltzman equation and one needs to employ the additional condition of fit (as discussed in sec.~\ref{sec:Hydro}), the conditions specifying the Eckart frame can be used). There are different strategies for treating this problem, e.g.,  \cite{Ferziger1972Book} (see also the appendix in Ref.~\cite{Dommes2020PhRvD}). However, in the lowest-order variational approximation it is instructive to employ the Stephan-Maxwell scheme eqs.~(\ref{eq:Xa_SM}) and express $\bm{d}_{(a)}$ through the diffusion fluxes. This is easily achieved by writing
\begin{equation}\label{eq:phivar_diffusion}
    \Phi_a(\bm{p}_1)=\bm{p}_1\bm{w}_{(a)}
\end{equation}
with constant vectors $\bm{w}_{(a)}$ which have the meaning of the diffusion velocity since in this case
\begin{equation}
    \Delta \bm{j}_{(a)}=n_a\bm{w}_{(a)}.
\end{equation}
Integrating the linearized transport equation multiplied with $\bm{p}_1$ results in
\begin{eqnarray}
    &n_a&\left(\bm{d}_{(a)}+\frac{h_a}{hn}\left[\bm{J}\times\bm{B} \right]\right)=\int_{\bm{p}_1}\bm{p}_1I_a[\left\{\Phi_b\right\}]\nonumber\\&&+q_a\left[\Delta\bm{j}_{(a)}\times \bm{B}\right]=\sum_bJ_{ab}(\bm{w}_b-\bm{w}_a)\nonumber\\
    &&+n_aq_a[\bm{w}_a\times\bm{B}],\label{eq:Ohm}
\end{eqnarray}
where 
\begin{equation}\label{eq:Jab_var}
    J_{ab}=\frac{T^2m_a^{*2}m_b^{*2}}{12\pi^6}\int_w\langle q^2 {\cal Q}_{ab}\rangle
\end{equation}
are the momentum transfer rates (or friction coefficients).
In this simplest case the friction coefficients $J_{ab}$
do not depend on the magnetic field. Going beyond eq.~(\ref{eq:phivar_diffusion}) in principle results in tensor structure of the coefficients $J_{ab}$, characteristic for the vector transport coefficients (sec.~\ref{sec:tensor}). 
\new{The equation of motions (\ref{eq:Ohm}) serve as a part of the input for the studies of the magnetic field evolution in NS cores, e.g.~\cite{Goldreich1992ApJ,Shalybkov1995MNRAS,Beloborodov2016ApJ,Passamonti2017MNRAS,Gusakov2017PhRvD,Castillo2020MNRAS}. When the diffusion driving forces $\bm{d}_{(a)}$ contain only the electric field contribution (no chemical gradients), sometimes it is desirable to invert eq.~(\ref{eq:Ohm}) and find the relation $\bm{J}=\hat{\sigma}\bm{E}$, where $\hat{\sigma}$ is the electrical conductivity tensor. Electrical conductivity depends on the momentum transfer rates $J_{ab}$ in a non-trivial way, see, e.g., \cite{Iakovlev1991Ap&SS}. A problem of inversion of eq.~(\ref{eq:Ohm}) is discussed in \ref{sec:econd} where also a practical expression for the electrical conductivity for a specific case of $npe\mu$ matter is given.}

In traditional Fermi liquids, the squared transition matrix element ${\cal Q}_{ab}$ does not depend on $\omega=wT$. Then integrals over $w$ in eqs.~(\ref{eq:trmat_kappa_ab}), (\ref{eq:trmat_eta_ab}) and (\ref{eq:Jab_var}) can be taken analytically and one obtains instead of eqs.~(\ref{eq:Lambda_ab_kappa})--(\ref{eq:Jab_var}) the following expressions:
\begin{subequations}\label{eq:trmat_kappa static}
\begin{eqnarray}
    \Lambda^\kappa_{ab} &=& \frac{ T^2 m_a^{*2}m_b^{*2}}{5\pi^2 p_{\mathrm{F}a}^2}
    \left\langle\left(1+\frac{q^2}{4p_{\mathrm{F}a}^2}\right){\cal Q}_{ab}\right\rangle,\\
    {\Lambda'}^{\kappa}_{ab} &=& -\frac{ T^2 m_a^{*2}m_b^{*2}}{5\pi^2 p_{\mathrm{F}a}^2}
     \left\langle\left(\frac{\bm{p}_1(\bm{p}_{2}+\bm{p}_{2'})}{2p_{\mathrm{F}a}p_{\mathrm{F}b}}\right){\cal Q}_{ab}\right\rangle.
\end{eqnarray}
\end{subequations}
\begin{subequations}\label{eq:trmat_eta_static}
\begin{equation}
    \Lambda^\eta_{ab} = \frac{ T^2 m_a^{*2}m_b^{*2}}{4\pi^2 p_{\mathrm{F}a}^2}
    \left\langle\frac{q^2}{p_{\mathrm{F}a}^2}\left(1-\frac{q^2}{4p_{\mathrm{F}a}^2}\right){\cal Q}_{ab}\right\rangle,
\end{equation}
\begin{equation}
    {\Lambda'}^{\eta}_{ab} = -\frac{T^2 m_a^{*2}m_b^{*2}}{4\pi^2 p_{\mathrm{F}a}^2}
     \left\langle\frac{q^2}{p_{\mathrm{F}a}^2}\left(\frac{\bm{p}_1(\bm{p}_{2}+\bm{p}_{2'})}{2p_{\mathrm{F}a}p_{\mathrm{F}b}}\right){\cal Q}_{ab}\right\rangle,
\end{equation}
\end{subequations}
\begin{equation}\label{eq:Jab_var_long}
    J_{ab}=\frac{T^2m_a^{*2}m_b^{*2}}{36\pi^4}
    \langle q^2 {\cal Q}_{ab}\rangle.
\end{equation}
In general, as stated above, in this case it is possible to obtain exact solutions of the system of transport equations analytically \cite{Anderson1987,PethickSchwenk2009PhRvC}, but the properties of electromagnetic interactions in NS cores spoil this picture.

Therefore, in order to find the transport coefficients of the magnetized multicomponent liquid inside the NS cores one needs to be supplied by the quasiparticle spectra (effective masses) and the squared matrix element ${\cal Q}_{ab}(\omega,q,\phi)$. The discussion of the latter \new{one} is the subject of the next section.

\section{Microphysics of quasiparticle  collisions in NS cores}\label{sec:interactions}
We assume here that the NS cores contain light leptonic component (electrons and muons) and heavy baryon component. For the transport properties of the quark NS core see, e.g., review in \cite{Schmitt2018} and references therein.
Leptons interact between \new{themselves} and with charged baryons via the electromagnetic forces, while baryons  mainly interact by the strong forces. The presence of the dense medium modifies the properties of both these interaction channels. When collisions of charged baryons are considered, in principle the interference between the electromagnetic and strong interaction channels needs to be taken into account \cite{Machleidt2011PhR}. However, \new{it was found that the charged baryons 
usually give negligible contribution to the transport coefficients \cite{FlowersItoh1979ApJ,Shternin2013PhRvC,Shternin2020PhRvD,Shternin2021Univ}}. Therefore, we neglect the interference contribution and consider the electromagnetic and strong channels separately for simplicity.

\subsection{Electromagnetic interactions}\label{sec:em}
The properties of the electromagnetic interaction are strongly affected by the character of the plasma screening. In the relativistic plasma the magnetic part of the interaction becomes dominant which in the leading order is screened dynamically. 
This changes considerably the behavior of the transport cross-sections in cold relativistic plasmas resulting in their non-Fermi liquid temperature dependence. This was realized in Refs.~\cite{Baym1990PhRvL,Heiselberg1992NuPhA,Heiselberg:1993cr} 
and applied to the nucleonic NS core matter by 
\cite{ShterninYakovlev2007,ShterninYakovlev2008,Shternin2008JETP,Stetina2019arXiv}, see \cite{Schmitt2018} for a review.

The transition matrix at the one-loop level is given as
\begin{equation}\label{eq:matel_em}
T_{ab}\to M_{ab}=4\pi\alpha_f\left(\frac{{\cal J}_{a}^0{\cal J}_{b}^{0}}{q^2+\Pi_l(\omega,q)}-\frac{\bm{\mathcal{J}}_{a,t}\bm{\mathcal{J}}_{b,t}}{q^2+\Pi_t(\omega,q)}\right),
\end{equation}
where $\alpha_f$ is the fine-structure constant, $\bm{\mathcal{J}}_{a,t}$ is the transverse with respect to $\bm{q}$ component of the spatial part of the transition current, ${\cal J}^0_a$ is the timelike component of the transition current,  and $\Pi_l(\omega,q)$  and $\Pi_t(\omega,q)$ are the longitudinal and transverse polarization operators of the photon in medium, respectively. When $a=b$, the exchange contribution needs to be taken into account according to 
\begin{equation}\label{eq:matel_em_exch}
    T_{ab}(12\to 1'2')\to M_{ab}(12\to 1'2')-M_{ab}(12\to 2'1'),
\end{equation}
which leads to appearance of the exchange-interference terms in the squared matrix element. However \new{these interference terms} lead only to  small corrections to the transport cross-sections (due to dominance of the small-angle collisions) and can be neglected \cite{ShterninYakovlev2007,ShterninYakovlev2008}.

Remember that the relevant energy transfer in the collisions is of the order of temperature due to Pauli blocking. 
Therefore it is enough to take the polarization operators in the limits  ($\omega, q \ll p_{\mathrm{F}a}$) and $\omega v_{\mathrm{F}a}\ll q$. In this case the longitudinal polarization operator is simply the squared Debye mass

\begin{equation}\label{eq:Pil_norm}
    \Pi_l(\omega,q)=q_l^2\equiv \frac{4\alpha_f}{\pi}\sum_a Z_a^2 m_a^* p_{\mathrm{F}a},
\end{equation}
where $Z_a$ is the quasiparticle charge number so that $q_a=Z_a e$. In NS cores, $Z_a=\pm 1$. Notice the appearance of the quasiparticle effective mass in eq.~(\ref{eq:Pil_norm}). In contrast, the transverse screening is not static, and in the leading order it is governed by the Landau damping. The transverse polarization operator is 
\begin{equation}\label{eq:Pit_norm}
    \Pi_t(\omega,q)=i\frac{\pi}{4}\frac{\omega}{q} q_t^2\equiv i\frac{\omega}{q} \alpha_f\sum_a Z_a^2 p_{\mathrm{F}a}^2\equiv i\Lambda^3/q,
\end{equation}
where the characteristic momentum scale $\Lambda$ is  introduced, which obeys $\Lambda\ll q_{l}\ll p_{\mathrm{F}a}$.
At high temperatures ($T\gtrsim q_l$), the static limit breaks down and one needs to consider more general structure of the polarization operator \cite{Heiselberg:1993cr}. Notice that the detailed analysis of the impact of various approximations to $\Pi_l$ and $\Pi_t$ is performed by \cite{Stetina2018PhRvC,Stetina2019arXiv}, where it is found that the so-called hard dense loop result for the in-medium polarization  functions is always a good approximation in the NS core context. 

In order to calculate ${\cal Q}_{ab}$, it is necessary to take the spin trace of the squared matrix element eq.~(\ref{eq:matel_em}). The electromagnetic transition current in free space is given by
\begin{equation}
    {\cal J}^\mu_{a}=\overline{u}_a(p_{1'})
    \left(F_{a1}(-q^2)\gamma^\mu+i\frac{F_{a2}(-q^2)}{2m_a}\sigma^{\mu\nu}q_\nu\right) u_a(p_1),\label{eq:em_trans_current}
\end{equation}
where $\sigma^{\mu\nu} =i/2[\gamma^\mu,\gamma^\nu]$, $F_{a1}(-q^2)$ and $F_{a2}(-q^2)$ are electromagnetic form-factors, $m_a$ is the bare particle mass, and the $u_a(p_1)$  is the Dirac bispinor normalized as
\begin{equation}\label{eq:DiracNorm}
    \overline{u}_a u_a = \frac{m_a}{\varepsilon_a}.
\end{equation}
This normalization is used in order to keep the notations for transition  matrix elements close in relativistic and non-relativistic cases.

Defining
\begin{equation}
    G^{\mu\nu}_a=\frac{1}{2}\sum\limits_{\mathrm{spins}} {\cal J}^\mu_a {\cal J}^\nu_a
\end{equation}
one obtains
\begin{eqnarray}
    G^{\mu\nu}_a&=&\frac{F_{a1}^2(-q^2)+F_{a2}^2(-q^2)\,q^2/(4m_a^2)}{4\varepsilon_a^{2}}
    P_a^\mu P_a^\nu
    \nonumber\\
    &+&\frac{F_{a,m}^2(-q^2)}{4\varepsilon^{2}_a}\left(Q^2g^{\mu\nu}-Q^\mu Q^\nu\right),\label{eq:Gmunu}
\end{eqnarray}
where $F_{a,m}(-q^2)=F_{a1}(-q^2)+F_{a2}(-q^2)$ is the Sachs magnetic form-factor,  $P^\mu_a=p_{1}^\mu+p_{1'}^\mu$ and $Q^\mu=p_{1'}^\mu-p_1^\mu$.  Notice that for leptons the form factors are $F_{\ell 1}=F_{\ell m}=1$, $F_{\ell 2}=0$ (\new{neglecting QED corrections}), while for proton $F_{p1}(0)=1$, $F_{pm}(0)=2.79$, and for the neutron $F_{n1}(0)=0$, $F_{nm}(0)=-1.91$.

\new{
The squared transition matrix element is then 
\begin{eqnarray}\label{eq:Q-G}
{\cal Q}_{ab}&=&16\pi^2\alpha_f^2\left(\frac{G^{00}_a G^{00}_b}{\left|q^2+\Pi_l(\omega,q)\right|^2}\right.\nonumber\\
&&-2\mathrm{Re}\ \frac{G_{a}^{k0} G_{b}^{k0}}{(q^2+\Pi_l(\omega,q))(q^2+\Pi_t(\omega,q))^*}\nonumber\\
&&\left.+\frac{G^{kl}_a G^{kl}_b}{\left|q^2+\Pi_t(\omega,q)\right|^2}\right),
\end{eqnarray}
where we neglected exchange-interference term  for identical particles ($a=b$) and took into account that 
$\bm{\mathcal{J}}_{a,t}\approx \bm{\mathcal{J}}_{a}$ in the limit $\omega v_{\mathrm{F}a}\ll 1$ since $Q=(0,\bm{q})$, $Q^2=-q^2$, and the spatial component of $P_{a}$ is transverse to $\bm{q}$. 
}

Equation~(\ref{eq:Gmunu}) is obtained for the free non-interacting particles and can be modified by the Fermi-liquid effects. For degenerate leptons which form almost ideal gas, it is enough to use eq.~(\ref{eq:Gmunu}). Since we are interested in the transition probabilities in the vicinity of the Fermi surface, we can set $\varepsilon_a=m^*_a$. 

The situation is different for baryons. However, the main contribution to the scattering probabilities comes from the small-angle collisions, where $q\ll p_{\mathrm{F}a}$. In non-relativistic limit (applicable well for protons in NS cores) this also means that $q\ll m_a$.  For charged baryons (protons) this allows one to neglect the magnetic term containing $F^2_{a,m}$ and do not discuss its renormalization in a first approximation. At the same level of approximation, one can neglect the kinematic structure of the form-factors and the anomalous magnetic moment interaction and set  \new{$F_{a1}(-q^2)\approx F_{a1}(0)=1$, $F_{a2}(-q^2)\approx F_{a2}(0)=0$}. Then the same expressions for lepton and lepton-baryon scattering \new{can be used}. Finally, notice that the transition  current is the matrix element of the velocity operator. That means that in vicinity of the Fermi surface one should substitute $\varepsilon_a\to m_a^*$ in denominator in eq.~(\ref{eq:Gmunu}) as in the case of free particles. These considerations are not valid for neutrons  which are chargeless particles. However their electromagnetic interaction with other charged particles is small and can be neglected.

\new{Therefore, using eqs.~(\ref{eq:Gmunu})--(\ref{eq:Q-G}),} one finds the following expression for ${\cal Q}_{ab}$
\begin{eqnarray}\label{eq:Qci_em}
{\cal Q}_{ab}&=&16\pi^2\alpha_f^2Z_a^2 Z_b^2
\left(\frac{L_l}{\left|q^2+\Pi_l(\omega,q)\right|^2}\right.\nonumber\\
&&-2\mathrm{Re}\ \frac{v_{\mathrm{F}a} v_{\mathrm{F}b} L_{tl}}{(q^2+\Pi_l(\omega,q))(q^2+\Pi_t(\omega,q))^*}\nonumber\\
&&\left.+\frac{v_{\mathrm{F}a}^2 v_{\mathrm{F}b}^2 L_t}{\left|q^2+\Pi_t(\omega,q)\right|^2}\right),
\end{eqnarray}
where the kinematic numerators are
\begin{subequations}\label{eq:Ls}
\begin{eqnarray}
L_l&=&\left(1-\frac{q^2}{4m_{a}^{*2} }\right)\left(1-\frac{q^2}{4m_{b}^{*2} }\right),\label{eq:Ll}\\
L_{tl}&=&\sqrt{\left(1-\frac{q^2}{4p_{\mathrm{F}a}^2}\right)\left(1-\frac{q^2}{4p_{\mathrm{F}b}^2}\right)}{\cos\phi},\label{eq:Ltl}\\
L_t&=&\left(1-\frac{q^2}{4p_{\mathrm{F}a}^2}\right)\left(1-\frac{q^2}{4p_{\mathrm{F}b}^2}\right){\cos^2\phi}\nonumber\\
&&+\frac{q^2}{4p_{\mathrm{F}a}^2}+\frac{q^2}{4p_{\mathrm{F}b}^2}.\label{eq:Lt}
\end{eqnarray}
\end{subequations}
The eqs.~(\ref{eq:Qci_em})--(\ref{eq:Ls}) needs to be substituted into the expressions for the various transport matrix components, eqs.~(\ref{eq:trmat_kappa_ab}), (\ref{eq:trmat_eta_ab}), and (\ref{eq:Jab_var}). We need to perform three integrations (over $q,\phi$, and $w$). The integration over $\phi$ is trivial since the angular weighting factors in `direct' transport matrices eqs.~(\ref{eq:Lambda_ab_kappa}), (\ref{eq:Lambda_ab_eta}), and (\ref{eq:Jab_var}) do not depend on $\phi$, while the weightnig factor in the non-diagonal transport matrices ${\Lambda'}^k_{ab}$ in eqs.~(\ref{eq:Lambda1_ab_kappa}) and (\ref{eq:Lambda1_ab_eta}) in contrast are proportional to $\cos\phi$. Therefore the longitudinal ($L_{l}$) and transverse ($L_t$) channels contribute to transport matrices $\Lambda^k_{ab}$, $J_{ab}$, so one can decompose $\Lambda^k_{ab}=\Lambda^{k,l}_{ab}+\Lambda^{k,t}_{ab}$, while the non-diagonal matrices ${\Lambda'}^k_{ab}$ contain only the mixed ($L_{lt}$) channel. After the $\phi$ integration, eq.~(\ref{eq:Qci_em}) contains only the polynomial in $q^2$ terms in the numerator. The integrals over $q$ (see  eq.~[\ref{eq:ang_int})] will depend on the following rational integrals for three terms in eq.~(\ref{eq:Qci_em})
\begin{equation}
    I_l^{(n)}=\int_0^{q_m} \frac{q^n \mathrm{d} q}{|q^2+\Pi_l|^2}=\int_0^{q_m} \frac{q^n \mathrm{d} q}{|q^2+q_{l}^2|^2},
\end{equation}
\begin{equation}
    I_t^{(n)}=\int_0^{q_m} \frac{q^n \mathrm{d} q}{|q^2+\Pi_{t}|^2}=\int_0^{q_m} \frac{q^{n+2} \mathrm{d} q}{q^6+\Lambda^6},
\end{equation}
\begin{eqnarray}
    I_{tl}^{(n)}&=&\int_0^{q_m} \mathrm{Re} \frac{q^n \mathrm{d} q}{(q^2+\Pi_l)(q^2+\Pi_t)^*}\nonumber\\
    &=&\int_0^{q_m} \frac{q^{n+4} \mathrm{d} q}{(q^2+q_{l}^2)(q^6+\Lambda^6)},
\end{eqnarray}
which all can be taken analytically. We do not give these lengthy expressions here\footnote{Explicit expressions for some of these integrals are given in \ref{sec:app:pract}.}, but notice that since the characteristic transverse screening momentum $\Lambda$ is small, only the leading terms in $\Lambda$ are needed in practice. Then
\begin{equation}
    I_t^{(0)}= \frac{\pi}{6\Lambda^3}+{\cal O}(\Lambda^{-1}),\quad I_t^{(2)}= \frac{\pi}{3\Lambda}+{\cal O}(\Lambda),
\end{equation}
\begin{equation}
    I_{tl}^{(0)}= \frac{\pi^2}{3q_{l}^2\Lambda}+{\cal O}(\Lambda^0),
\end{equation}
but for the mixed integral and $n\geq 2$, the transverse screening can be neglected and
\begin{equation}
    I_{tl}^{(n)}=I_l^{(n)}+q_{l}^2I_l^{(n-2)}+{\cal O}(\Lambda),\quad n\geq 2.
\end{equation}

The longitudinal part of the squared matrix element does not depend on $\Lambda$ at all. The only dependence on $w$ in expressions is contained in the $\Lambda$-dependent terms. When the leading contribution in $\Lambda$ is identified, the integral over $w$ can be performed analytically. Collecting all results, we find for the transport matrices for the thermal conductivity problem
\begin{equation}
    \Lambda^\kappa_{ab}=\Lambda^{\kappa,t}_{ab}+\Lambda^{\kappa,l}_{ab},
\end{equation}
where the transverse contribution is
\new{
\begin{equation}
    \Lambda^{\kappa,t}_{ab}=\frac{24\zeta(3) T}{\pi^3 q_t^2} \alpha_f^2 Z_a^2Z_b^2p_{\mathrm{F}b}^2,\label{eq:Lambda_kappa_t}
\end{equation}
}
where $\zeta(3)$ is the Riemann zeta-function, 
while for the longitudinal contribution one obtains rather lengthy, but simple expression
\begin{eqnarray}
    \Lambda^{\kappa,l}_{ab}&=&\frac{16\pi\alpha_f^2 Z_a^2Z_b^2 T^2m_a^{*2}m_b^{*2}}{5p_{\mathrm{F}a}^2}\left(I_l^{(0)}\right.\nonumber\\
    &+&\left.\frac{1}{4}\left(\frac{1}{p_{\mathrm{F}a}^2}-\frac{1}{m_a^{*2}}-\frac{1}{m_b^{*2}}\right)I_l^{(2)}\right.\nonumber\\
    &-&\left.\frac{1}{16}\left(\frac{1}{p_{\mathrm{F}a}^2m_a^{*2}}-\frac{1}{m_a^{*2}m_b^{*2}}+\frac{1}{p_{\mathrm{F}a}^2 m_b^{*2}}\right)I_l^{(4)}
    \right.\nonumber\\
    &+&\left.\frac{1}{64p_{\mathrm{F}a}^2m_a^{*2} m_b^{*2}}
    I_l^{(6)}
    \right).\label{eq:Lambda_kappa_l}
\end{eqnarray}
In practice it is enough to restrict the calculations to the first line in  eq.~(\ref{eq:Lambda_kappa_l}). 
\new{Keeping explicitly only the leading term in $q_l$, one gets} 
\begin{equation}
    \Lambda^{\kappa,l}_{ab}=\frac{4\pi^2\alpha_f^2 Z_a^2Z_b^2 T^2m_a^{*2}m_b^{*2}}{5p_{\mathrm{F}a}^2q_l^3}+{\cal O}(q_l^{-1}),
\end{equation}
however it is not recommended to use this limit, since the correction induced by the exact expression for $I_l^{(0)}$ can be significant. Other integrals in eq.~(\ref{eq:Lambda_kappa_l}) are written for completeness.

The non-diagonal term for thermal conductivity, eq.~(\ref{eq:Lambda1_ab_kappa}), contains only the mixed ($tl$) contribution. Keeping the leading order in $\Lambda$ results in
\begin{equation} 
    {\Lambda'}^{\kappa}_{ab}=\left(\frac{4}{\pi}\right)^{1/3}\Gamma(11/3)\zeta(8/3) \alpha_f^2Z_a^2Z_b^2\frac{m_a^*m_b^* p_{\mathrm{F}b}T^{5/3}}{p_{\mathrm{F}a}q_{l}^2q_t^{2/3}}, \label{eq:Lambda_kappa_tl}
\end{equation}
where $\Gamma(11/3)$ is the Euler gamma-function,
and this result is exact in $q_l$.

Similarly, for the shear viscosity transport matrices,
\begin{equation}
    \Lambda^\eta_{ab}=\Lambda^{\eta,t}_{ab}+\Lambda^{\eta,l}_{ab},
\end{equation}
where the transverse contribution is
\begin{equation}\label{eq:Lambda_eta_t}
    \Lambda^{\eta,t}_{ab}=2\left(\frac{4}{\pi}\right)^{1/3}\Gamma(8/3)\zeta(5/3)\alpha_f^2 Z_a^2Z_b^2\frac{p_{\mathrm{F}b}^2 T^{5/3}}{p_{\mathrm{F}a}^2 q_t^{2/3}}.
\end{equation}
Notice the different power of temperature in comparison with  eq.~(\ref{eq:Lambda_kappa_t}). This is a result of an additional $q^2$ factor in eq.~(\ref{eq:Lambda_ab_eta}) which leads to different contribution of plasma screening to the final result. The longitudinal contribution with account for all kinematic corrections is given by
\begin{eqnarray}
    \Lambda^{\eta,l}_{ab}&=& \frac{4\pi\alpha_f^2Z_a^2Z_b^2 T^2 m_a^{*2}m_b^{*2}}{p_{\mathrm{F}a}^4}\left(I_{l}^{(2)}\right.\nonumber\\
    &-&\left.\frac{1}{4}\left(\frac{1}{p_{\mathrm{F}a}^2}+\frac{1}{m_a^{*2}}+\frac{1}{m_b^{*2}}\right)I_l^{(4)}\right.\nonumber \\
    &+&\left.\frac{1}{16}\left(\frac{1}{p_{\mathrm{F}a}^2m_a^{*2}}+\frac{1}{m_a^{*2}m_b^{*2}}+\frac{1}{p_{\mathrm{F}a}^2 m_b^{*2}}\right)I_l^{(6)}
    \right.\nonumber\\
    &-&\left.\frac{1}{64p_{\mathrm{F}a}^2m_a^{*2} m_b^{*2}}
    I_l^{(8)}
    \right).\label{eq:Lambda_eta_l}
\end{eqnarray}
The leading contribution is given by the first  line in eq.~(\ref{eq:Lambda_eta_l}). \new{Retaining the leading order in $q_l$}, eq.~(\ref{eq:Lambda_eta_l}) reads
\begin{equation}
    \new{\Lambda^{\eta,l}_{ab}}= \frac{\pi^2\alpha_f^2Z_a^2Z_b^2 T^2 m_a^{*2}m_b^{*2}}{p_{\mathrm{F}a}^4 q_l}+{\cal O}(q_l).
\end{equation}
In contrast to the thermal conductivity case, for the shear viscosity problem the non-diagonal matrix element ${\Lambda'}^\eta_{ab}$ does not depend on the transverse screening in the leading order and has the same order in $q_l$ as the longitudinal part, namely
\begin{eqnarray}
    {\Lambda'}^{\eta}_{ab}&=& \frac{4\pi\alpha_f^2Z_a^2Z_b^2 T^2 m_a^*m_b^{*}p_{\mathrm{F}b}}{p_{\mathrm{F}a}^3}\left(I_{tl}^{(2)}\right.\\
    &-&\left.\frac{1}{4}\left(\frac{1}{p_{\mathrm{F}a}^2}+\frac{1}{p_{\mathrm{F}b}^2}\right)I_{tl}^{(4)}\right.\\
    &+&\left.\frac{1}{16p_{\mathrm{F}a}^2p_{\mathrm{F}b}^2}I_{tl}^{(6)}\right),\label{eq:Lambda_eta_tl}
\end{eqnarray}
and, \new{keeping} the leading order in $q_l$,
\begin{equation}
    \new{{\Lambda'}^{\eta}_{ab}}=\frac{2\pi^2\alpha_f^2Z_a^2Z_b^2 T^2 m_a^*m_b^{*}p_b}{p_{\mathrm{F}a}^3 q_l}+{\cal O}(q_l).
\end{equation}

Finally, performing the same integration for the momentum transfer rates in eq.~(\ref{eq:Jab_var}), one finds
\begin{equation}\label{eq:Jab_decomp}
    J_{ab}=J_{ab}^l+J_{ab}^t,
\end{equation}
i.e. there is no mixed longitudinal-transverse contributions in this case.
The transverse term $J_{ab}^t$ in eq.~(\ref{eq:Jab_decomp}) is
\begin{equation}\label{eq:Jab_t}
    J_{ab}^t=\frac{n_a p_{Fa}}{3}\Lambda_{ab}^{\eta,t}
\end{equation}
and the longitudinal one is
\begin{eqnarray}
    J_{ab}^l&=&\frac{4}{9\pi} \alpha_f^2Z_a^2Z_b^2 T^2 m_a^{*2}m_b^{*2}\left(I_{l}^{(2)}\right.\nonumber\\
    &-&\left.\frac{1}{4}\left(\frac{1}{m_a^{*2}}+\frac{1}{m_b^{*2}}\right)I_l^{(4)}\right.\nonumber\\
    &+&\left.\frac{1}{16m_a^{*2} m_b^{*2}}
    I_l^{(6)}
    \right).\label{eq:Jab_l}
\end{eqnarray}
These contributions behave similarly as the transport matrix for the shear viscosity problem. Indeed, according to eqs.~(\ref{eq:Lambda_ab_eta}) and (\ref{eq:Jab_var}), if the kinematical corrections in eq.~(\ref{eq:Lambda_ab_eta}) can be neglected (extremely weak screening case), one finds $n_ap_{\mathrm{F}a} \Lambda^\eta_{ab} = 3J_{ab}$.
\new{Keeping, as above,} the leading order in $q_l$, the longitudinal contribution to $J_{ab}$ reads
\begin{equation}
    \new{J_{ab}^l}=\frac{\alpha_f^2Z_a^2Z_b^2 T^2 m_a^{*2}m_b^{*2}}{9 q_l}+{\cal O}(q_l).
\end{equation}

In practice, the transverse contributions to $\Lambda^\kappa_{ab}$, $\Lambda^\eta_{ab}$, and $J_{ab}$ always dominate over the longitudinal contributions. 
This leads to the non-Fermi-liquid behaviour of the transport matrices and hence of the transport coefficients \cite{Heiselberg:1993cr,Schmitt2018}.
However, \new{while} for the thermal conductivity one can just neglect the longitudinal contribution, for the shear viscosity 
and momentum transfer rates this dominance is not dramatic and both contributions should be kept at realistic values of temperature. Moreover, it is not enough to retain only the leading order in $q_l$ and the reasonable approximation is to keep the integral $I_n^l$ with lowest order $n$ for each expression above. 

\new{In principle, there are also contributions to the transport matrix due to lepton---neutron scatterings, namely $\Lambda^k_{en}$ and $\Lambda^k_{\mu n}$ terms. They arise from the interaction of the lepton  with the anomalous neutron magnetic moment [$F_{a2}$ term in eq.~(\ref{eq:em_trans_current})]. It is much less effective than lepton-proton interaction. Thus we expect $\Lambda^k_{en}$ and $\Lambda^k_{\mu n}$ to give negligible contributions to viscosity and thermal conductivity and do not consider these transport matrix elements in this work. The only quantities, for which the lepton---neutron interaction is important, are the friction coefficients $J_{en}$ and $J_{\mu n}$. These coefficients can be found in section~VII of the Ref.~\cite{Dommes2020PhRvD} and references therein.}

Practical expressions for calculating the electromagnetic contribution to 
transport coefficients based on the results of this section are given in \ref{sec:app:pract}.

\subsection{Strong interactions}\label{sec:strong}

Description of collisions in  the strong sector is hampered by the poorly known impact of many-body effects on the baryon interactions at supranuclear densities. Different approaches to the problem in the context of the NS cores transport are reviewed in \cite{Schmitt2018}. Remember, that the necessary transport matrices $\Lambda_{ab}^k$ depend on the in-medium effective masses of quasiparticles and the scattering matrix elements ${\cal Q}_{ab}$.

 One approach is to forget about all in-medium modifications or include them only in the effective masses and employ the free-space scattering matrix elements. The simplest model of the strong interaction is the free one-pion exchange model (FOPE). For completeness we briefly discuss it in sec.~~\ref{sec:ope}, since in principle it can serve as a starting model for the in-medium modifications.
 
It is well-known, however, that the free one-pion exchange model in the first Born approximation overestimates the scattering cross-sections so it does not serve as a good approximation for calculation of ${\cal Q}_{ab}$. Therefore the next level of approximation is to extract ${\cal Q}_{ab}$ directly from the experimental data on the scattering experiments or from the accurate free-space calculations that reproduce the experimental data. The latter approach was elaborated in a range of studies \cite{FlowersItoh1979ApJ,Yakovlev1991Ap&SS,Baiko2001AA,ShterninYakovlev2008,Shternin2008JETP}, we consider it in sec.~\ref{sec:cross}.

Of course, this is a `poor man solution' to the 
problem of in-medium transport cross-sections. Both effective masses and scattering probabilities need to be calculated from the same microscopic model (and the same as the EOS). Unfortunately, model dependence of results is quite substantial, especially at large densities (e.g., \cite{Schmitt2018,Shternin2020PhRvD}). Several approaches were proposed for calculation of the in-medium scattering rates in NS cores \cite{Schmitt2018}. In pure neutron matter approximation these include thermodynamic $T$-matrix approach \cite{Sedrakian1994PhLB} or effective quasiparticle interaction \cite{Wambach1993NuPhA}. Effective interaction construction on a basis of the variational methods and subsequent application to NS matter was performed in \cite{Benhar2010PhRvC,BenharValli2007PhRvL,CarboneBenhar2011JPhCS}. The calculations within the Brueckner-Hartree-Fock (BHF) theory \cite{Baldo1999Book} were also carried out \cite{Benhar2010PhRvC,Zhang2010PhRvC,Shternin2013PhRvC,Shternin2017JPhCS,Shternin2020PhRvD,Shternin2021Univ}. Frequently a partial wave expansion of the scattering problem is used and we show how this expansion enters the calculations in sec.~\ref{sec:gmat}.

A different approach is known as the medium-modified one-pion exchange (MOPE) model derived in the framework of the Landau-Migdal Fermi-liquid theory for nuclear matter \cite{Migdal1990PhR}. In this model the modification (softening) of the pionic mode by the in-medium effects is explicitly introduced in the theory. 
At present, up to our knowledge, the calculations of transport coefficients of dense nuclear matter within this model  exist for pure neutron matter only. For thermal conductivity this was done by  \cite{Blaschke2013PhRvC} and for shear viscosity by  \cite{Kolomeitsev2015PhRvC}. The review of the MOPE model is outside the scope of the present manuscript.

\subsubsection{Kinematics of collisions}\label{sec:strong_kinematics}
For the collisions mediated by the strong interactions, the static case ($\omega=0$) can be employed and one needs to calculate the angular averages given in eqs.~(\ref{eq:trmat_kappa static})--(\ref{eq:Jab_var_long}). It turns out that it is convenient to use the $(Pq)$ representation of angular averages given in eq.~(\ref{eq:ang_av_Pq}). The total momentum  $P$ of the colliding pair with respect to surrounding medium can be used to parameterize the in-medium effects. On the other hand, the scattering is frequently described in the center of mass (c.m.) frame by defining the relative $p$ momenta of the colliding pair  [$\bm{p}=\frac{1}{2}(\bm{p}_1-\bm{p}_2)$, see fig.~\ref{fig:angles}], and the center of mass scattering angle $\theta_{\mathrm{cm}}$ given by (see fig.~\ref{fig:angles}) \begin{equation}\label{eq:theta_cm}
  \cos \theta_{\mathrm{cm}}=1-\frac{q^2}{2p^2}.
\end{equation}
Notice that the total and relative momenta are related via
\begin{equation}\label{eq:Pp}
    P^2+4p^2=2(p_{\mathrm{F}a}^2+p_{\mathrm{F}b}^2),
\end{equation}
in addition the following useful equality holds
\begin{equation}
    \bm{p}_1(\bm{p}_2+\bm{p}_{2'})=P^2-p_{\mathrm{F}a}^2-p_{\mathrm{F}b}^2+\frac{q^2}{2}.
\end{equation}
Then the angular weighting factors in eqs.~(\ref{eq:trmat_kappa static})--(\ref{eq:Jab_var_long}) are observed to be polynomial in $P$ and $q$. This suggests that it is convenient to introduce the averages\footnote{Notice that the angular averages in \cite{Shternin2013PhRvC,Shternin2017JPhCS} differ by a factor of 2 from eq.~(\ref{eq:Qij}) since this factor in eq.~(\ref{eq:ang_av_Pq}) was taken out in those references, which is not convenient here.} \cite{Shternin2013PhRvC}
\begin{equation}\label{eq:Qij}
    {\cal Q}_{ab}^{(ij)}\equiv\left\langle {\cal Q}_{ab} P^i q^j\right\rangle.
\end{equation}

The relevant transport matrices in eqs.~(\ref{eq:trmat_kappa static})--(\ref{eq:Jab_var_long}) in terms of averages $Q_{ab}^{(ij)}$ are given as \cite{Shternin2017JPhCS}
\begin{subequations}\label{eq:Lambda_Pq}
\begin{eqnarray}
    \Lambda_{ab}^\kappa&=&\frac{  T^2 m_a^{*2}m_b^{*2}}{5\pi^2 p_{\mathrm{F}a}^2}\left({\cal Q}_{ab}^{(00)}+\frac{1}{4p_{\mathrm{F}a}^2}{\cal Q}_{ab}^{(02)}\right),\label{eq:Lambdak_ab_Pq}\\
    {\Lambda'}_{ab}^\kappa&=&\frac{  T^2 m_a^{*2}m_b^{*2}}{10\pi^2 p_{\mathrm{F}a}^3p_{\mathrm{F}b}}\nonumber\\
    &&\times\left((p_{\mathrm{F}a}^2+p_{\mathrm{F}b}^2) {\cal Q}_{ab}^{(00)}\right.\nonumber\\
    &&\left.-{\cal Q}_{ab}^{(20)}-\frac{1}{2}{\cal Q}_{ab}^{(02)}\right),\label{eq:Lambdak1_ab_Pq}\\
    \Lambda^\kappa_{aa}+{\Lambda'}^\kappa_{aa}&=&\frac{  T^2 m_a^{*4}}{10\pi^2 p_{\mathrm{F}a}^4}\left(4p_{\mathrm{F}a}^2 {\cal Q}_{aa}^{(00)}-{\cal Q}_{aa}^{(20)}\right)
    ,\label{eq:Lambdak_aa_Pq}\\
   \Lambda_{ab}^\eta&=&\frac{ T^2 m_a^{*2}m_b^{*2}}{4\pi^2 p_{\mathrm{F}a}^4}\left({\cal Q}_{ab}^{(02)}-\frac{1}{4p_{\mathrm{F}a}^2}{\cal Q}_{ab}^{(04)}\right),\label{eq:Lambdah_ab_Pq}\\
    {\Lambda'}_{ab}^\eta&=&\frac{ T^2 m_a^{*2}m_b^{*2}}{8\pi^2 p_{\mathrm{F}a}^5p_{\mathrm{F}b}}\nonumber\\
    &&\times\left((p_{\mathrm{F}a}^2+p_{\mathrm{F}b}^2) {\cal Q}_{ab}^{(02)}\right.\nonumber\\
    &&\left.-{\cal Q}_{ab}^{(22)}-\frac{1}{2}{\cal Q}_{ab}^{(04)}\right),\label{eq:Lambdah1_ab_Pq}\\
     \Lambda_{aa}^\eta+{\Lambda'}_{aa}^\eta &=&
     \frac{ T^2 m_a^{*4}}{8\pi^2 p_{\mathrm{F}a}^6}\left(4p_{\mathrm{F}a}^2{\cal{Q}}^{(02)}_{aa}-{\cal Q}^{(04)}_{aa} \right.\nonumber\\
    &&\left.- {\cal Q}^{(22)}_{aa}\right),\label{eq:Lambdah_aa_Pq}\\
    J_{ab} &=&\frac{T^2m_a^{*2}m_b^{*2}}{36\pi^4}{\cal Q}^{(02)}_{ab}.\label{eq:Jab_Pq}
\end{eqnarray}
\end{subequations}
In eqs.~(\ref{eq:Lambdak_aa_Pq}) and (\ref{eq:Lambdah_aa_Pq}) we combined the direct and primed transport matrix elements for like species since it is this form in which they couple to $\widetilde{\lambda}_a^k$ in eqs.~(\ref{eq:kinvar_system})--(\ref{eq:kinvar_system1}). 

\subsubsection{Free one-pion exchange.}\label{sec:ope}

The simplest model of the strong $NN$-interaction is the one pion exchange (OPE) model. It appeals to a simple phenomenological low-energy Lagrangian of the $\pi N$ interaction
\begin{equation}\label{eq:Lagr_piN}
    \mathcal{L}_{\pi N} = g_{ab} \varphi_\pi \Bar{\psi}_a \gamma^5 \psi_b ,
\end{equation}
where $\psi_a$ and $\psi_b$ are the wavefunctions of incoming and outcoming nucleons $a$ and $b$, and $\varphi_\pi$ is the pion field (charged or neutral, depending on what the nucleons $a$ and $b$ are). For $\pi pp$ and $\pi nn$-scatterings, $g_{pp} = - g_{nn} \approx 13.1$ 
and for $\pi np$-scattering $g_{np} = g_{pp}\sqrt{2}$ (see, e.g., sec. 4.1.1. in ref.~\cite{Machleidt2011PhR}). In this section we restrict ourselves to the free OPE interaction in the first Born approximation. In this case the transition matrix element for $NN$-scattering  is [cf. eq.~(\ref{eq:matel_em})]
\begin{equation}\label{eq:Tab_piN}
    T_{ab}(12\to 1'2') = M_{ab}(12\to 1'2') - M_{ab}(12\to 2'1').
\end{equation}

The direct and exchange parts of the matrix element are\footnote{We employ the bispinor normalization eq.~(\ref{eq:DiracNorm}).}
\begin{subequations}\label{eq:Mab_piN}
\begin{multline}\label{eq:Mab_piN-dir}
    M_{ab}(12\to 1'2') = g_{aa} g_{bb} \Bar{u}_a(p'_1) \gamma^5 u_a(p_1) \\ 
        \times D_\pi(p'_1-p_1) \, \bar{u}_b(p'_2) \gamma^5 u_b(p_2),
\end{multline}
\begin{multline}\label{eq:Mab_piN-exch}
    M_{ab}(12\to 2'1') = g_{ab}^2 \Bar{u}_b(p'_2) \gamma^5 u_a(p_1) \\
        \times D_\pi(p'_2-p_1) \, \bar{u}_a(p'_1) \gamma^5 u_b(p_2),
\end{multline}
\end{subequations}
where $D_\pi$ is the pion propagator. Consequently, 
\begin{subequations}\label{eq:Qab_piN}
\begin{equation}\label{eq:Qab_piN-Q}
    {\cal Q}_{ab} = \frac{\mathcal{A}^2(q^2) + \mathcal{B}^2(q'^2) + \mathcal{A}(q^2)\mathcal{B}(q'^2)}{16 (1+\delta_{ab}) \varepsilon_a^2 \varepsilon_b^2},
\end{equation}
where 
\begin{equation}\label{eq:Qab_piN-Adef}
    \mathcal{A} = g_{aa} g_{bb} q^2 D_\pi(-q^2),
\end{equation}
\begin{multline}\label{eq:Qab_piN-Bdef}
    \mathcal{B} = g_{ab}^2 \left[ q'^2 + (m_a-m_b)^2 \right. \\ 
    \left. - (\varepsilon_a - \varepsilon_b)^2 \right] D_\pi\!\left((\varepsilon_a - \varepsilon_b)^2-q^2\right).
\end{multline}
\end{subequations}
In the latter three formulas $\bm{q}$ and $\bm{q'}$ are the 3-momentum transfers defined in sec.~\ref{sec:collisions}. 

The formula for ${\cal Q}_{ab}$ could be simplified for non-relativistic collisions of nucleons. In this case one can set $m_a=m_b=m_N$%
\new{, where we adopt $m_N= 939\,$MeV as a typical nucleon mass, }%
and $\varepsilon_a \approx \varepsilon_b$. In-vacuum pion propagator reads $D_\pi(-q^2) = (-q^2-m_\pi^2)^{-1}$, which is of course a crude estimate. Then ${\cal Q}_{ab}$ reduces to a standard form of the squared matrix element of one pion exchange nuclear potential
\begin{multline}\label{eq:Qab_piN_nonrel}
    {\cal Q}_{ab} = \frac{1}{16(1+\delta_{ab})\varepsilon^2_a\varepsilon^2_b} \left[ 
    \frac{g_{aa}^2g_{bb}^2q^4}{(q^2 + m_\pi^2)^2} \right.\\
    \left.  
    + \frac{g^4_{ab} q'^4}{(q'^2+m_\pi^2)^2} + \frac{g_{aa}g_{bb}g^2_{ab} q^2 q'^2}{(q^2+m_\pi^2)(q'^2+m_\pi^2)}
    \right].
\end{multline}

We can now calculate the averages ${\cal Q}_{ab}^{(ij)}$ in eq.~(\ref{eq:Qij}) and hence the OPE transport matrices in eq.~(\ref{eq:Lambda_Pq}). After transferring eq.~(\ref{eq:Qab_piN_nonrel}) to the ($Pq$) variables by employing the relation
\begin{equation}
   q^2+{q'}^2=4p^2=2(p_{\mathrm{F}a}^2+p_{\mathrm{F}b}^2)- P^2,
\end{equation} 
the integrations, in principle, this can be performed analytically, but the results are by no means illuminating. 
We therefore do not give the explicit expressions here, and illustrate the results in sec.~\ref{sec:discuss}.

\subsubsection{In-vacuum cross-sections}\label{sec:cross}
In this section we restrict ourselves to the nuclear composition of the NS core matter. In other words, we consider only scattering within the $np$ \new{subsystem}.
Neglecting relativistic corrections, one relates the differential cross-section to the squared matrix element as
\begin{equation}
    \frac{\mathrm{d}\sigma_{ab}}{\mathrm{d}\Omega}(E_\mathrm{lab},\theta_{\mathrm{cm}}) = \frac{m_N^2}{16\pi^2}(1+\delta_{ab}){\cal Q}_{ab},
\end{equation}
where $\cos\theta_{\mathrm{cm}}$ is given by eq.~(\ref{eq:theta_cm}) and \new{$E_\mathrm{lab} = 2 p^2/m_N$}. Assuming that the differential cross-sections are known, one can use them to calculate the transport matrices in eq.~(\ref{eq:ang_av_Pq}). 

Using this approach, Baiko et al. \cite{Baiko2001AA} defined for neutron-neutron collisions in thermal conductivity problem
\begin{equation}\label{eq:Lambdakappa_Bai}
    \Lambda^\kappa_{nn}+{\Lambda'}^\kappa_{nn}=\frac{64 T^2 m_n^{*4}}{5m_N^2 p_{\mathrm{F}n}} S_{nn2}
\end{equation}
where $S_{nn2}$ is an effective transport cross-section with a dimension of area,
\begin{equation}\label{eq:Sn2}
    S_{nn2}
    = \frac{m_N^2}{128\pi^2 p_{\mathrm{F}n}^3}  \left\langle (q^2+q'^2){\cal Q}_{nn} \right\rangle.
\end{equation}
We use here the variables $q$ and $q'$ inside the angular brackets to make a clear connection with ref.~\cite{Baiko2001AA} \new{who also used the notation $S_{n2}$ for $S_{nn2}$. Notice that in vacuum $S_{nn2}$  depends only on one argument, the neutron Fermi-momentum $p_{\mathrm{F}n}$}.

Baiko et al. \cite{Baiko2001AA} were interested in the neutron contribution to the transport coefficients and protons were considered as passive scatterers. While this is a good approximation for the non-magnetized case, the proton transport, in principle, can be significant in presence of the magnetic field. If we neglect the charge dependence of \new{nucleon interactions} (i.e. consider the electromagnetic forces separately), then eqs.~(\ref{eq:Lambdakappa_Bai})--(\ref{eq:Sn2}) can be also used for the proton-proton collisions by substituting $p_{\mathrm{F}n}\to p_{\mathrm{F}p}$, namely
\begin{equation}\label{eq:Lambdakappa_pp_Bai}
    \Lambda^\kappa_{pp}+{\Lambda'}^\kappa_{pp}=\frac{64 T^2 m_p^{*4}}{5m_N^2 p_{\mathrm{F}p}} S_{pp2},
\end{equation}
\new{where $S_{pp2}=S_{nn2}|_{p_{\mathrm{F}n}\to p_{\mathrm{F}p}}$.}

For the neutron-proton scattering in thermal conductivity problem one has \cite{Baiko2001AA}
\begin{equation}
    \Lambda^\kappa_{np} = \frac{64 T^2 m_{n}^{*2} m_{p}^{*2}}{5m_N^2 p_{\mathrm{F}n}} (S_{np1}+S_{np2}),
\end{equation}
where two effective $np$ cross-sections are defined:
\begin{equation}\label{eq:Sp1}
    S_{np1}=\frac{m_N^2}{64\pi^2p_{\mathrm{F}n}} \langle {\cal Q}_{np} \rangle
\end{equation}
and
\begin{equation}\label{eq:Sp2}
    S_{np2}= \frac{m_N^2}{256\pi^2 p_{\mathrm{F}n}^3} \langle {q^2\cal Q}_{np} \rangle. 
\end{equation}
Second of them, eq.~(\ref{eq:Sp2}), also enters the expression for the momentum transfer rates \cite{Shternin2008JETP}
\begin{equation}
    J_{np}=J_{pn}=\frac{64}{9\pi^2} T^2 \frac{m_n^{*2}m_p^{*2}}{m_N^2} p_{\mathrm{F}n}^3 S_{np2}.
\end{equation}

The same effective cross-sections define the transposed ($pn$) matrix elements for the thermal conductivity problem
\begin{equation}
    \Lambda^\kappa_{pn}=\frac{64 T^2 m_n^{*2} m_p^{*2} p_{\mathrm{F}n}}{5m_N^2 {p_{\mathrm{F}p}}^2} \left(S_{np1}+\frac{p_{\mathrm{F}n}^2}{p_{\mathrm{F}p}^2}S_{np2}\right).
\end{equation}
This spurious asymmetry results from the asymmetric definition of the effective transport cross-sections 
in \cite{Baiko2001AA}.

The shear viscosity problem was treated in a similar way in \cite{ShterninYakovlev2008}. Equation~(\ref{eq:Lambdah_aa_Pq}) for $nn$ or $pp$ collisions is rewritten as ($a=n,\,p$)
\begin{equation}
    \Lambda^\eta_{aa}+{\Lambda'}^\eta_{aa}=\frac{64 T^2 m_a^{*4}}{m_N^2 p_{\mathrm{F}a}} S_{aa4}
\end{equation}
with
\begin{equation}\label{eq:Sn4}
    S_{aa4}
    =\frac{m_N^2}{16\pi^2}  \frac{1}{(2p_{\mathrm{F}a})^5}\langle q^2{q'}^2{\cal Q}_{aa}\rangle.
\end{equation}
Here we modified slightly the definition of $S_{aa4}$ \new{(with $a=n$)} compared to the quantity $S_{\mathrm{nn}}^{\mathrm{SY08}}$ introduced in \cite{ShterninYakovlev2008}, namely  $S_{nn4}=S_{\mathrm{nn}}^{\mathrm{SY08}}/12$. 
\new{Notice that like $S_{nn2}$ and $S_{pp2}$ above, $S_{nn4}$ and $S_{pp4}$ are actually the values of the same function $S_{aa4}(p_{\mathrm{F}a})$ evaluated at  $p_{\mathrm{F}n}$ or $p_{\mathrm{F}p}$, respectively.
}

For the $np$ scattering in the shear viscosity problem we can write
\begin{equation}
    \Lambda^\eta_{np}=\frac{64 T^2 m_n^{*2} m_p^{*2}}{m_N^2 p_{\mathrm{F}n}} (S_{np2}-S_{np4}),
\end{equation}
where
\begin{equation}\label{eq:Sp4}
    S_{np4} = \frac{m_N^2}{16\pi^2} \frac{1}{2(2p_{\mathrm{F}n})^5}\langle q^4 {\cal Q}_{np}\rangle.
\end{equation}
Shternin and Yakovlev \cite{ShterninYakovlev2008} did not separate $S_{np2}$ and $S_{np4}$, their $S_{\mathrm{np}}^{\mathrm{SY08}}=6(S_{np2}-S_{np4})$.

In \cite{ShterninYakovlev2008} like in \cite{Baiko2001AA} only neutron transport was considered. However the $pn$ components of the transport matrix can be expressed via the functions $S_{p2}$ and $S_{p4}$ as
\begin{equation}
    \Lambda^\eta_{pn}=\frac{64 T^2 m_n^{*2}m_p^{*2}}{m_N^2 p_{\mathrm{F}p}} \left(\frac{p_{\mathrm{F}n}^3}{p_{\mathrm{F}p}^3}S_{np2}-\frac{p_{\mathrm{F}n}^5}{p_{\mathrm{F}p}^5}S_{np4}\right).
\end{equation}

The primed elements of the transport matrices which mix neutron and proton transport were not considered in \cite{Baiko2001AA,ShterninYakovlev2008}. In order to keep a close analogy with their notations we express
\begin{equation}
    {\Lambda'}^\kappa_{np}=\frac{p_{\mathrm{F}p}^2}{p_{\mathrm{F}n}^2}{\Lambda'}^\kappa_{pn}=\frac{64 T^2 m_n^{*2} m_p^{*2}}{5m_N^2 p_{\mathrm{F}p}} S'_{np2}
\end{equation}
and
\begin{equation}
    {\Lambda'}^\eta_{np}=\frac{p_{\mathrm{F}p}^4}{p_{\mathrm{F}n}^4}{\Lambda'}^\eta_{pn}=\frac{64 T^2 m_n^{*2} m_p^{*2}}{m_N^2 p_{\mathrm{F}p}} S'_{np4},
\end{equation}
where
\begin{equation}\label{S1_p2}
    S'_{np2} = \frac{m_N^2}{16\pi^2} \frac{1}{8p_{\mathrm{F}n}^3}\left\langle\left( (p_{\mathrm{F}n}^2+p_{\mathrm{F}p}^2) -P^2-\frac{q^2}{2}\right){\cal Q}_{np}\right\rangle
\end{equation}
and
\begin{equation}\label{eq:S1_p4}
    S'_{np4} = \frac{m_N^2}{16\pi^2} \frac{1}{32p_{\mathrm{F}n}^5}\left\langle q^2\left( (p_{\mathrm{F}n}^2+p_{\mathrm{F}p}^2) -P^2-\frac{q^2}{2}\right){\cal Q}_{np}\right\rangle.
\end{equation}

The functions \new{$S_{nn2}$, $S_{np1}$, $S_{np2}$, $S_{nn4}$, and $S_{np2}-S_{np4}$} were fitted in refs.~\cite{Baiko2001AA} and \cite{ShterninYakovlev2008} on the basis of the free-space cross-sections obtained in refs.~\cite{LiMachleidt1993PhRvC,LiMachleidt1994PhRvC} based in turn on the realistic Bonn potential \cite{Machleidt1987PhR} which accurately reproduces observed $NN$ scattering data and properties of a few-nucleon systems. Available range of fitting parameters in \cite{Baiko2001AA,ShterninYakovlev2008} is insufficient for consideration of the $pp$ collisions, therefore we refitted these functions (and fitted the complementary ones) on a larger parameter space. Specifically, we took the free-space cross-section calculated  in  \cite{Shternin2013PhRvC} based on the realistic free-space Argonne v18 potential \cite{Wiringa1995PhRvC}. Resulting fit expressions are  given in \ref{sec:app:pract}.

\subsubsection{Partial wave expansion}\label{sec:gmat}
In many cases when the quasiparticle scattering matrix is obtained from the microscopic theory, it is given in the partial wave basis in the center of mass frame of the scattering baryons. Let us denote the in-medium scattering matrix by $G_{ab}$ in this case \new{instead of $T_{ab}$} (having in mind the $G$-matrix of the Brueckner-Hartree-Fock theory as a prototype). The partial wave basis depends on the total momentum $P$ and the relative momentum $p$ of the colliding pair  and the angular momentum quantum numbers. Namely, the partial wave state is $|P,p; J\ell S M_J\rangle$, where $S$ is the pair total spin, $\ell$ is the pair orbital momentum, $J$ is its total angular momentum, and $M_J$ is the total angular momentum projection. 

Let $G^{JS}_{\ell\ell'}(P,p)$ be the matrix element of the scattering matrix which is diagonal in $P$, $p$, $J$ and $S$ and do not depend on $M_J$.
Then the quantity ${\cal Q}_{ab}$ can be expanded in the series in Legendre polynomials
${\cal P}_L \left(\cos \theta_{\rm cm}\right)$:
\begin{equation}\label{eq:Q_legendre}
  {\cal Q}_{ab}(q,P)=\frac{1}{1+\delta_{ab}}\sum_{L} {\cal Q}^{(L)}_{ab}(P) {\cal
  P}_L\left(\cos \theta_{\rm cm}\right),
\end{equation}
where $\cos\theta_{\mathrm{cm}}$ is given in eq.~(\ref{eq:theta_cm}) and the  coefficients of expansion are related to the matrix elements of the transition amplitude in the partial wave basis as  \cite{Shternin2013PhRvC} 
\begin{eqnarray}
  {\cal Q}_{ab}^{(L)}(P)&=& \frac{1}{16 \pi^2}\sum i^{\ell'-\ell+\bar{\ell}-\bar{\ell}'}
  \Pi_{\ell\ell'}\Pi_{\bar{\ell}\bar{\ell}'}\Pi^2_{J\bar{J}}
  C^{L'0}_{\ell' 0 \bar{\ell}'0} C^{L0}_{\ell0\bar{\ell}0}\nonumber\\
  &&\times
  \left\{\begin{array}{ccc}
    \bar{\ell} & S & \bar{J}\\
    J & L & \ell
  \end{array}\right\}
 \left\{\begin{array}{ccc}
    \bar{\ell}' & S & \bar{J}\\
    J & L & \ell'
  \end{array}\right\}\nonumber\\
  &&\times
  \left(1+\delta_{ab}(-1)^{S+\ell}\right)\left(1+\delta_{ab}(-1)^{S+\bar{\ell}}\right)\nonumber\\
  &&\times
  G^{JS}_{\ell\ell'}(P,p) \left(G^{\bar{J} S}
  _{\bar{\ell}\bar{\ell}'}(P,p)\right)^*\label{eq:Q_L}.
\end{eqnarray}
Here  $\Pi_{fg}\equiv\sqrt{(2f+1)(2g+1)}$,
terms in curly brackets are 6$j$-symbols of the quantum angular momentum theory \cite{Varshalovich},
 terms in the third line take into account exchange contributions, and the pair index $ab$ is omitted at the scattering matrix elements for brevity.
The summation in eq.~(\ref{eq:Q_L}) is carried over
all angular momenta and spin variables, except $L$.

The medium effects enter eqs.~(\ref{eq:Q_legendre})--(\ref{eq:Q_L}) via the dependence of the scattering matrix on the total pair momentum $P$ which breaks the translation invariance. 
The expansion (\ref{eq:Q_legendre}) \new{together with eq.~(\ref{eq:theta_cm}) allow} us to perform the integration over $q$ in eq.~(\ref{eq:Qij}) analytically with the help of the following relation  \cite{Shternin2013PhRvC}
\begin{eqnarray}
  &&\int\limits_0^{q_m} \frac{q^j {\rm d} q}{\sqrt{q_m^2-q^2}}{\cal
  P}_L\left(1-\frac{q^2}{2p^2}\right)=\frac{q_m^j}{2}B\left(\frac{j+1}{2},\frac{1}{2}\right)\nonumber\\
  &&\times\ 
  {}_3F_{2}\left(-L,L+1,\frac{j+1}{2};1,\frac{j}{2}+1;\frac{q_m^2}{4p^2}\right),
\end{eqnarray}
where $B\left(\frac{j+1}{2},\frac{1}{2}\right)$ is the
beta-function, and ${}_3F_2$ is the generalized hypergeometric
function. It reduces to the $L-1$ order
polynomial in $q_m^2/(4p^2)$, as its first argument, $-L$, is a
negative integer. The remaining integral over $P$ in 
eq.~(\ref{eq:Qij}) in general should be performed numerically.

\section{Discussion}
\label{sec:discuss}
\begin{figure*}
  \includegraphics[width=\textwidth]{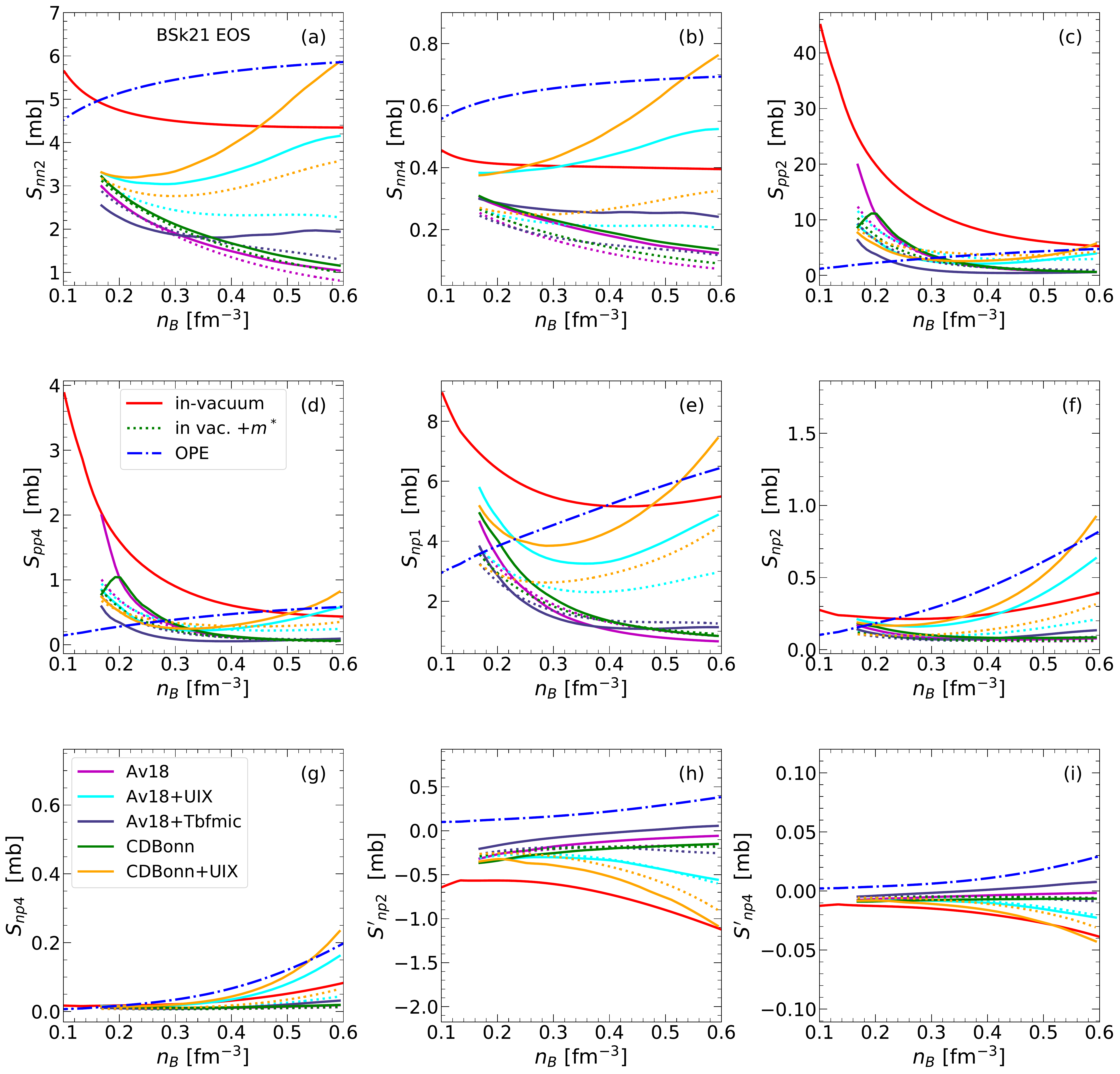}
\caption{Effective transport cross-sections defined in sec.~\ref{sec:cross} as functions of density for the BSk21 EOS. Blue dash-dotted line show OPE results, red solid line shows the in-vacuum cross-sections. Coloured solid lines correspond to different microscopic models of nuclear interactions as coded in the legend in panel (g) and described in the text. Dotted lines show in-vacuum effective transport cross-sections combined with the nucleon effective masses for the microscopic interactions (with the same color coding as the respective solid lines), see text for details.}
\label{fig:AllS_bsk21}       
\end{figure*}

Section~\ref{sec:collisions} identifies  key quantities that are needed from the microscopic theory for calculating the diffusive  transport coefficients  \new{in} the NS core in the weak-coupling regime. Namely, these are the effective masses $m_a^*$ of the quasiparticles presented in the matter and the values of transport matrices $\Lambda^k_{ab}$, ${\Lambda'}^k_{ab}$ , which are found from the angular averaging of quasiparticle scattering probabilities. With these quantities available, the transport coefficients of the non-magnetized matter can be easily found from the solution of the system eq.~(\ref{eq:kinvar_system1}) or, in case of the momentum transfer rates,  directly from the corresponding eqs.~(\ref{eq:Jab_var}) or (\ref{eq:Jab_var_long}). Magnetic field does not modify this procedure significantly if the irreducible spherical tensor formalism is used and the transport coefficients are expressed via the complex functions (\ref{eq:Ltilde}). The inclusion of the magnetic field in this formalism is achieved by a simple addition of the imaginary  diagonal matrix to the system eq.~(\ref{eq:kinvar_system1}).

\subsection{Transport matrices and momentum transfer rates}\label{sec:lambda_discuss}
The electromagnetic part of the transport matrices can be calculated in the universal form, independent on the particular EOS of the dense matter inside the NS cores. The necessary microscopic ingredients are the particle fractions and the quasiparticle effective masses at Fermi surface. The lepton effective masses are 
ones for free particles, $m_\ell^*=\sqrt{m_\ell^2+p_{\mathrm{F}\ell}^2}$ for $\ell=e,\ \mu$. The proton (or other charged baryon) effective masses should be provided along with the EOS.  For instance, some of the EOSs in the CompOSE database already contain the information about $m^*$. Otherwise one needs to assume some typical value of the proton effective mass, say $m^*_p=0.8 m_N$.
In this case the electromagnetic contribution to eq.~(\ref{eq:kinvar_system}) can be readily calculated.  For convenience, we give  practical expressions for the electromagnetic contribution in~\ref{sec:app:pract}.

For the strong sector the situation is much less certain. Ideally, the averages (\ref{eq:Qij}) and hence the elements of the transport matrices need to be calculated on the basis of the same microphysics as the EOS. In reality this is rarely accessible for a modelling practioners and they need to extract transport coefficients and EOS from unrelated sources. 

Let us illustrate this uncertainty following \cite{Shternin2017JPhCS,Shternin2020PhRvD}. Assume that we take some EOS of the dense matter and want to calculate the transport coefficients. For certainty, we take the BSk21 EOS from the  Brussels-Skyrme EOS family \cite{Potekhin2013A&A} which has a convenient analytical parametrization of the particle fractions. In fig.~\ref{fig:AllS_bsk21} we plot the effective transport cross-sections defined in sec.~\ref{sec:cross} as functions of density for various  models of nuclear interaction. Blue dash-dotted lines show the results of the free OPE model in the first Born approximation of sec.~\ref{sec:ope}, while red solid lines show the results of the in-vacuum cross-sections calculated as described in sec.~\ref{sec:cross}.  Clearly the free OPE model in Born approximation, as expected, is a bad approximation for the effective transport cross-sections. 

Other solid lines in fig.~\ref{fig:AllS_bsk21} show the results of \cite{Shternin2020PhRvD} who calculated effective nucleon transport mean free paths on a grid of baryon densities and proton fractions ($x_p>0.05$ and $n_B<0.6\,$fm$^{-3}$) for five nuclear interactions models in the non-relativistic BHF approach. Each interaction results in its own EOS and, specifically, in its own proton fraction for NS core matter in  the beta equilibrium. These proton fractions are also different from those for the BSk21 EOS.

The NN interactions considered in \cite{Shternin2020PhRvD} are based on two realistic two-body potentials: the Argonne v18 (Av18 for short) potential \cite{Wiringa1995PhRvC} and the  charge-dependent Bonn (CD-Bonn for short) potential \cite{Machleidt2001PhRvC} and two models for the three-body interactions, i.e. \new{the} phenomenological Urbana IX (UIX for short) model \cite{Carlson1983NuPhA}  and \new{the} microscopic three-body force  (Tbfmic for short) model based on the meson-nucleon theory of the nucleon interaction \cite{Li2008PhRvC,Li2012PhRvC}. Combinations of these interactions result in five interaction models (see \cite{Shternin2020PhRvD} for details) which we denote as Av18, CDBonn, Av18+UIX, CDBonn+UIX, and Av18+Tbfmic. \new{In these calculations, effective masses and in-medium scattering matrices were calculated in a self-consistent way within the BHF framework and included in transport mean free paths.}

One sees a considerable scatter in the effective transport cross-sections in fig.~\ref{fig:AllS_bsk21}, especially at high $n_B$. For most of those, the in-vacuum interaction (red solid lines) provides a poor approximation. \new{One of the exceptions} is the $S_{np2}$ cross-sections [panel (f) in fig.~\ref{fig:AllS_bsk21}] related to the friction coefficient $J_{np}$ and can be used in calculations (see, e.g., \cite{Dommes2020PhRvD}). 

Somewhat better results can be achieved by combining the effective transport cross-sections with the results of the effective mass calculations for the microscopic models (e.g., \cite{Baldo2014PhRvC}). The latter \new{ones} are available more frequently. We illustrate this with the dotted lines in fig.~\ref{fig:AllS_bsk21} that show the combinations \new{$m_a^{*2}m_b^{*2}S_{\alpha}/m_N^4$} where $m_{a,b}^*$ 
are the effective masses for five microscopic interactions considered and effective transport cross-sections $S_{\alpha}$ with $\alpha=nn2$, $nn4$, $pp2$, $pp4$, $np1$, $np2$, $np4$, $np2'$, and $np4'$ are calculated with the in-vacuum interaction. 
\new{The difference of these results from the complete calculations (colored solid lined) is solely due to in-medium $\mathcal{Q}_{ab}$ and not in $m_{a,b}^*$. We observe, that
the agreement with the complete calculations is poor, but better than that for pure in-vacuum interactions.} Thus it can be recommended to use the fits for the in-vacuum cross-sections given in \ref{sec:app:pract} together with microscopic values of effective masses, if available. If not, one can use some typical values, e.g., $m_n^*=m_p^*=0.8 m_N$, see below.

Notice that here only five interaction models within a singe many-body  method (non-relativistic BHF) \new{are} explored. In reality, the uncertainty in microscopic treatment can be considerably larger (e.g., when the calculations are performed within the MOPE model \cite{Blaschke2013PhRvC,Kolomeitsev2015PhRvC}).

\begin{table}
\new{
\caption{Summary of EOSs used in the paper. First part corresponds to models taken from CompOSE database, second for additional model not in CompOSE.}
\label{tab:eos}
\begin{tabular*}{\columnwidth}{@{\extracolsep{\fill}}llll@{}}
\hline
Name & Composition & Type & Refs.
\\
\hline
\multicolumn{4}{c}{CompOSE}\\
RG(KDE0v) & $npe\mu$& Skyrme & \cite{Gulminelli2015PhRvC,Agrawal2005PhRvC}\\
RG(KDE0v1) & $npe\mu$& Skyrme & \cite{Gulminelli2015PhRvC,Agrawal2005PhRvC}\\
RG(Rs) & $npe\mu$& Skyrme & \cite{Gulminelli2015PhRvC,Friedrich1986PhRvC}\\
RG(SK255) & $npe\mu$& Skyrme & \cite{Gulminelli2015PhRvC,Agrawal2005PhRvC}\\
RG(SK272) & $npe\mu$& Skyrme & \cite{Gulminelli2015PhRvC,Agrawal2005PhRvC}\\
RG(SKa) & $npe\mu$& Skyrme & \cite{Gulminelli2015PhRvC,Koehler1976NuPhA}\\
RG(SKb) & $npe\mu$& Skyrme & \cite{Gulminelli2015PhRvC,Koehler1976NuPhA}\\
RG(SKI\ 2---5) & $npe\mu$& Skyrme & \cite{Gulminelli2015PhRvC,Reinhard1995NuPhA.584}\\
RG(SKI 6) & $npe\mu$& Skyrme & \cite{Gulminelli2015PhRvC,Nazarewicz1996PhRvC}\\
RG(SKMp) & $npe\mu$& Skyrme & \cite{Gulminelli2015PhRvC,Bennour1989PhRvC}\\
RG(SKOp) & $npe\mu$& Skyrme & \cite{Gulminelli2015PhRvC,Reinhard1999PhRvC}\\
RG(SLy2) & $npe\mu$& Skyrme & \cite{Gulminelli2015PhRvC,ChabantPhd1995}\\
RG(SLy9) & $npe\mu$& Skyrme & \cite{Gulminelli2015PhRvC,ChabantPhd1995}\\
DS(CMF)-2,4,6 & $npe$ & RMF & \cite{Dexheimer2008ApJ,Dexheimer2017PASA}\\
BL & $npe\mu$ & BHF & \cite{Bombaci2018A&A}\\
\hline
\multicolumn{4}{c}{not in CompOSE}\\
DDME2 & $npe\mu$ & RMF & \cite{Fortin2016PhRvC}\\
NL3$\omega\rho$ & $npe\mu$ & RMF & \cite{Fortin2016PhRvC}\\
SLy4 & $npe\mu$ & Skyrme & \cite{Douchin2001A&A}\\
BHF & $npe\mu$ & BHF & \cite{Sharma2015A&A}\\
BSk 20$-$21 & $npe\mu$ & Skyrme & \cite{Potekhin2013A&A}\\
BSk 22, 24$-$26 & $npe\mu$ & Skyrme & \cite{Pearson2018MNRAS}\\
APR & $npe\mu$ & Var. & \cite{Akmal1998PhRv}\\
HHJ I---IV & $npe\mu$ & Phen. & \cite{Fortin2016PhRvC,Heiselberg1999Ap,Heiselberg2000PhR}\\
PAL I---III & $npe\mu$ & Phen. & \cite{Prakash1988PhRvL,HPY2007Book}\\
PAL-IV & $npe\mu$ & Phen. & \cite{Page1992ApJ,HPY2007Book}\\
\hline
\end{tabular*}
}
\end{table}

\begin{figure}
  \includegraphics[width=\columnwidth]{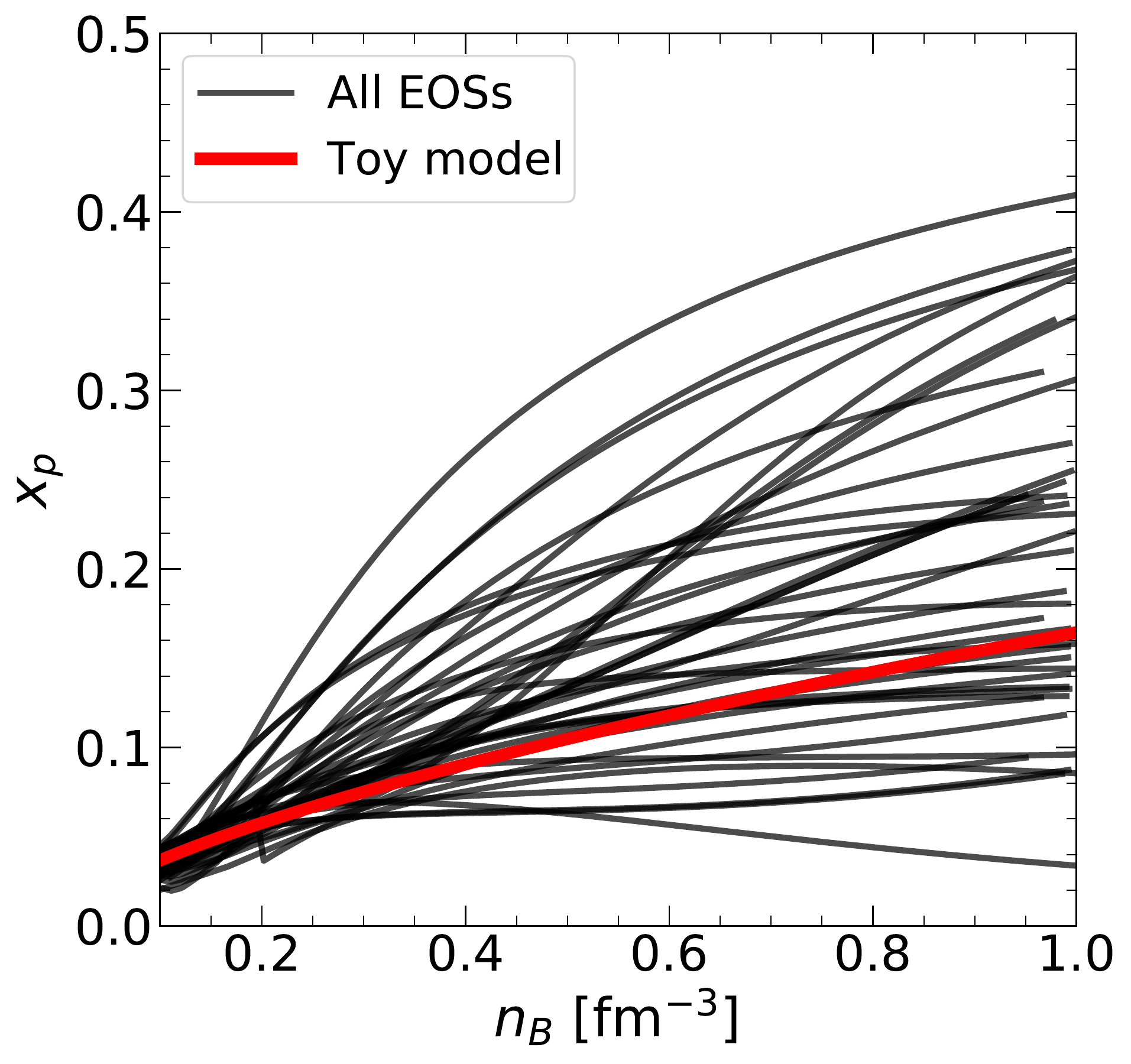}
\caption{Proton fraction $x_p$ versus density for all EOSs considered in the paper (black solid lines). Thick red line shows the toy model of eq.~(\ref{eq:xpp_ddo}). }
\label{fig:xp_eos}       
\end{figure}

Let us focus now on the dependence of the in-vacuum transport cross-sections on EOSs of NS core matter. To this end we downloaded several EOSs from the 
`Cold neutron star EOS' section of the CompOSE database. Specifically we restricted ourselves to the models with purely nucleonic  composition, and also to those where the composition is stored in the database. These include 16 models of the unified Skyrme group  \cite{Gulminelli2015PhRvC} denoted as `RG(Name)' in CompOSE, where `Name' is the  label of the interaction \new{model}, three relativistic mean %
field \new{(RMF) } models from \cite{Dexheimer2008ApJ,Dexheimer2017PASA}, denoted as DS(CMF)-2, DS(CMF)-4, and DS(CMF)-J6, and the BHF calculations based on the chiral perturbation theory from \cite{Bombaci2018A&A}.

We extended our EOS bank beyond the current CompOSE database. Namely, we added two EOSs denoted by DDME2 and NL3$\omega\rho$ in \cite{Fortin2016PhRvC}, the Skyrme-based EOS SLy4 from \cite{Douchin2001A&A} and the microscopic BHF EOS from \cite{Sharma2015A&A}. 

We \new{also} employed several models from the BSk family which have convenient analytical representations. Specifically, we used the models BSk20$-$22 and BSk24$-$26  \cite{Potekhin2013A&A,Pearson2018MNRAS}. We also added a celebrated variational EOS APR 
\cite{Akmal1998PhRv} and four EOSs built upon the anaytical parameterization of the APR EOS of \cite{Heiselberg1999Ap,Heiselberg2000PhR}, specifically those denoted by APR I$-$IV  in \cite{Gusakov2005MNRAS} or HHJ I$-$IV in \cite{Fortin2016PhRvC}.
Finally we included four analytical PAL models from \cite{Prakash1988PhRvL,Page1992ApJ}, which differ by the functional form of the density dependence of the symmetry energy, see appendix D in \cite{HPY2007Book} for details.

\begin{figure*}
  \includegraphics[width=\textwidth]{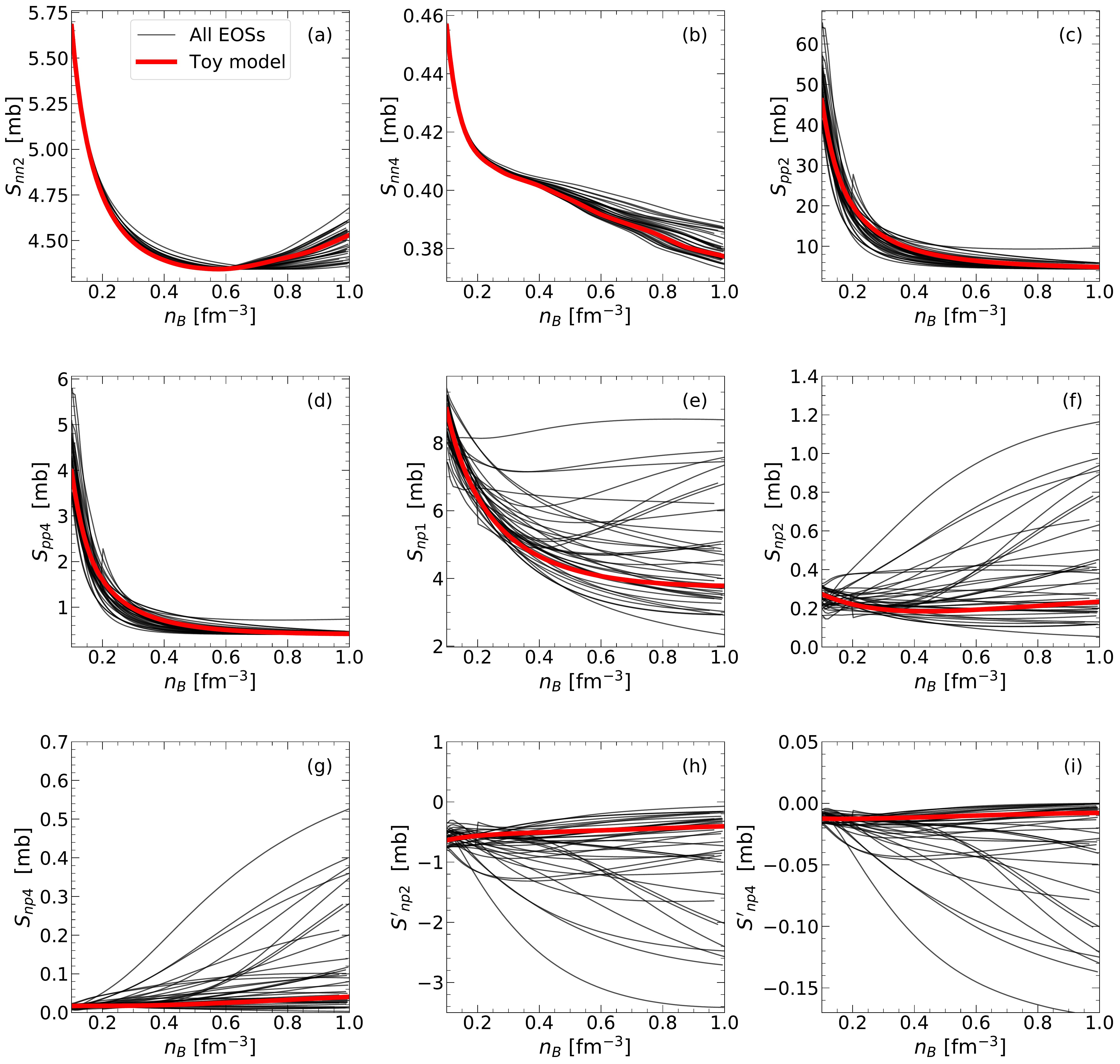}
\caption{Effective transport cross-sections defined in sec.~\ref{sec:cross} as functions of density for all EOS described in the text and the in-vacuum interaction (black solid lines). Thick red solid lines show the results for the toy model with the proton fraction from eq.~(\ref{eq:xpp_ddo}).}
\label{fig:AllS}       
\end{figure*}

In total, our EOS bank contains 39 models \new{summarized in tab.~\ref{tab:eos}}. The proton fractions $x_p$ as functions of the baryon density $n_B$ are given in fig.~\ref{fig:xp_eos} by the black solid lines.\footnote{A prominent jump in one of the black curves seen in fig.~\ref{fig:xp_eos} corresponds to the phase transition in the original APR EOS \cite{Akmal1998PhRv}.} Notice that some EOSs result in quite large (and probably unrealistic) proton fractions for beta-equilibrium matter. 
In the same figure with the thick red line we plot the approximate expression 
\begin{equation}\label{eq:xpp_ddo}
    x_p=0.05 \left(\frac{n_B}{n_0}\right)^{0.65},
\end{equation}
where $n_0=0.16$~fm$^{-3}$, proposed in \cite{Ofengeim2017PhRv} as a qualitative estimate of the proton fraction inside a beta-equilibrium nucleonic NS core. We will use this model as a toy model below.

\begin{figure*}
  \includegraphics[width=\textwidth]{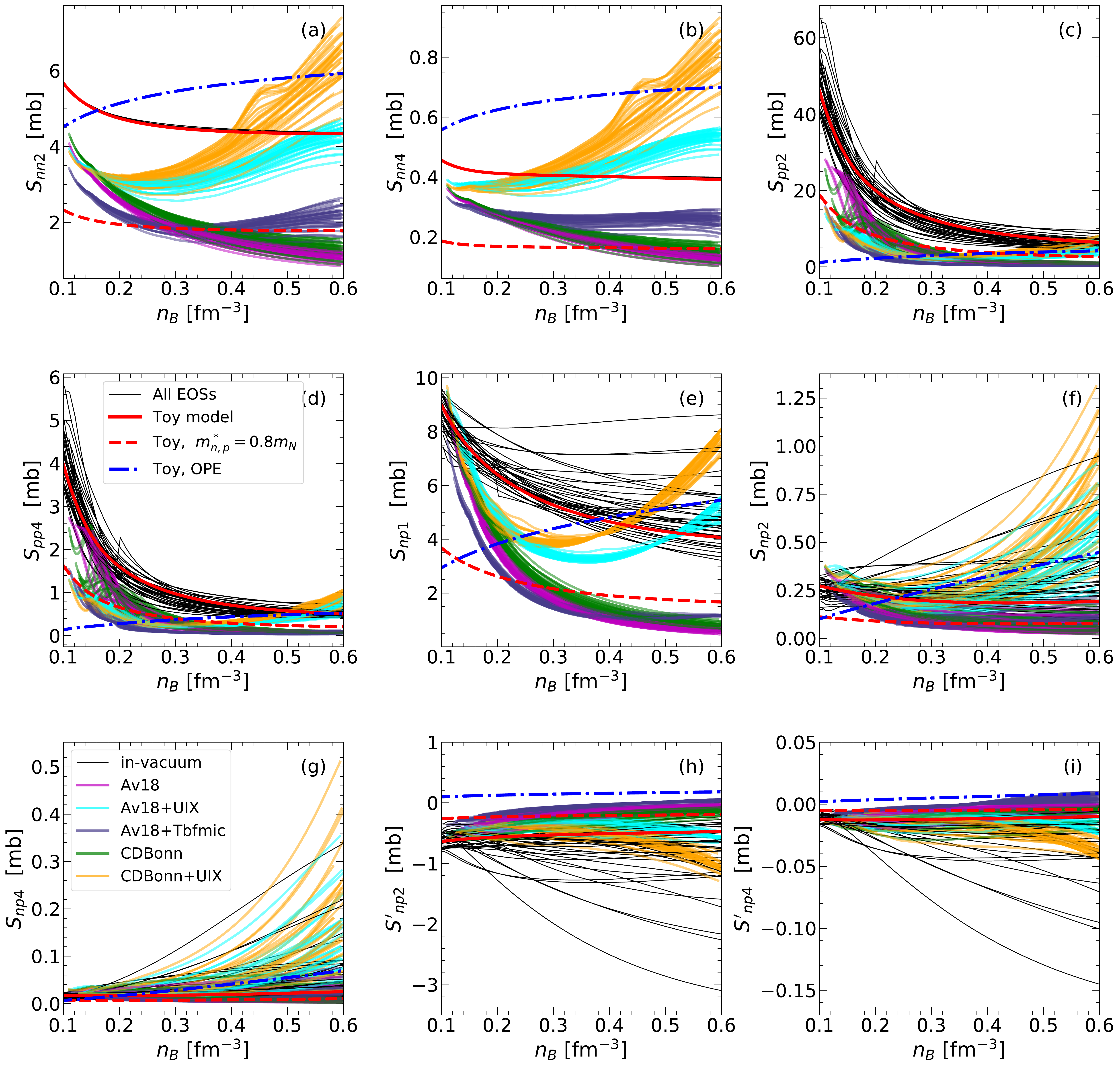}
\caption{Same as fig.~\ref{fig:AllS} with addition of the results for various microscopic potentials applied to all EOSs considered in the paper (cf. fig.~\ref{fig:AllS_bsk21} for BSk21 EOS). Different microscopic interactions are coded as in fig.~\ref{fig:AllS_bsk21}. Thick lines show the toy model (\ref{eq:xpp_ddo}) results for the OPE (blue dash-dotted), in-vacuum (red solid), and in-vacuum with constant $m^*_{n,p}=0.8 m_N$ (red dashed).   }
\label{fig:AllS_pota}       
\end{figure*}

In fig.~\ref{fig:AllS} we plot the same effective transport cross-sections as in fig.~\ref{fig:AllS_bsk21} for the in-vacuum interaction for all EOS considered here (black lines). Thick red lines correspond to the `toy model' composition of eq.~(\ref{eq:xpp_ddo}). One observes that the effective transport cross-sections for the collisions of like particles $S_{nn2}$, $S_{nn4}$, and to a lesser extent, $S_{pp2}$, $S_{pp4}$ are described by the toy model relatively well, in contrast to the cross-sections that involve $np$ collisions. This is a direct consequence of the large scatter of the proton fraction within the EOS bank, see fig.~\ref{fig:xp_eos}. \new{Relatively small scatter of $S_{pp2}$ and $S_{pp4}$ around the toy model is because the main phase-space dependence is accounted for by the normalization factors in definitions (\ref{eq:Sn2}) and (\ref{eq:Sn4}).} Notice that the furthest curves from the thick red line are those with the largest proton fractions.
Nevertheless the toy model is expected to give an acceptable estimate for the transport coefficients of nucleonic NS core matter when more sophisticated calculations are not available.

For the sake of illustration we combine the results shown in figs.~\ref{fig:AllS_bsk21} and \ref{fig:AllS} in fig.~\ref{fig:AllS_pota}. Here we compare the in-vacuum effective transport cross-sections (black lines) with the effective transport cross-sections for the microscopic models from  \cite{Shternin2020PhRvD} (colored lines). The OPE \new{cross-sections} for the toy model (\ref{eq:xpp_ddo}) \new{are} given by the blue dash-dotted lines, while the in-vacuum cross-sections for \new{this model}  are shown with the thick red lines. In addition, with the thick red dashed lines we show the effective transport cross-sections \new{for in-vacuum interaction but} assuming $m_n^*=m_p^*=0.8 m_N$. Figure~\ref{fig:AllS_pota} looks rather cumbersome and basically illustrates that until the microscopic calculations of transport properties are available alongside the EOSs, all `simple' calculations are mere qualitative estimates rather than the quantitative ones.
We can range the levels of approximation as toy model plus $m_{n,p}^*=0.8m_N$, in-vacuum fits for the actual composition plus $m_{n,p}^*=0.8m_N$ and  in-vacuum fits with somehow known $m_{n,p}^*$. Below we continue with first two options.

\begin{figure*}
  \includegraphics[width=\textwidth]{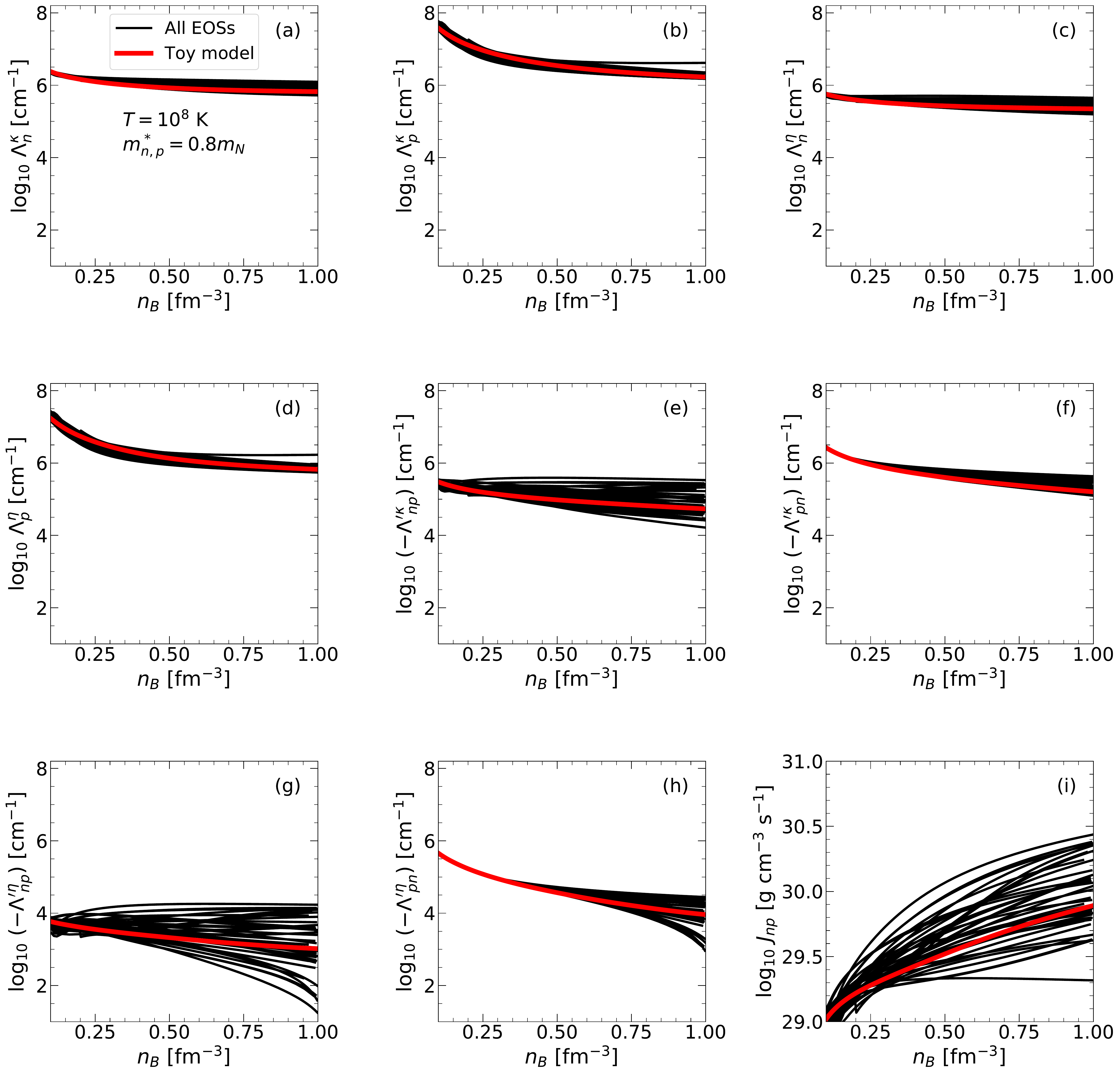}
\caption{ Transport matrix elements for strong interaction in the $np$ subsystem (panels a--h) and the momentum transfer rate $J_{np}$ (panel i) as functions of baryon number density. Results are shown for all EOS considered in the paper  and in-vacuum interaction  (black solid lines). Thick red line shows the toy model (\ref{eq:xpp_ddo}) result. In all cases, $T=10^8$~K and $m^*_{n,p}=0.8 m_N$.
}
\label{fig:AllL}       
\end{figure*}

In fig.~\ref{fig:AllL}, panels (a)--(h) we plot (in the logarithmic scale) the elements of the transport matrices which enter the system of eqs.~(\ref{eq:kinvar_system1}) for the $np$ subsystem and in the panel (i) we plot the neutron-proton momentum transfer rate $J_{np}$ which is given by the similar angular average expression, see sec.~\ref{sec:strong_kinematics}. All quantities are calculated within the in-vacuum interaction model for $T=10^8$~K and $m^*_n=m^*_p=0.8 m_N$. Remember that each $\Lambda_{ab}$ in fig.~\ref{fig:AllL} scales as $\Lambda_{ab}\propto m^{*2}_a m^{*2}_b T^2$. As in fig.~\ref{fig:xp_eos}, black solid lines show results for a complete EOS bank described here, while thick red lines correspond to the toy model composition. We see that the typical transport mean free paths for nucleons due to collisions mediated by strong interactions in NS cores are of the order of $10^{-6}\ T_8^{-2}$~cm, where $T_8=T/(10^8~\mathrm{K})$.

\begin{figure*}
  \includegraphics[width=\textwidth]{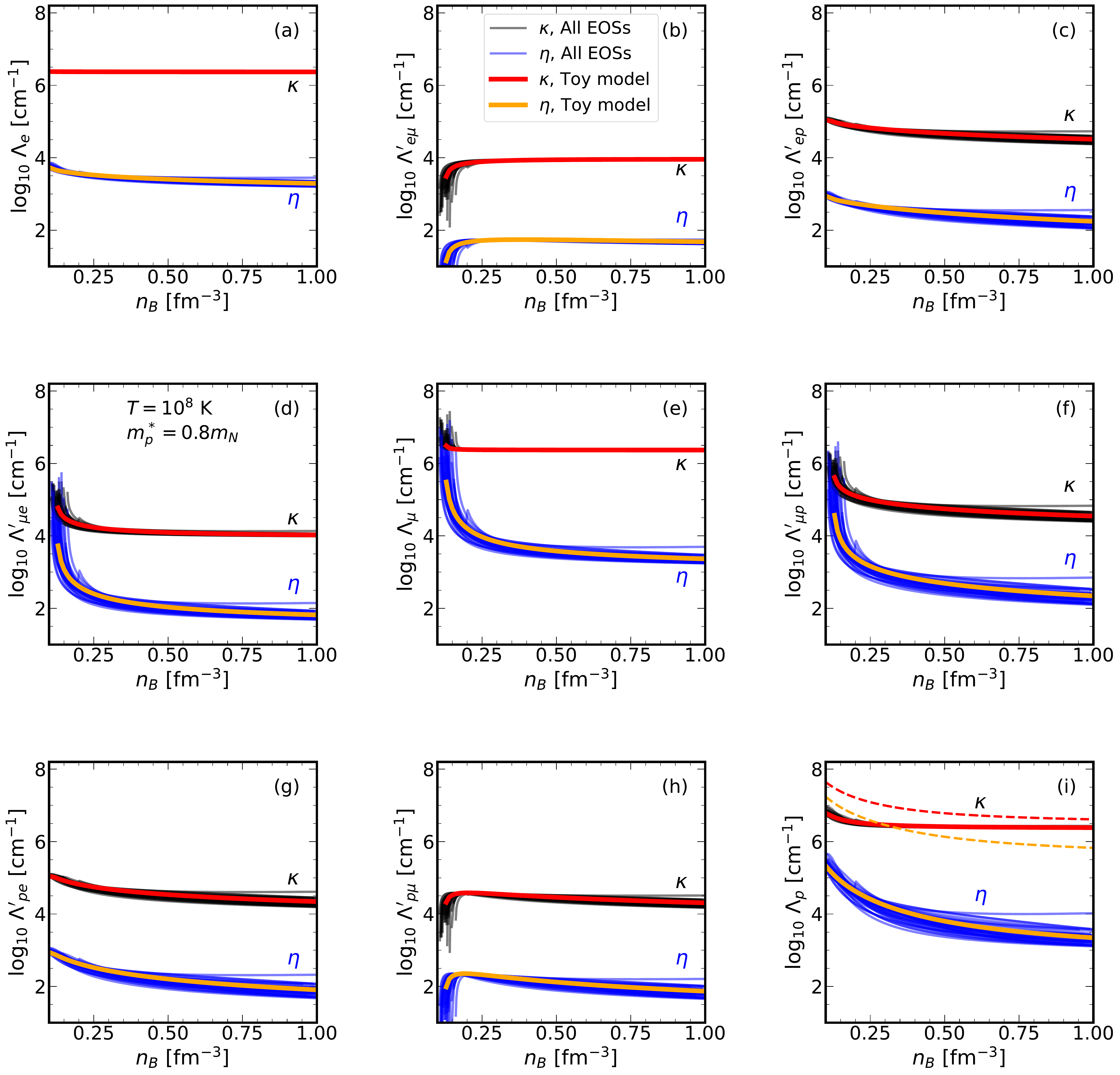}
\caption{ Transport matrix elements for electromagnetic interaction in the $e\mu p$ subsystem as functions of baryon number density. Results for all EOS considered in the paper are shown with   black solid lines for thermal conductivity and blue solid lines for shear viscosity, as marked in the plot. Thick red and  orange line shows the toy model e (\ref{eq:xpp_ddo}) results for $\kappa$ and $\eta$, respectively. Dashed red and orange lines in panel (i) show total $\Lambda_p=(\Lambda_p)^{\mathrm{strong}}+(\Lambda_p)^{\mathrm{em}}$ for thermal conductivity and shear viscosity for the toy model, respectively.
In all cases, $T=10^8$~K and $m^*_{p}=0.8 m_N$.
}
\label{fig:AllLem}       
\end{figure*}

For comparison, in Fig.~\ref{fig:AllLem} we plot the electromagnetic contributions to the transport matrices in the $e\mu p$ subsystem as function of $n_B$. Black and blue lines correspond to thermal conductivity ($k=\kappa$) and shear viscosity ($k=\eta$) problems, respectively, for all EOSs considered here. Thick red and orange lines show the toy model results for $\kappa$ and $\eta$ transport matrices, respectively. Since the transport matrices for shear viscosity, in general, contain additional factor $q^2$ in angular averages (see sec.~\ref{sec:em}), the values of $\Lambda^\eta$ are  couple orders of magnitude smaller than $\Lambda^\kappa$. Remember that due to dominance of the dynamically-screened transverse part of the electromagnetic interaction, the temperature dependence of $\Lambda^k_{ab}$ plotted in fig.~\ref{fig:AllLem} differs from the Fermi-liquid law $\Lambda^k_{ab}\propto T^2$. Approximate scaling for the diagonal components is $\Lambda^\kappa_a\propto T$ and   $\Lambda^\eta_a\propto T^{5/3}$, although in the latter case this scaling works worse as discussed in sec.~\ref{sec:em}. One observes that the diagonal components of the transport matrix for electromagnetic interaction are always much larger than the non-diagonal components \new{which can} be neglected in the first approximation. Protons interact both via the strong and electromagnetic forces therefore in fig.~\ref{fig:AllLem}(i) we plot with dashed lines the total $\Lambda^\kappa_p$ (red dashed lines) and $\Lambda^\eta_p$ (red orange lines). While strong interactions dominate the effective proton mean free path for the shear viscosity problem, this is not so for the thermal conductivity problem, where the contributions from the electromagnetic and strong sectors are comparable. Figures~\ref{fig:AllL} and  \ref{fig:AllLem} show that it is always a good approximation to treat proton as the passive scatterers. 

\begin{figure*}
  \includegraphics[width=\textwidth]{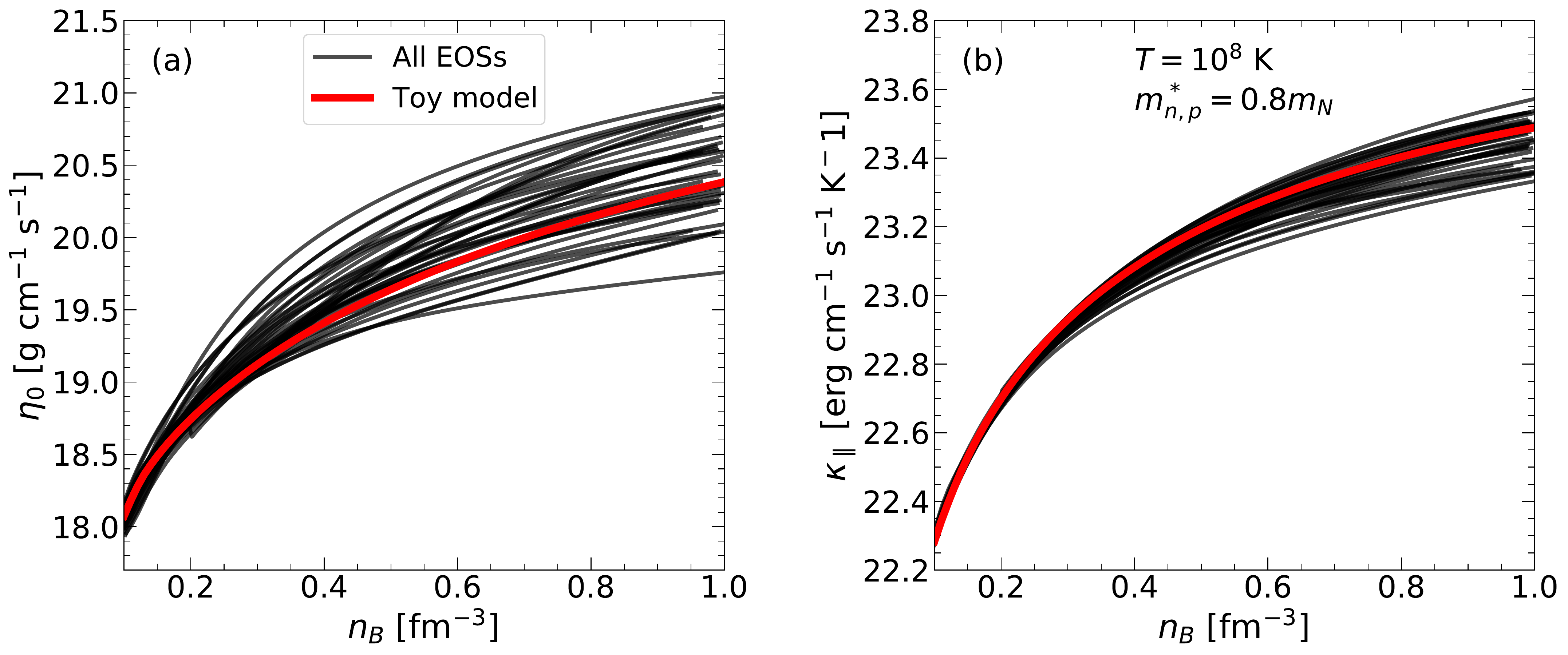}
\caption{Longitudinal component of the shear viscosity $\eta_0$ (panel a) and longitudinal component of the thermal conductivity $\kappa_\parallel$ (panel b) as functions of density for all EOSs considered in the paper (black solid lines) and the toy model  (\ref{eq:xpp_ddo}) (thick red solid line). In all cases, $T=10^8$~K and $m^*_{n,p}=0.8 m_N$.}
\label{fig:etakappa_par}       
\end{figure*}

\subsection{Shear viscosity and thermal conductivity at $\bm{B}=0$.}\label{sec:kin0_discuss}
Now we have all in hand to  calculate the shear viscosity and the thermal conductivity in the NS cores. Figure~\ref{fig:etakappa_par} shows the parallel components of the shear viscosity, $\eta_0$ [panel (a)] and thermal conductivity  $\kappa_\parallel$ [panel (b)] which do not depend on $\bm{B}$. These \new{components are equal to the isotropic} transport coefficients in absence of the magnetic field. As for fig.~\ref{fig:AllL}, we set $T=10^8$~K and $m^*_n=m^*_p=0.8 m_N$ and plot the results for all EOSs  with black lines and the result for the toy model composition with the thick red line. As the electromagnetic contribution and strong sector contribution have different temperature dependencies, there is no general simple temperature scaling for these coefficients (however, see below). One observes qualitatively similar behavior of the transport coefficients with density for all considered EOSs. 

\begin{figure*}
  \includegraphics[width=\textwidth]{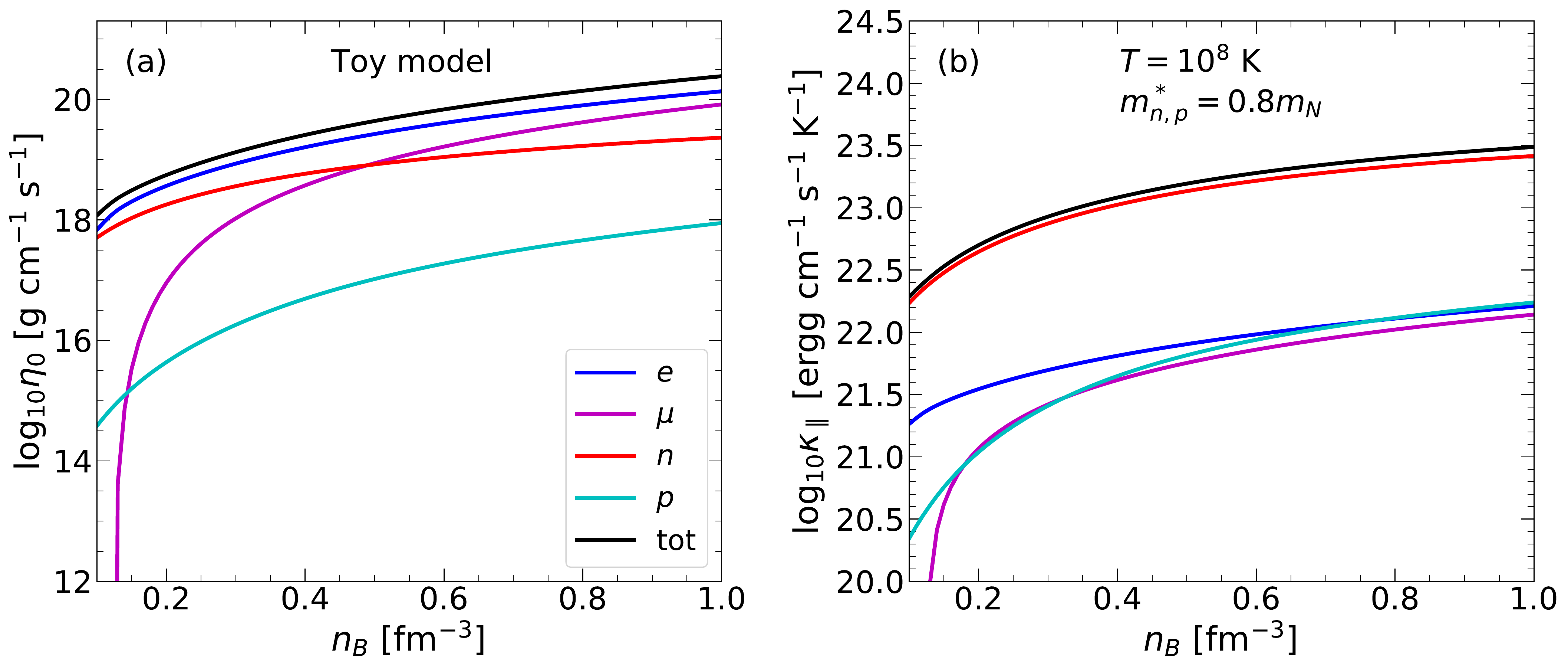}
\caption{Partial contributions to the longitudinal component of the shear viscosity $\eta_0$ (panel a) and longitudinal component of the thermal conductivity $\kappa_\parallel$ (panel b) as functions of density for the toy model  (\ref{eq:xpp_ddo}) and the $npe\mu$ composition. Total contribution and partial contributions are color coded as indicated in the legend on the panel (a). In this calculation, $T=10^8$~K and $m^*_{n,p}=0.8 m_N$.}
\label{fig:etakappa0_partial}       
\end{figure*}

The partial contributions to $\eta_0$ and $\kappa_\parallel$ are shown in fig.~\ref{fig:etakappa0_partial} in panels (a) and (b), respectively, for the toy model composition, $T=10^8$~K, and $m_{n}^*=m_p^*=0.8 m_N$. Other, more realistic EOSs, show similar behavior. Notice that the thermal conductivity, fig.~\ref{fig:etakappa0_partial}(b), is almost completely determined by the neutron contribution \cite{Schmitt2018}. This is due to the strong dominance of the transverse channel of the electromagnetic interaction which suppresses the lepton thermal conductivity and also makes it temperature-independent, sec.~\ref{sec:em}. The neutron contribution, in contrast, scales as $T^{-1}$.
Thus, only at high temperatures, $T\gtrsim 3\times 10^9$~K, leptons start to play some role in \new{heat conduction in the} non-superfluid matter  \cite{ShterninYakovlev2007,Schmitt2018}. The neutron thermal conductivity is determined by the $\Lambda_{n}^\kappa$ transport matrix element shown in fig.~\ref{fig:AllL}(a). Its relatively small scatter over EOSs transforms to the relatively small scatter for $\kappa_\parallel$ in fig.~\ref{fig:etakappa_par}(b). It is clear that the modification of the microscopic interaction results in large changes in $\kappa_\parallel$ \cite{Shternin2020PhRvD}. Anyway, the important fact for practical applications is that the thermal conductivity  in NS cores is large, effectively washing out the temperature gradients. 

The situation is in some sense opposite for the shear viscosity. According to fig.~\ref{fig:etakappa0_partial}(a), leptons give the main contribution to $\eta_0$, while neutron contribution is less important \cite{Shternin2020PhRvD,Schmitt2018}. \new{The difference with the thermal conductivity case is due to considerably larger transport lepton mean free paths for the shear viscosity problem than for the thermal conductivity (in other words $\Lambda^\eta_{\ell}\ll \Lambda^\kappa_{\ell}$) due to different kinematics, see fig.~\ref{fig:AllLem} and the discussion in sec.~\ref{sec:lambda_discuss}.} 
Notice, that 
\new{the conclusion $\eta_\ell \gg \eta_n$ }
does not depend on temperature, since the temperature dependencies of the lepton and neutron contributions are close. Namely, the neutron shear viscosity scales as $T^{-2}$, while the non-Fermi liquid effects in the electromagnetic interactions modify this scaling for leptons at most to $T^{-5/3}$ at lowest temperatures \cite{ShterninYakovlev2008,Schmitt2018} [see also fig.~\ref{fig:etakappa_ddo}(a) below]. The gross result of the realistic microscopic interaction, at least for the models considered here, is the further reduction of the neutron contribution to $\eta_0$ \cite{Shternin2013PhRvC,Shternin2020PhRvD}. In this respect, a considerable scatter in the results for various EOSs seen in fig.~\ref{fig:etakappa_par}(a) results from the scatter in proton (and hence lepton) fractions, see fig.~\ref{fig:xp_eos}, and not \new{from the} scatter in $S_\alpha$. This means that the knowledge of the proton effective mass and the particle fractions is enough for reliable calculation of the shear viscosity coefficient $\eta_0$ (i.e., shear viscosity in the absence of magnetic field) in non-superfluid NS core matter.

\begin{figure*}
  \includegraphics[width=\textwidth]{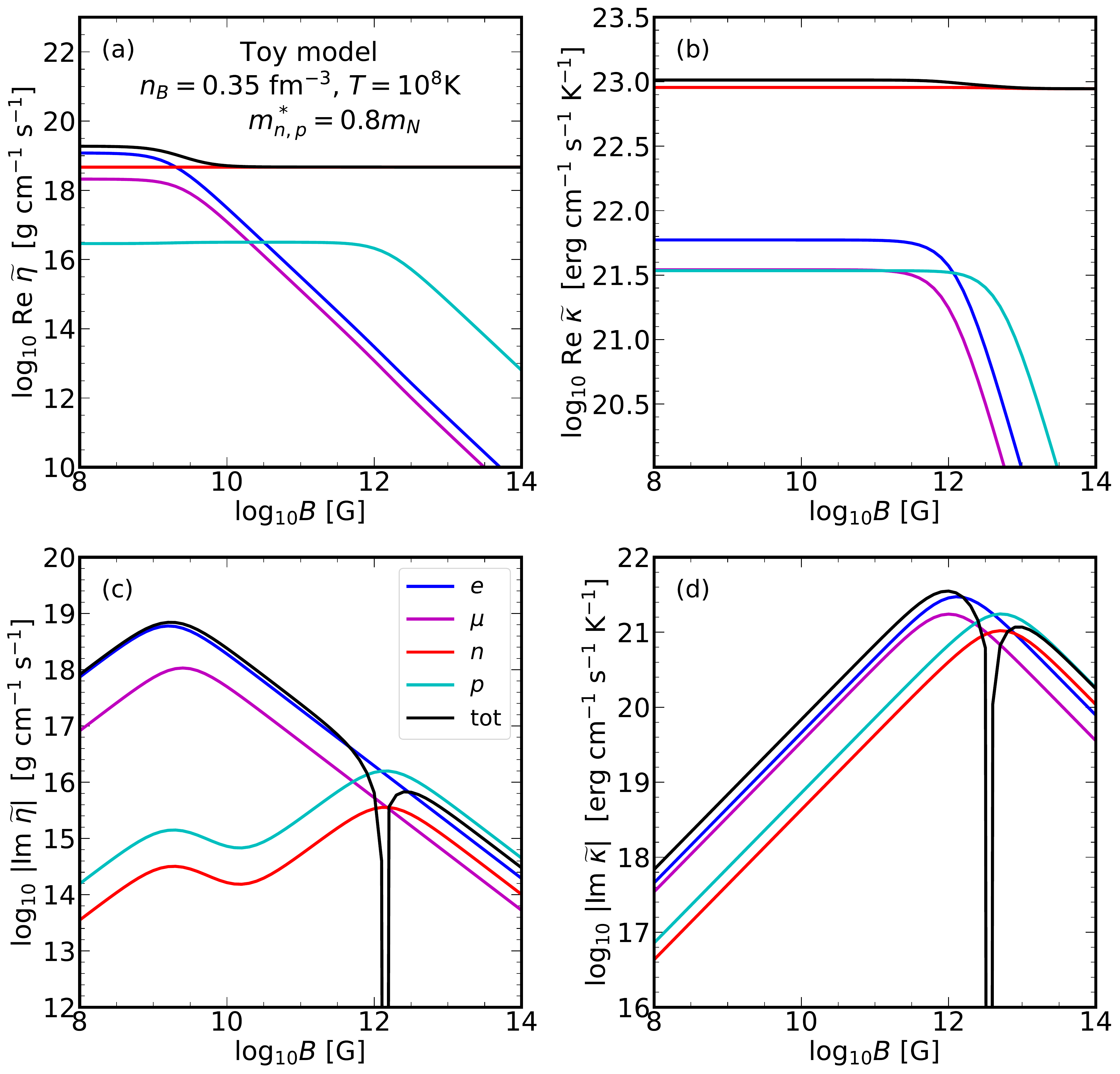}
\caption{Partial contributions to the complex shear viscosity $\widetilde{\eta}$ and thermal conductivity $\widetilde{\kappa}$ coefficients as functions of the magnetic field. Panels a--d show, respectively, $\mathrm{Re}\widetilde{\eta}(B)$, $|\mathrm{Im}\widetilde{\eta}(B)|$,
$\mathrm{Re}\widetilde{\kappa}(B)$, and 
$|\mathrm{Im}\widetilde{\kappa}(B)|$. Results are plotted for the toy model (\ref{eq:xpp_ddo}), $npe\mu$ composition, $T=10^8$~K, $n_B=0.35$~fm$^{-3}$ and $m^*_{n,p}=0.8 m_N$. Black solid line marked $B=0$ in panel (a) and lines marked $B<10^{12}$~G in panel (b) also represent the longitudinal components $\eta_0$ and $\kappa_\parallel$, respectively. Dashed line marked $e\mu\ (B=0)$ in panel (b) shows the electron and muon contribution to  $\kappa_\parallel$. Discontinuities in the Hall contributions (panels c and d) indicate the sign change of these components, see text for details.
}
\label{fig:etakappaB_ddo}       
\end{figure*}

\subsection{Shear viscosity and thermal conductivity of magnetized NS cores.}\label{sec:kinB_discuss}
We now turn to the transport coefficients in presence of the magnetic field. Remember, that at the lowest variational order  there is no magnetic field dependence and no tensor structure in the momentum transfer rates $J_{ab}$ (this is of course not true for the electric conductivity tensor that can be obtained by inverting the generalized Ohm law (\ref{eq:Ohm}), see, e.g., \cite{Iakovlev1991Ap&SS}). Therefore we refer the reader to ref.~\cite{Shternin2020PhRvD} and sec. VII of ref.~\cite{Dommes2020PhRvD} for the detailed discussion of the momentum transfer rates in the non-superfluid $npe\mu$ NS cores which extends the discussion here. 

\new{In the magnetic field, components of the thermal conductivity and shear viscosity tensors 
 are described by the complex functions $\widetilde{\kappa}(B)$ and $\widetilde{\eta}(B)$  (secs. \ref{sec:tensor} and \ref{sec:collisions})}. We have already discussed the longitudinal contributions $\kappa_\parallel=\widetilde{\kappa}(B=0)$ and $\eta_0=\widetilde{\eta}(B=0)$. In fig.~\ref{fig:etakappaB_ddo} we plot real and imaginary parts of $\widetilde{\kappa}(B)$ and $\widetilde{\eta}(B)$ 
and corresponding partial contributions from different particle species as functions of  $B$. 
\new{Recall that the Cartesian components of the thermal conductivity are $\kappa_\perp=\mathrm{Re}\ \widetilde{\kappa}(B)$ and $\kappa_\wedge=-\mathrm{Im}\ \widetilde{\kappa}(B)$, while \new{for the shear viscosity} $\eta_1=\mathrm{Re}\ \widetilde{\eta}(2B)$, $\eta_3=-\mathrm{Im}\ \widetilde{\eta}(2B)$, $\eta_2=\mathrm{Re}\ \widetilde{\eta}(B)$, and $\eta_4=-\mathrm{Im}\ \widetilde{\eta}(B)$.}
As above, we use the toy model composition at a characteristic baryon density $n_B=0.35$~fm$^{-3}$. We also set $T=10^8$~K and $m^*_n=m^*_p=0.8 m_N$. The behavior of the transport coefficients can be understood by comparing fig.~\ref{fig:etakappaB_ddo} with   fig.~\ref{fig:kinHall} which is plotted in the relaxation-time approximation.

Indeed, the results of sec.~\ref{sec:em} for electromagnetic collisions suggest that the non-diagonal (primed) components of the  transport matrices for electromagnetic interactions  are significantly smaller than the diagonal components, as seen in  fig.~\ref{fig:AllLem}. Therefore in the first approximation one can neglect mixing between leptons and other species and the lepton mean free paths can be determined from the simple equation of the effective relaxation time form, namely
\begin{eqnarray}\label{eq:lambda_ell}
    \widetilde{\lambda}^k_\ell&=&\frac{1}{\Lambda^k_\ell+i\omega_{\mathrm{BF}\ell}/v_{\mathrm{F}\ell}},\quad \ell=e,\,\mu
\end{eqnarray}
for $k=\kappa,\,\eta$. Therefore lepton transport coefficients behave as shown in fig.~\ref{fig:kinHall} where the  effective Hall parameters are $x_{\mathrm{Hall}\ell}=\omega_{\mathrm{BF}\ell} \left(\Lambda^k_\ell v_{\mathrm{F}\ell}\right)^{-1}$. In a strong sector situation is not so simple. Dashed lines in fig.~\ref{fig:AllLem}(i) can be viewed as an estimate for the real part of inverse proton mean free paths. Comparing it with the non-diagonal components of the transport matrices ${\Lambda'}^k_{np}$ in fig.~\ref{fig:AllL}, one can assume that protons do not influence the real part of the equation for neutron effective mean free path $\widetilde{\lambda}^k_n$ (but not the imaginary part, since neutrons are electrically neutral, see below). Therefore, one can write for neutrons the relaxation-time type expression \begin{equation}\label{eq:lambda_n}
    \mathrm{Re}\ \widetilde{\lambda}^k_n=\frac{1}{\Lambda^k_n},
\end{equation}
which does not depend on the magnetic field. According to fig.~\ref{fig:AllL}, inverse neutron mean free path $\Lambda^k_n$ [see eq.~(\ref{eq:lambda_n})] can be comparable to the non-diagonal components of the transport matrices ${\Lambda'}^k_{pn}$, therefore this mixing can affect proton mean free paths. In this approximation, one finds
\begin{equation}\label{eq:lambda_p}
    \mathrm{Re}\ \widetilde{\lambda}^k_p=\mathrm{Re}\ \frac{1-{\Lambda'}^k_{pn}/\Lambda^k_n}{\Lambda^k_p+i\omega_{\mathrm{BF}p}/v_{\mathrm{F}p}}.
\end{equation}
Equation (\ref{eq:lambda_p}) has the effective relaxation-time form with the Hall parameter
$x_{\mathrm{Hall}p}=\omega_{\mathrm{BF}p} \left(\Lambda^k_p v_{\mathrm{F}p}\right)^{-1}$, but with numerator different from unity.

Consider now the transverse components of the shear viscosity  shown in fig.~\ref{fig:etakappaB_ddo}(a). At low magnetic field ($B\lesssim 3\times 10^9$~G for conditions in fig.~\ref{fig:etakappaB_ddo}) leptons are not magnetized, i.e. $x_{\mathrm{Hall}{e,\mu}}<1$. In this case, $\mathrm{Re}\ \widetilde{\eta}(B)\approx \eta_0\approx \eta_{0e\mu}$, see fig.~\ref{fig:etakappaB_ddo}(a). When $x_{\mathrm{Hall}{e,\mu}}$ reaches unity and starts to increase with increasing $B$, leptons become strongly magnetized and their contribution to $\mathrm{Re}\ \widetilde{\eta}$ (i.e., to the transverse components of the shear viscosity tensor) rapidly diminishes [see blue and magenta lines in fig.~\ref{fig:etakappaB_ddo}(a)]. At this point ($B\gtrsim  10^{10}$~G for conditions in fig.~\ref{fig:etakappaB_ddo}), $\mathrm{Re}\ \widetilde{\eta}$ is fully determined by neutrons according to eq.~(\ref{eq:lambda_n}). Protons become magnetized at larger fields when $x_{\mathrm{Hall}p}\gtrsim 1$ , but their contribution to $\mathrm{Re}\ \widetilde{\eta}$ is negligible. 

A simpler behavior is observed for the transverse component of the thermal conductivity, $\mathrm{Re}\ \widetilde{\kappa}$, plotted in fig.~\ref{fig:etakappaB_ddo}(b). According to fig.~\ref{fig:etakappa0_partial}(b), the lepton contribution to the thermal conductivity is negligible (at least at the selected temperature, see below) and their magnetization results in almost no changes in $\mathrm{Re}\ \widetilde{\kappa}$ [fig.~\ref{fig:etakappaB_ddo}(b), black line]. Notice that the lepton Hall parameters for thermal conductivity reach unity at much larger magnetic fields $B$ than those for shear viscosity. This is due to smaller effective transport mean free paths for thermal conductivity due to stronger dominance of the transverse channel of electromagnetic interactions.

Consider now the Hall components, $\mathrm{Im}\ \widetilde{\eta}$ and  $\mathrm{Im}\ \widetilde{\kappa}$, plotted in fig.~\ref{fig:etakappaB_ddo}(c) and fig.~\ref{fig:etakappaB_ddo}(d), respectively.  Until protons are magnetized, the Hall components are fully determined by leptons and follow the relaxation-time expressions (\ref{eq:lambda_ell}). They reach maxima at $x_{\mathrm{Hall}{e,\mu}}\approx 1$, cf. fig.~\ref{fig:kinHall}. 
 According to eqs.~(\ref{eq:eta3_strong}) and (\ref{eq:kappa_hall_rta_lim}), asymptotic values of Hall components of shear viscosity and thermal conductivity at $x_{\mathrm{Hall}{e,\mu}}\gg 1$ (but $x_{\mathrm{Hall}p}\ll 1$) are given by the universal expressions
\begin{subequations}\label{eq:emu_hall_lim}
\begin{eqnarray}
    \mathrm{Im}\,\widetilde{\eta}&\approx&\mathrm{Im}\,\widetilde{\eta}_{e\mu}= -\frac{n_ep_{Fe}^2+n_\mu p_{F\mu}^2}{5e B},\label{eq:eta_emu_hall_lim}\\
    \mathrm{Im}\,\widetilde{\kappa}&\approx&\mathrm{Im}\,\widetilde{\kappa}_{e\mu}= -\frac{\pi^2T}{3}\frac{n_e+n_\mu}{e B}.\label{eq:kappa_emu_hall_lim}
\end{eqnarray}
\end{subequations}
In figs.~\ref{fig:etakappaB_ddo}c and \ref{fig:etakappaB_ddo}d we plot the Hall components in the logarithmic scale, so their sign is not shown. The Hall contribution  of leptons have negative sign due to negative charges of leptons, while the proton contribution is positive. 
Figures \ref{fig:etakappaB_ddo}a and \ref{fig:etakappaB_ddo}c show that at $x_{\mathrm{Hall}e,\mu}\approx 1$, the Hall components of the shear viscosity are comparable to the transverse components.
For thermal conductivity, $\mathrm{Im}\,\kappa$ is always much smaller than $\mathrm{Re}, \kappa$ (figs. \ref{fig:etakappaB_ddo}b and \ref{fig:etakappaB_ddo}d). 

From eq.~(\ref{eq:kinvar_system}) for electrically neutral ($q_n=0$) neutrons, for $k=\eta,\,\kappa$ one obtains
\begin{equation}\label{eq:im_lambda_n}
    \mathrm{Im}\,\widetilde{\lambda}^k_{n}=-\frac{{\Lambda'}^k_{np}}{{\Lambda}^k_{n}}\ 
    \mathrm{Im}\,\widetilde{\lambda}^k_{p},
\end{equation}
i.e. despite being charge neutral, neutrons contribute  to the Hall components of thermal conductivity or shear viscosity due to their interaction with protons. Depending on the sign of $(\Lambda')^{k}_{np}$ which can be either positive or negative, the neutron contribution to the total $\mathrm{Im}\,\widetilde{\eta}$ or $\mathrm{Im}\,\widetilde{\kappa}$  can be positive or negative. For the specific case shown in fig.~\ref{fig:etakappaB_ddo}, both proton and neutron contributions are positive and larger that those for leptons, leading to change in sign of this component at $x_{\mathrm{Hall}p}\gtrsim 1$, which is seen as a discontinuities in total Hall components shown in fig.~\ref{fig:etakappaB_ddo}(c) and fig.~\ref{fig:etakappaB_ddo}(d).

Proton contribution to the Hall components of transport coefficients, and hence the neutron contribution as well show more complicated behavior around $x_{\mathrm{Hall}e,\mu}\sim 1$ in figs.~\ref{fig:etakappaB_ddo}c, \ref{fig:etakappaB_ddo}d than follows from a simple relaxation-time approximation. This can be understood as follows. Taking into account eq.~(\ref{eq:im_lambda_n}), the equation for $\mathrm{Im}\, \widetilde{\lambda}^k_p$ reads
\begin{multline}
    \left(
    \Lambda^k_p- \frac{{\Lambda'}^k_{np}{\Lambda'}^k_{pn}}{\Lambda^k_n} 
    \right)\mathrm{Im}\ \widetilde{\lambda}^k_p=-\frac{\omega_{\mathrm{BF}p}}{v_{\mathrm{F}p}}\,\mathrm{Re}\ \widetilde{\lambda}_p\\
    -{\Lambda'}^k_{pe}\,\mathrm{Im}\ \widetilde{\lambda}^k_e-{\Lambda'}^k_{p\mu}\,\mathrm{Im}\ \widetilde{\lambda}^k_\mu.\label{eq:im_lambda_p}
\end{multline}
Assuming that $\widetilde{\lambda}^k_e$, $\widetilde{\lambda}^k_\mu$, and $\mathrm{Re}\ \widetilde{\lambda}^k_p$ are determined by eqs.~(\ref{eq:lambda_ell}) and (\ref{eq:lambda_p}), the imaginary part $\mathrm{Im}\ \widetilde{\lambda}^k_p$ can be calculated from eq.~(\ref{eq:im_lambda_p}).
Until $x_{\mathrm{Hall}e,\mu,p}<1$, $\mathrm{Im}\ \widetilde{\lambda}^k_\ell\propto \omega_{\mathrm{BF}\ell}$ and one has  $\mathrm{Im}\ \widetilde{\lambda}^k_p\propto B$. When leptons become magnetized, $x_{\mathrm{Hall}e,\mu}\sim1$, but still  $x_{\mathrm{Hall}p}<1$, lepton contribution to right-hand side of eq.~(\ref{eq:im_lambda_p}) drops down that is clearly visible in cyan and red lines in fig.~\ref{fig:etakappaB_ddo}(c). For thermal conductivity, this effect is not visible in fig.~\ref{fig:etakappaB_ddo}(d) since the protons and leptons have similar Hall parameters here.
Finally, when all charged particles are  magnetized,  $\mathrm{Im}\ \widetilde{\lambda}^k_\ell\propto \omega_{\mathrm{BF}\ell}^{-1}$, $\mathrm{Re}\ \widetilde{\lambda}^k_p\propto \omega_{\mathrm{BF}p}^{-2}$ so that $\mathrm{Im} \widetilde{\lambda}^k_p\propto B^{-1}$, but with the proportionality coefficient which contains combination of the components of  transport matrices. Therefore the total Hall components of shear viscosity and thermal conductivity do not reach universal asymptotics in the large Hall parameter regime analogous to eqs.~(\ref{eq:emu_hall_lim}). 
Almost complete cancellation of the \new{lepton and neutron contributions} to $\mathrm{Im}\,\widetilde{\kappa}$  seen at large $B$ in fig.~\ref{fig:etakappaB_ddo}(d) is a chance coincidence which breaks down at other densities.

\begin{figure*}
  \includegraphics[width=\textwidth]{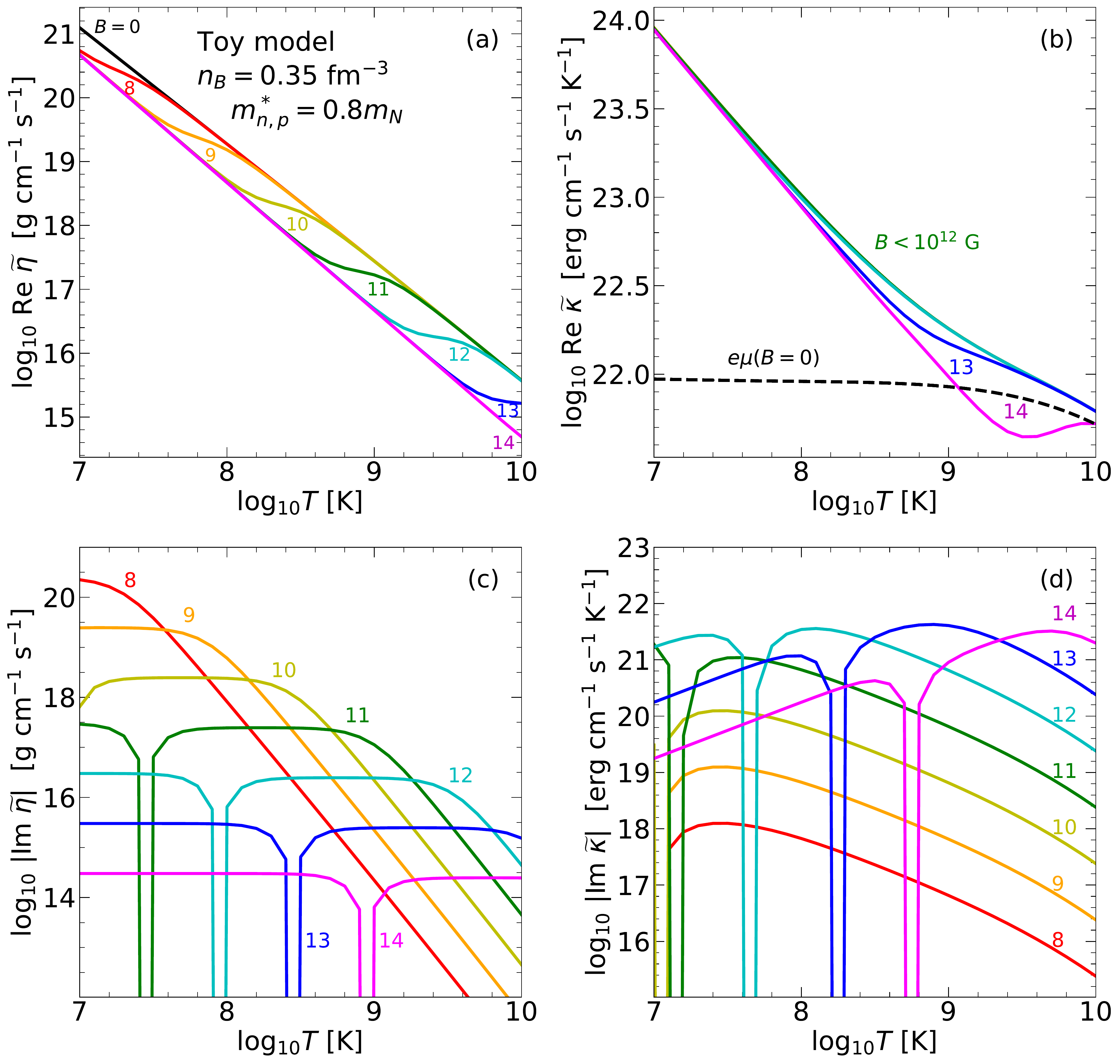}
\caption{Temperature dependence of the  complex shear viscosity $\widetilde{\eta}$ and thermal conductivity $\widetilde{\kappa}$ coefficients for various values of the magnetic field $B$. Different values of $B$ are shown in different colors and indicated by $\log_{10} B [\mathrm{G}]$ near the respective curves. 
Panels a--d show, respectively, $\mathrm{Re}\widetilde{\eta}(B)$, $|\mathrm{Im}\widetilde{\eta}(B)|$,
$\mathrm{Re}\widetilde{\kappa}(B)$, and 
$|\mathrm{Im}\widetilde{\kappa}(B)|$. Results are plotted for the toy model  (\ref{eq:xpp_ddo}), $npe\mu$ composition, $n_B=0.35$~fm$^{-3}$ and $m^*_{n,p}=0.8 m_N$. }
\label{fig:etakappa_ddo}       
\end{figure*}

Finally, in fig.~\ref{fig:etakappa_ddo} we explore the temperature dependence of  $\widetilde{\eta}$ and $\widetilde{\kappa}$ for  $B=10^8,\, 10^9,\dots, 10^{14}$~G; the values of $\log_{10} B~[\mathrm{G}]$ are shown near the respective curves. As in fig.~\ref{fig:etakappaB_ddo}, we use the toy model composition at $n_B=0.35$~fm$^{-3}$ and $m_n^*=m_p^*=0.8 m_N$.

In fig.~\ref{fig:etakappa_ddo}(a) we show $\mathrm{Re}\ \widetilde{\eta}$ component of the shear viscosity tensor. According to the discussion above, at $B=0$ [black solid line in fig.~\ref{fig:etakappa_ddo}(a) marked $B=0$], leptons give the dominant contribution to the shear viscosity \cite{Shternin2020PhRvD}.  At low temperatures, the shear viscosity obeys the $\eta\propto T^{-5/3}$ asymptotic instead of the standard Fermi-liquid result $\eta\propto T^{-2}$ due to the dominance of the transverse channel of the electromagnetic collisions, see eq.~(\ref{eq:Lambda_eta_t}). At higher temperatures, the contribution of the longitudinal channel to the lepton shear viscosity is significant and the slope of the $\eta(T)$ dependence lies in between the $\propto T^{-5/3}$ and $\propto T^{-2}$ powerlaws. At each given $B>0$ the behavior of $\mathrm{Re}\,\widetilde{\eta}(B)$ is similar. Since the effective mean free paths increase with decrease 
of temperature, the effective Hall parameters for leptons also increases. Until $x_{\mathrm{Hall}e\mu}=1$, magnetic field does not influence the transport significantly and $\mathrm{Re}\,\widetilde{\eta}(B)\approx \widetilde{\eta}(B=0)$, see fig.~\ref{fig:etakappaB_ddo}. With lowering temperature, leptons become magnetized and their contribution to $\mathrm{Re}\,\widetilde{\eta}$ decreases until the transverse shear viscosity components become completely determined by neutrons. In fig.~\ref{fig:etakappa_ddo}(a) this is the line corresponding to the highest $B=10^{14}$~G plotted. At this point, $\mathrm{Re}\,\widetilde{\eta}\approx \mathrm{Re}\,\widetilde{\eta}_n\propto T^{-2}$ and does not depend on $B$ (see fig.~\ref{fig:etakappaB_ddo}).

The Hall components of the shear viscosity, i.e. $\mathrm{Im}\,\widetilde{\eta}$, are plotted in fig.~\ref{fig:etakappa_ddo}(c). This figure can be understood in the same way as fig.~\ref{fig:etakappaB_ddo}(c). At a given $B$, starting from large temperatures, $\mathrm{Im}\,\widetilde{\eta}$ is governed by leptons and basically follow the relaxation-time approximation law eq.~(\ref{eq:eta_rta_deg}). When leptons are magnetized,  $\mathrm{Im}\,\widetilde{\eta}$ is given by (\ref{eq:eta_emu_hall_lim}) and does not depend on temperature until the magnetization of protons occur. After that, $\mathrm{Im}\,\widetilde{\eta}$ changes sign rapidly (in present model) and become determined by all particle species. At low temperatures, $\mathrm{Im}\,\widetilde{\eta}$ stays constant, which is however not universal and depends on the transport scattering cross-sections as follows from above [see eq.~(\ref{eq:im_lambda_p})].

Less interesting situation is observed for the thermal conductivity.
In fig.~\ref{fig:etakappa_ddo}(b) we show $\mathrm{Re}\ \widetilde{\kappa}$ component of the thermal conductivity tensor. Since at $B=0$ thermal conductivity is governed by neutrons [figs.~\ref{fig:etakappa0_partial}(b) and \ref{fig:etakappaB_ddo}(b)], magnetic field does not influence it considerably and for $B<10^{12}$~G for conditions shown in  fig.~\ref{fig:etakappa_ddo}(b), $\mathrm{Re}\ \widetilde{\kappa}(B)\approx \mathrm{Re}\ \widetilde{\kappa}(B=0)=\kappa_n$. The lepton contribution to thermal conductivity at $B=0$ is shown with dashed line in fig.~\ref{fig:etakappa_ddo}(b). At lower temperatures it is temperature-independent and start to transit to the standard Fermi-liquid behavior $\kappa\propto T^{-1}$ at large $T$. 
At large $B\gtrsim 10^{12}$~G, the lepton magnetization starts at high temperatures, where the lepton contribution to  the total thermal conductivity is sizable. Therefore, for $B\gtrsim 10^{13}$~G one observes the drop in $\mathrm{Re}\ \widetilde{\kappa}$ representing basically  the neutron contribution. In any case, thermal conductivity perpendicular to the magnetic field stays large. 

The behavior of the Hall component of thermal conductivity, $\mathrm{Im}\ \widetilde{\kappa}$, shown in  fig.~\ref{fig:etakappa_ddo}(d) is \new{qualitatively similar} 
to the behavior of $\mathrm{Im}\ \widetilde{\eta}$. \new{In the limit of large magnetization, the} only difference is the appearance of temperature in the limiting expressions eq.~(\ref{eq:kappa_hall_rta_lim}) and (\ref{eq:kappa_emu_hall_lim}) leading to $\mathrm{Im}\ \widetilde{\kappa} \propto T$ limiting behavior at low temperatures. \new{In the opposite limit of weak magnetization (large $T$),  $\mathrm{Im}\ \widetilde{\kappa}$ is determined by leptons and basically have relaxation-time approximation behavior according to eqs.~(\ref{eq:kappa_rta}) an (\ref{eq:lambda_ell}). The change of the slope of $\mathrm{Im}\ \widetilde{\kappa}(T)$ curve seen in fig.~\ref{fig:etakappa_ddo}(d) reflects the change of  the $\Lambda^\kappa_\ell$ temperature dependence 
[seen also with the dashed line in fig.~\ref{fig:etakappa_ddo}(b)]. This is a result of increasing contribution of the longitudinal plasmon exchange to lepton scattering at large $T$. }

\section{Conclusion}
\label{sec:conclude}
In this paper we reviewed the  framework for calculating the first-order transport coefficients (shear viscosity, thermal conductivity, and momentum transfer rates) of the multicomponent non-superfluid NS core matter in the presence of the strong magnetic field. In this case, transport coefficients become tensor quantities which can be most economically calculated utilizing the spherical tensor formalism. In this approach, each tensorial transport coefficient is described by a single complex function of the magnetic field as discussed in sec.~\ref{sec:tensor}.

At the simplest variational level, thermal conductivity and shear viscosity are given by eqs.~(\ref{eq:kappa_var}) and (\ref{eq:eta_var}), where the effective generalized complex mean free paths are found from the system of linear algebraic equations (\ref{eq:kinvar_system1} or (\ref{eq:kinvar_system1}). The core of this system is the transport matrix elements ${\Lambda}_{ab}^k$ and ${\Lambda'}_{ab}^k$ ($k=\kappa,\, \eta$, and $a,\, b$ indicate particle species) which are expressed by the angular averages of the scattering probabilities with certain weighting factors depending on the transport problem in question. The definitions of these matrix elements are given in eqs.~(\ref{eq:trmat_kappa_ab})--(\ref{eq:trmat_eta_ab}).
The momentum transfer rates are also expressed via the corresponding transport matrix element, eq.~(\ref{eq:Jab_var_long}).

The transport matrices depend on the interactions between the constituents of the  NS core liquid and need to be calculated from the microscopic theory. Leptons interact via the electromagnetic forces, the corresponding `electromagnetic' contribution to transport matrices can be calculated in the relatively model-independent way, provided composition and effective masses of charged baryons are known. In contrast, baryons interact mainly through the strong force channel. In this case calculations of EOS and the transport properties should be performed in consistent way. Ideally, public EOS databases such as CompOSE, should also provide the transport matrices for each EOS.
When such calculations are not available, one is forced to use the approximate approaches. 

We applied the developed formalism to the npe$\mu$ composition of the NS cores.  Utilizing 39 EOS models and five microscopic interactions (not consistent with these EOSs), we investigated the uncertainty range that can be expected in realistic calculations. 
We then compared these results with approximate approach based on the in-vacuum cross-section of the nucleon scattering.
We expect that these results (supplemented with the nucleon effective masses if available) give qualitatively correct picture of transport properties \new{of the} NS core matter. We provide (\ref{sec:app:pract}) the practical fitting expressions that can be used for any EOS of dense matter. In addition we provide the `toy model' fits based on the crude estimate (\ref{eq:xpp_ddo}) for the proton fraction that allow one to estimate the transport matrices and hence the transport coefficients at a given baryon density.

Magnetic field more strongly affects the motion of light leptons than the baryonic component of the matter. Therefore magnetic field effects are more prominent for the shear viscosity, where leptons dominate the transport at $B=0$, than for the thermal conductivity, where the situation is opposite
 \cite{Schmitt2018}.
We find that the shear viscosity can be strongly influenced even by the moderate $B\gtrsim 10^{10}$~G. At sufficiently low temperatures, the shear viscosity in the plane transverse to magnetic field becomes fully determined by neutrons. In this sense it can vary significantly for the different models of the nuclear interaction. The `parallel' shear viscosity coefficient $\eta_0$ is not affected by the magnetic field and is dominated by the lepton contribution. 
Thermal conductivity is less affected by the magnetic field, basically the perpendicular component of the conductivity is affected in the high-temperature and high-magnetic field region of the parameter space studied. 

Notice that while these conclusions  were illustrated in the paper for the `toy' model under the approximation of the in-vacuum nucleon scattering, they are expected to be qualitatively valid if more accurate calculations are available.

In our study we ignored the possibility of inelastic collisions (i.e. reactions). The formalism outlined here can be extended to this case by considering  particle non-conserving collision integrals in the system of Fermi-liquid transport equations. Reactions lead to appearance of the corresponding terms in the entropy generation rate and corresponding scalar transport coefficients. These processes are in the mutual interaction with the bulk viscosity coefficients. Moreover in this case a dynamical problem needs to be considered. The reaction-driven bulk viscosity in the dynamical regime can be derived without relying on the kinetic equations, see, e.g., \cite{Schmitt2018} for a review.

We illustrated the formalism described in the paper exclusively for nucleonic NS core composition. In principle, NS cores can have richer composition.  For instance, the hyperonic NS cores are widely discussed, see, e.g.
\cite{Logoteta2021Univ} for a review.  
Transport coefficients for the hypernuclear NS cores can be in principle, calculated using the similar formalism as studied here. \new{For the non-magnetized case  this was done by \cite{Shternin2021Univ} for two models of hypernuclear interaction.}
\new{The momentum} transfer rates within the free-particle model and $npe\Sigma^{-}$ composition \new{were considered} in \cite{Yakovlev1991Ap&SS}. When several hyperon species are present, in principle, there is a possibility of strong inelastic collisions \new{(neglected in \cite{Shternin2021Univ})}. In this case, even when the matter is in the equilibrium state with respect to these reactions, they can contribute to the transport cross-sections. In the presence of inelastic collisions, the system  eq.~(\ref{eq:kinvar_system1}), which determine transport mean free paths,
has the same form, but the transport matrices receive additional contributions. These contributions can be determined following the lines of sec.~\ref{sec:collisions}, but the kinematic expressions for the transport matrix elements and angular averages will be more involved, because all four Fermi-momenta of the quasiparticles participating in collisions are different in this case.

In this study we did not consider the transport properties of the NS core matter in presence of superfluidity/superconductivity. It is known that baryons in NS core matter can form Cooper pairs due to  the presence of the attractive  character  of  some partial wave channels of the baryon interaction, see \cite{SedrakianClark2019EPJA} for a review. This pairing may lead to the existence of superflows and supercurrents. In this case the hydrodynamic equations become much more complicated, and multifluid consideration is necessary, see e.g., \cite{Andersson2021Univ,Rau2020PhRvD}, and references therein. Additional complications arise since the pairing in NS cores, like for the superfluid He$^3$, can have the anisotropic pattern. For instance, the  neutrons in NS core are thought to pair in the ${}^3$P$_2-{}^3$F$_2$ channel. Even in the superfluid case, the flows of the  normal part of the fluid are subject to the dissipative processes. In principle the relevant transport coefficients can be calculated based on the the kinetic theory for the anisotropic superfluids e.g.~\cite{Vollhardt1990}, although it is severely more complicated task than the non-superfluid problem.

Superfluidity modifies the transport coefficients on NS core matter in several ways, e.g.~\cite{Schmitt2018}, which we outline briefly.
The relevant excitations of the normal component in the superfluid matter are not the usual quasiparticles, but rather the the Bogoliubov quasiparticles, which spectra have a gap at the Fermi surface. In addition, the number of such quasiparticles does not necessarily conserve in collisions. The main effect of these modifications results in the exponential suppression of the scattering rates for the processes in which the superfluid species participate. Moreover, if the paired species are charged (e.g., protons) this modifies the properties of the electromagnetic screening. This affects the scattering of all charged species, for instance, leptons \cite{Schmitt2018,Shternin2020PhRvD}.
At low temperatures, another excitations, namely the superfluid phonons, can contribute to the transport properties of NS core matter. This contribution is reviewed in, e.g., \cite{Manuel2021Univ}.

Even more complicated is the case of the magnetized superfluid/superconducting matter. Under a certain conditions, the matter can be in the type II superconducting state where the magnetic field is confined in  the topological defects called Abrikosov vortices. This introduces another mesoscopic scale on the stage and additional terms in dissipative hydrodynamics, e.g. \cite{Dommes2021arXiv}. 

A detailed discussion of all these effects is \new{left for future studies}.

\begin{acknowledgements}
The authors are grateful to Dima Yakovlev for discussions.
\end{acknowledgements}

\appendix

\section{Linearized kinetic equations}\label{app}

Let us outline derivation of eq.~(\ref{eq:Iab_lm}) from eq.~(\ref{eq:Boltz_lin}). Functions $\Phi(\bm{p}_i)$ are expanded in spherical harmonics according to eq.~(\ref{eq:phi_spherical}). These spherical harmonics can be rotated to the body frame which has $\bm{p}_1$ as a polar axis via
\begin{equation}\label{eq:Ylm_transform}
    Y_{\ell m}(\hat{\bm{p}}_i)=\sum_{m'} \left({\cal D}^\ell_{mm'}(\hat{\bm{p}}_1,\varphi)\right)^* Y_{\ell m'}(\hat{\bm{p}}_i\hat{\bm{p}}_1,\varphi_i),
\end{equation}
where $\varphi$ is the azimuthal angle of the body frame system with respect to the laboratory system
$\varphi_i$ is the azimuthal angle of $\bm{p}_i$ in the body-frame coordinate system. Isotropy of the collision probability in eq.~(\ref{eq:Iab_lm}) means, in particular, that it does not depend on $\varphi$ however angular integration contains integration over $\varphi$.
Therefore only $m'=0$ term in eq.~(\ref{eq:Ylm_transform}) survives and one has
\begin{equation}
    Y_{\ell m}(\hat{\bm{p}}_i)\to Y_{\ell m}(\hat{\bm{p}}_1){\cal P}_{\ell }(\hat{\bm{p}}_i\hat{\bm{p}}_1)
\end{equation}
resulting in eq.~(\ref{eq:Iab_lm}).

Complete system of the linearized transport equations for the thermal conductivity or the shear viscosity of multicomponent Fermi-liquid is obtained by substituting eqs.~(\ref{eq:Phia_def}) and (\ref{eq:Ia_Iab_lm}) to eq.~(\ref{eq:kineq_irred}) and placing all quasiparticles on respective Fermi surfaces using energy-angular decomposition. It is convenient to substitute $x_2\to -x_2$ in the collision integral and exchange $x_2\leftrightarrow x_{2'}$ in terms containing  $\widetilde{\Psi}_b(x_{2'})$. Finally, one obtains
\begin{eqnarray}
-f_F'(x_1)&&\widetilde{X}_{k}(x_1)=\frac{T^2 m_a^{*2}}{4\pi^4 p_{\mathrm{F}a}^2 \widetilde{D}_k(p_{\mathrm{F}a})}
 \int \mathrm{d} x_{1'} \mathrm{d} x_{2} \mathrm{d} x_{2'} \nonumber\\
 &\times&f_F(x_1)f_F(-x_{1'})f_F(-x_{2})f_F(-x_{2'})\nonumber\\
 &\times&\delta(x_1-x_2-x_{1'}-x_{2'})\nonumber\\
&\times& \left(\widetilde{D}_k(p_{\mathrm{F}a})\widetilde{\Psi}^{k}_a(x_1) \lambda_a^{(k,\mathrm{eff})} \sum_{b} m_b^{*2}\left\langle{\cal Q}_{ab}\right\rangle\right.\nonumber\\
&-& \left.\widetilde{D}_k(p_{\mathrm{F}a})\widetilde{\Psi}^{k}_a(x_{1'}) \lambda_a^{(k,\mathrm{eff})} \sum_{b} m_b^{*2}\left\langle{\cal Q}_{ab} {\cal P}_\ell(\hat{\bm{p}}_1\hat{\bm{p}}_{1'})\right\rangle\right.\nonumber\\
&&- \left.\sum_{b} m_b^{*2}\widetilde{D}_k(p_{\mathrm{F}b})\widetilde{\Psi}^{k}_b(x_{2}) \lambda_b^{(k,\mathrm{eff})} \right.\nonumber\\
&&\times\left.\left\langle{\cal Q}_{ab} \left({\cal P}_\ell(\hat{\bm{p}}_1\hat{\bm{p}}_{2'}) -(-1)^{\xi_k} {\cal P}_\ell(\hat{\bm{p}}_1\hat{\bm{p}}_{2})\right)\right\rangle
\right)\nonumber\\
&&+if'_F(x_1) \frac{\omega_{\mathrm{BF}a}}{v_{\mathrm{F}a}}\lambda_a^{(k,\mathrm{eff})} \widetilde{\Psi}^{k}_a(x_1),\label{eq:linkineq_fullgen}
\end{eqnarray}
where the symmetry of the transport problem due to $x$ inversion is used (hence appearance of the phase $(-1)^\xi_k$) and angular \new{integrations} are encapsulated in the angular brackets according to eq.~(\ref{eq:ang_av}).
This equation for $\ell=1,2$ contains the following angular averages that should be provided by the microscopic theory
\begin{eqnarray}\label{eq:angav_general}
    &&\left\langle {\cal Q}_{ab}\right\rangle,\quad \left\langle {\cal Q}_{ab} P_1(\hat{\bm{k}}_1 \hat{\bm{k}}_{1'})\right\rangle,\quad  \left\langle {\cal Q}_{ab} P_2(\hat{\bm{k}}_1 \hat{\bm{k}}_{1'})\right\rangle,\nonumber\\ &&  \left\langle {\cal Q}_{ab} (P_1(\hat{\bm{k}}_1 \hat{\bm{k}}_{2})\pm P_1(\hat{\bm{k}}_1 \hat{\bm{k}}_{2'}))\right\rangle,\nonumber \\
    &&
    \left\langle {\cal Q}_{ab} (P_2(\hat{\bm{k}}_1 \hat{\bm{k}}_{2'})-P_2(\hat{\bm{k}}_1 \hat{\bm{k}}_{2}))\right\rangle.
\end{eqnarray}
Notice that the \new{averages} $\langle {\cal Q}_{aa}\rangle$ (i.e. without angular factors) do not enter the simplest variational solution, but is necessary for solution of (\ref{eq:linkineq_fullgen}). If the squared matrix element ${\cal Q}_{ab}$ does not depend on $w=x_{1'}-x_1$, two of three integrations over $x_i$ can be performed analytically and eq.~(\ref{eq:linkineq_fullgen}) reduces to much simpler integral equation given in \cite{Anderson1987} that allows for exact solution. In principle, one can \new{integrate} over $x_2$ and $x_{2'}$ in the terms containing $\widetilde{\Psi}_a$ in eq.~(\ref{eq:linkineq_fullgen}), but in general two energy integrations remain in terms containing $\Psi_b(x_2)$.

After the solution of the system of kinetic equations (\ref{eq:linkineq_fullgen}), the complex thermal conductivity and shear viscosity coefficients are given by (cf. eqs.~(\ref{eq:kappa_var})--(\ref{eq:eta_var}))
\begin{eqnarray}
    \widetilde{\kappa}_a&=&\frac{\pi^2 n_a T}{3p_{\mathrm{F}a}}\lambda_{\mathrm{eff}}^{\kappa,a} \times \frac{3}{\pi^2}\int\mathrm{d}x (-f'_F(x)) x^2 \widetilde{\Psi}_a^\kappa(x),\label{eq:kappa_psi}\\
    \widetilde{\eta}_a&=&\frac{n_ap_{\mathrm{F}a}}{5}  \lambda_{\mathrm{eff}}^{\eta,a}\times \int\mathrm{d}x (-f'_F(x)) \widetilde{\Psi}_a^\eta(x).\label{eq:eta_psi}
\end{eqnarray}

For the diffusion problem the general transport equation is even more cumbersome, since it contains the summation over $k$. Moreover when solving these equations one needs to carefully account for the conditions of fit that get rid of the kernel solutions \cite{Ferziger1972Book}. We do not write these equations here; the formal equations can be easily written via the methods described in sec.~\ref{sec:LandauFL}. The respective equations for the normal Fermi-liquid case  where $w_{ab}$ does not depend on $\omega$ can be found in \cite{Anderson1987} (without the magnetic field).

\section{Practical formulae for nuclear NS core matter}\label{sec:app:pract}
In this appendix we give practical expressions for calculating transport coefficients considered in the paper, namely, thermal conductivity $\hat{\kappa}$, shear viscosity $\hat{\eta}$ tensors, and momentum transfer rates $J_{ab}$, $a\neq b$, in the nucleonic ($npe\mu$) NS cores.

Independent components of the thermal conductivity and shear viscosity tensors are determined by the complex functions $\widetilde{\kappa}(B)$ and $\widetilde{\eta}(B)$, respectively, as 
\begin{subequations}
\begin{eqnarray}
\kappa_\parallel&=& \widetilde{\kappa}(B=0),\\
\kappa_\perp &=& \mathrm{Re}\,\widetilde{\kappa}(B),\\ \kappa_{\wedge}&=& - \mathrm{Im}\,\widetilde{\kappa}(B),
\end{eqnarray}
\end{subequations}
\begin{subequations}
\begin{eqnarray}
\eta_0&=&\widetilde{\eta}(B=0),\\ \eta_1&=&\mathrm{Re}\,\widetilde{\eta}(2B),\\ \eta_3&=&-\mathrm{Im}\,\widetilde{\eta}(2B),\\
\eta_2&=&\mathrm{Re}\,\widetilde{\eta}(B), \\
\eta_4&=&-\mathrm{Im}\,\widetilde{\eta}(B).
\end{eqnarray}
\end{subequations}

Complex functions $\widetilde{\kappa}$, $\widetilde{\eta}$ are sums of the partial contributions, $a=n,\,p,\,e,\,\mu$
\begin{equation}
    \widetilde{\kappa}=\sum_a \widetilde{\kappa}_a,
\end{equation}
\begin{equation}
    \widetilde{\eta}=\sum_a \widetilde{\eta}_a,
\end{equation}
where
\begin{eqnarray}
    \widetilde{\kappa}_a&=&
    5.66\times10^{22} \left(\frac{n_a}{n_0}\right)^{2/3} T_8 \frac{\widetilde{\lambda}^\kappa_a}{10^{-6}~\mathrm{cm}}
    ~\mathrm{erg}~\mathrm{cm}^{-1}\mathrm{s}^{-1}\mathrm{K}^{-1},\nonumber\\&&\label{app:kappa_var}\\
    \widetilde{\eta}_a&=&
    5.66\times10^{17} \left(\frac{n_a}{n_0}\right)^{4/3}  \frac{\widetilde{\lambda}^\eta_a}{10^{-6}~\mathrm{cm}} ~\mathrm{g}~\mathrm{cm}^{-1}\mathrm{s}^{-1} ,\label{app:eta_var}
\end{eqnarray}
$n_0=0.16$~fm$^{-3}$, $T_8=T/(10^8~\mathrm{K})$ and $\widetilde{\lambda}^k_a$ with $k=\kappa,\,\eta$ are generalized complex effective mean free paths of quasiparticles.

Effective mean free paths are found from the solution of a system of linear equations ($k=\kappa,\,\eta$)
\begin{equation}\label{app:kinvar_system}
    1=\left(\Lambda^k_{a}+i\frac{\omega_{\mathrm{BF}a}}{v_{\mathrm{F}a}}\right)\widetilde{\lambda}^k_a+ \sum_{b\neq a}{\Lambda'}^{k}_{ab}\widetilde{\lambda}^k_b,
\end{equation}
where magnetic field enters through the inverse gyroradii
\begin{equation}\label{app:omegaB}
    \frac{\omega_{\mathrm{BF}a}}{v_{\mathrm{F}a}}=\frac{q_a B}{p_{\mathrm{F}a} c} = 
    \new{9.04\times 10^5\ }
    Z_a B_{12} \left(\frac{n_a}{n_0}\right)^{-1/3}~\mathrm{cm}^{-1},
\end{equation}
where $B_{12}=B/(10^{12}~\mathrm{G})$ and charge numbers $Z_a=0,\ \pm 1$ in $npe\mu$ matter.

Coefficients $\Lambda^k_a$ in eq.~(\ref{app:kinvar_system}) are
\begin{equation}
    \Lambda^k_a=\sum_{b} \Lambda^k_{ab}+{\Lambda'}^k_{aa}.
\end{equation}

Transport matrix elements $\Lambda^{k}_{ab}$, ${\Lambda'}^{k}_{ab}$ in general contain contributions from all interaction channels between the particle species $a$ and $b$. In the model considered here, leptons ($\ell=e,\,\mu$) participate in electromagnetic interactions, neutrons in strong interactions, and protons in both. In latter case one needs to sum both contributions. Below we give the practical expressions for the transport matrix elements  and related momentum transfer rates $J_{ab}$ for electromagnetic and strong channels separately. In this Appendix, $x_a\equiv n_a/n_B$, where $n_B$ is the total baryon density.

\subsection{Electromagnetic channel}\label{app:em}
In the electromagnetic sector, $a,b=e,\,\mu,\, p$. Expressions below are based on the results described in detail in sec.~\ref{sec:em}.

Practical expressions for the screening wavenumbers  are
\begin{subequations}
\begin{equation}\label{app:ql}
q_{l}= 0.27~\left(\frac{n_B}{n_0}\right)^{1/6}\sqrt{\sum_a Z_a^2 x_a^{1/3}\frac{m_a^*}{m_N}}~\mathrm{fm}^{-1},
\end{equation}
\begin{equation}\label{app:ql}
q_{t}= 0.16 ~\left(\frac{n_B}{n_0}\right)^{1/3} \sqrt{\sum_a Z_a^2 x_a^{2/3}}~\mathrm{fm}^{-1}.
\end{equation}
\end{subequations}
The effective mass for  leptons is $m_\ell^*=\sqrt{m_\ell^2+p_{\mathrm{F}\ell}^2/c^2}$, while for protons one needs to use the Landau effective mass on the Fermi surface here. 

In what follows we normalize screening momenta over fm$^{-1}$: $\overline{q}_l=q_l/(\mathrm{fm}^{-1})$, $\overline{q}_t=q_t/( \mathrm{fm}^{-1}).$

We need the following integrals
\begin{subequations}
\begin{equation}
    I_l^{(0)}=\frac{1}{2\overline{q}_l^3}\left[\arctan \frac{q_m}{q_l}+\frac{q_m q_l}{q_m^2+q_l^2}\right],
\end{equation}
\begin{equation}
    I_l^{(2)}=\frac{1}{2\overline{q}_l}\left[\arctan \frac{q_m}{q_l}-\frac{q_m q_l}{q_m^2+q_l^2}\right],
\end{equation}
\begin{equation}
    I_{tl}^{(2)}=I_{l}^{(2)}+\overline{q}_l^2I_l^{(0)}=\frac{1}{\overline{q}_l} \arctan \frac{q_m}{q_l},
\end{equation}
\end{subequations}
where $ q_m=2\min (p_{\mathrm{F}a},p_{\mathrm{F}b})$.

Now the transport matrix elements for the thermal conductivity are
\begin{equation}
    \left[\Lambda^\kappa_{ab}\right]^{\mathrm{em}}=\Lambda^{\kappa,t}_{ab}+\Lambda^{\kappa,l}_{ab},
\end{equation}
where the transverse contribution is
\new{
\begin{equation}
    \Lambda^{\kappa,t}_{ab}=
    6.10\times 10^4\ 
    T_8 x_b^{2/3}\left(\frac{n_B}{n_0}\right)^{2/3}\overline{q}_t^{-2}~\mathrm{cm}^{-1} ,\label{app:Lambda_kappa_t}
\end{equation}
}
and the longitudinal contribution is 
\begin{eqnarray}
    \Lambda^{\kappa,l}_{ab}&=&
    1.86\times 10^3\ 
    T_8^2 \left(\frac{m_a^*m_b^*}{m_N^2}\right)^2 \left(\frac{n_B}{n_0}\right)^{-2/3}\nonumber\\&\times& x_a^{-2/3}  I_l^{(0)}~\mathrm{cm}^{-1}.\label{app:Lambda_kappa_l}
\end{eqnarray}

Non-diagonal coefficients in eq.~(\ref{app:kinvar_system}) for thermal conductivity are
\begin{equation} \label{app:Lambda_kappa_tl}
    \left[{\Lambda'}^{\kappa}_{ab}\right]^{\mathrm{em}}=
    \new{3.65
    \times 10^3} \frac{T_8^{5/3}}{\overline{q}_l^2\overline{q}_t^{2/3}} \frac{m_a^*m_b^*}{m_N^2} \frac{x_b^{1/3}}{x_a^{1/3}}~\mathrm{cm}^{-1}.
\end{equation}

Similar expressions for shear viscosity problem read
\begin{equation}
    \left[\Lambda^\eta_{ab}\right]^{\mathrm{em}}=\Lambda^{\eta,t}_{ab}+\Lambda^{\eta,l}_{ab},
\end{equation}
where 
\begin{equation}
    \Lambda^{\eta,t}_{ab}
    =200\ T_8^{5/3} \frac{x_b^{2/3}}{x_a^{2/3}}\overline{q}_t^{-2/3}~\mathrm{cm}^{-1}, \label{app:Lambda_eta_t}
\end{equation}
and
\begin{eqnarray}
    \Lambda^{\eta,l}_{ab}&=&
    \new{822\ }
    T_8^2 \left(\frac{m_a^*m_b^*}{m_N^2}\right)^2\left(\frac{n_B}{n_0}\right)^{-4/3}  x_a^{-4/3}  I_l^{(2)}~\mathrm{cm}^{-1}.\nonumber\\\label{app:Lambda_eta_l}
\end{eqnarray}
Non-diagonal elements are
\begin{eqnarray}\label{app:Lambda_eta_tl}
    \left[{\Lambda'}^{\eta}_{ab}\right]^{\mathrm{em}}&=& 
    \new{102\ }
    T_8^2  \frac{m_a^*m_b^*}{m_N^2}\left(\frac{n_B}{n_0}\right)^{-2/3} \frac{x_b^{1/3}}{x_a}I_{tl}^{(2)}~\mathrm{cm}^{-1}.\nonumber\\
\end{eqnarray}

The momentum transfer rates $J_{ab}$ appearing in the diffusion problems are calculated in a similar way:
\begin{equation}\label{app:Jab_decomp}
    \left[J_{ab}\right]^{\mathrm{em}}=J_{ab}^l+J_{ab}^t,
\end{equation}
where
\begin{subequations}
\begin{eqnarray}\label{app:Jab_t}
    J_{ab}^t&=&
    \new{1.89
    \times 10^{26}}\ T_8^{5/3}\left(\frac{n_B}{n_0}\right)^{4/3}\frac{ x_a^{2/3}x_b^{2/3}}{\overline{q}_t^{2/3}} ~\mathrm{g}~\mathrm{cm}^{-3}~\mathrm{s}^{-1},\nonumber\\
    \\
    J_{ab}^l&=&   7.78
    \times10^{26}\ T_8^2 \left(\frac{m_a^*m_b^*}{m_N^2}\right)^2I_{l}^{(2)}~\mathrm{g}~\mathrm{cm}^{-3}~\mathrm{s}^{-1}.\nonumber\\
    \label{app:Jab_l}
\end{eqnarray}
\end{subequations}

\subsection{Strong channel}
The strong interaction operates in the \new{subsystem of protons and neutrons.}
We define
\begin{equation}
    C_\Lambda=1.65\times 10^{6}\ 
    T_8^2 \left(\frac{n_B}{n_0}\right)^{-1/3}~\mathrm{cm}^{-1}.
\end{equation}

Then, for the thermal conductivity and $a=n,p$
\begin{subequations}\label{app:Lambdakappa_Bai}
\begin{eqnarray}
    \Lambda^\kappa_{aa}&+&{\Lambda'}^\kappa_{aa}=\frac{1}{5}C_\Lambda \left(\frac{m_a^*}{m_N}\right)^4  x_a^{-1/3} \frac{S_{aa2}}{\mathrm{mb}},\\
    \Lambda^\kappa_{np} &=& \frac{1}{5}C_\Lambda \left(\frac{m_p^*m_n^*}{m_N^2}\right)^2  x_n^{-1/3} \frac{S_{np1}+S_{np2}}{\mathrm{mb}},\\
    \Lambda^\kappa_{pn}&=& \frac{1}{5}C_\Lambda   \left(\frac{m_p^*m_n^*}{m_N^2}\right)^2  \frac{x_n^{1/3}}{x_p^{2/3}} \left(\frac{S_{np1}}{\mathrm{mb}}+\frac{x_n^{2/3}}{x_p^{2/3}}\frac{S_{np2}}{\mathrm{mb}}\right),\nonumber\\
    &&\\
    {\Lambda'}^\kappa_{np}&=&\frac{x_p^{2/3}}{x_n^{2/3}}{\Lambda'}^\kappa_{pn}=\frac{1}{5}C_\Lambda \left(\frac{m_p^*m_n^*}{m_N^2}\right)^2  x_p^{-1/3} \frac{S'_{np2}}{\mathrm{mb}}.\nonumber\\
\end{eqnarray}
\end{subequations}
For the shear viscosity
\begin{subequations}
\begin{eqnarray}
    \Lambda^\eta_{aa}&+&{\Lambda'}^\eta_{aa}=C_\Lambda \left(\frac{m_a^*}{m_N}\right)^4  x_a^{-1/3} \frac{S_{aa4}}{\mathrm{mb}},\\
    \Lambda^\eta_{np}&=&C_\Lambda \left(\frac{m_p^*m_n^*}{m_N^2}\right)^2  x_n^{-1/3} \frac{S_{np2}-S_{np4}}{\mathrm{mb}},\\
    \Lambda^\eta_{pn}&=&C_\Lambda \left(\frac{m_p^*m_n^*}{m_N^2}\right)^2  x_p^{-1/3} \left(\frac{x_n}{x_p}\frac{S_{np2}}{\mathrm{mb}}-\frac{x_n^{5/3}}{x_p^{5/3}}\frac{S_{np4}}{\mathrm{mb}}\right),\nonumber\\
    &&\\
    {\Lambda'}^\eta_{np}&=&\frac{x_p^{4/3}}{x_n^{4/3}}{\Lambda'}^\eta_{pn}=C_\Lambda \left(\frac{m_p^*m_n^*}{m_N^2}\right)^2  x_p^{-1/3} \frac{S'_{np4}}{\mathrm{mb}}.
\end{eqnarray}
\end{subequations}
The momentum transfer rates (here only $J_{np}$ coefficient exists)
\begin{equation}
    J_{np}=1.56\times 10^{30}\ T_8^2 \left(\frac{m_p^*m_n^*}{m_N^2}\right)^2\left(\frac{n_B}{n_0}\right) x_n \frac{S_{np2}}{\mathrm{mb}}.
\end{equation}

If the effective masses of nucleons are not known, we suggest to use some typical effective mass, e.g.,  $m_n^*=m_p^*=0.8 m_N$. 

We provide two types of fits for the transport cross-sections $S_\alpha$ in expressions above. First ones are for the toy model where particle fractions are given in eq.~(\ref{eq:xpp_ddo}). These fits give $S_\alpha(n_B)$. The second fits are valid for any particle fractions and give $S_\alpha(p_{\mathrm{F}n},\,p_{\mathrm{F}p})$. Both fits are based on the in-vacuum interaction.

\begin{table*}
\caption{Fit parameters for the eq.~(\ref{app:Sfit-toy}).}
\label{tab:Sfit-toy_params}
\begin{tabular*}{\textwidth}{@{\extracolsep{\fill}}llllllcc@{}}
\hline
cross-sec. & $a$ & $p$ & $b$ & $q$ & $c$ & rrms [\%] & max error [\%]
\\
\hline
$S_{nn2}$  & 0.912    & 1.23     & 0.00631  & 2.248   & 4.045    & 0.05 & 0.1 \\
$S_{pp2}$  & 25.6     & 1.23     & 0.404    & 0.932   & 0.000    & 0.7  & 3.1 \\
$S_{nn4}$  & 0.0151   & 2.42     & $-$0.00334 & 1.26    & 0.409    & 0.2  & 0.5 \\
$S_{pp4}$  & 3.04     & 1.12     & $-$2.00    & $-$0.336  & 1.10     & 1.1  & 2.2 \\
$S_{np1}$  & 7.03     & 0.518    & 0.0789   & 1.43    & 0.000    & 0.3  & 0.8 \\
$S_{np2}$  & 2.879    & 0.03806  & 0.02942  & 1.115   & $-$2.674   & 1.0  & 2.4 \\
$S_{np4}$  & 0.03562  & 0.0514   & 0.001068 & 1.802   & $-$0.02000 & 1.8  & 3.9 \\
$S'_{np2}$ & $-$0.1833  & 0.5109   & 0.001413 & 2.139   & $-$0.4024  & 0.2  & 0.4 \\
$S'_{np4}$ & $-$0.4658  & $-$0.03099 & 0.06761  & 0.2137  & 0.3855   & 1.3  & 4.0 \\
\hline
\end{tabular*}
\end{table*}

\begin{enumerate}

\item Fits for the toy model eq.~(\ref{eq:xpp_ddo}).

In this case all nine cross-sections $S$ could be fitted by the formula
\begin{equation}\label{app:Sfit-toy}
    S_\alpha(n_B) = a\left( \frac{n_0}{n_B} \right)^p + b\left( \frac{n_B}{n_0} \right)^q + c,
\end{equation}
where the appropriate values of fitting parameters $a$, $b$, $c$, $p$, and $q$ are listed in tab.~\ref{tab:Sfit-toy_params}. The fitting domain is $n_B = 0.08...1\,$fm$^{-3}$.

\item Fits for the general compositions.

In expressions below $p_{\mathrm{F}n}$ and $p_{\mathrm{F}p}$ are measured in fm$^{-1}$ and $P_{\rm min} = |p_{\mathrm{F}n}-p_{\mathrm{F}p}|$, $P_{\rm max}=p_{\mathrm{F}n}+p_{\mathrm{F}p}$. All functions are fitted on a grid $0.1<p_{\mathrm{F}n,p}<3.5$~fm$^{-3}$.

\begin{equation}\label{app:Sn2fit}
        S_{aa2} =7.82 \frac{1+0.576\bigl|p_{\mathrm{F}a}-1.05\bigr|^{1.47}}{\left( 1-0.198 p_{\mathrm{F}a} \right)\left(0.0349+p_{\mathrm{F}a}^{2.14}\right)
    }\,\mathrm{mb},
\end{equation}
rrms\footnote{
rrms = root mean square relative error.
}$=1.2\%$, max error$=2.9\%$ at $p_{\mathrm{F}a}=0.2$fm$^{-1}$.
\begin{eqnarray}\label{app:Sp1fit}
        p_{\mathrm{F}n}S_{np1} &=& 79.5(P_{\rm max} - P_{\rm min})\nonumber\\
        &\times&\frac{1-0.164P_{\rm max}+0.0258 P_{\rm max}^2}{P_{\rm max}^{1.52}} \nonumber\\
        &\times&
    \frac{\sqrt{1+0.637P_{\rm min}^4}}{1+1.35P_{\rm min}^2}\mathrm{mb},
\end{eqnarray}
where rrms$=4.2\%$, max error$=12\%$ at $p_{\mathrm{F}n} \approx 0.7\,\text{fm}^{-1}$, $p_{\mathrm{F}p} = 0.7\,\text{fm}^{-1}$.

\begin{eqnarray}\label{app:Sp2fit}
    p_{\mathrm{F}n}^3 S_{np2} &=& 5.158\ \frac{(P_{\rm max} - P_{\rm min})^{3.342} }{P_{\rm max}^{1.618}\left( 1+0.3691 P_{\rm max}^{1.985} \right)}\nonumber\\
    &\times&\left( 1+0.2072(P_{\rm max} - P_{\rm min}-1.505)^2 \right)\nonumber\\
    &\times& \left( 1+0.2279P_{\rm min}-0.6076P_{\rm min}^2\right.\nonumber\\
    &&\ +\ \left.0.4962P_{\rm min}^3-0.07539P_{\rm min}^4  \right) \mathrm{mb},\nonumber\\
\end{eqnarray}
 rrms$=2.6\%$, max error$=14\%$ at $p_{\mathrm{F}n} \approx 0.1\,\text{fm}^{-1}$, $p_{\mathrm{F}p} = 0.1\,\text{fm}^{-1}$.

\begin{equation}\label{app:Sn4fit}
    S_{aa4} =  2.84\frac{1+1.65(p_{\mathrm{F}a}-0.903)^2}{(1+1.70p_{\mathrm{F}a})\left((p_{\mathrm{F}a}+0.165)^2+0.103\right)}\,\mathrm{mb},
\end{equation}
rrms$=0.95\%$, max error$=2.1\%$ at $p_{\mathrm{F}a}=0.6$fm$^{-1}$.

\begin{eqnarray}
       p_{\mathrm{F}n}^5 S_{np4} &=& 0.621 \frac{(P_{\rm max} - P_{\rm min})^{5.33}}{P_{\rm max}^{1.64}}\nonumber\\
       &\times&
       \frac{\ 1+0.274(P_{\rm max} - P_{\rm min}-1.62)^2}{\left( 1+0.262P_{\rm max}^{2.12}\right)}\nonumber\\
    &\times& \left( 1+0.122P_{\rm min}-0.616K_{\rm min}^2\right.\nonumber\\
    &&\ +\ 0.538P_{\rm min}^3-\left.0.0860P_{\rm min}^4  \right)\,\mathrm{mb},\label{app:Sp4fit}
\end{eqnarray}
rrms$=4.0\%$, max error$=19\%$ at $p_{\mathrm{F}n} \approx 0.1\,\text{fm}^{-1}$, $p_{\mathrm{F}p} = 0.1\,\text{fm}^{-1}$.

\begin{multline}\label{app:S1_p2fit}
    -p_{\mathrm{F}n}^3 S'_{np2} = 1.858 \frac{(P_{\rm max}-P_{\rm min})^{3.250}}{P_{\rm max}^{2.415}}\\
    \times \left( -1+7.383 P_{\rm max}-1.695P_{\rm max}^2 \right.\\
    +0.2441P_{\rm max}^3-0.01657P_{\rm max}^4 + 9.700P_{\rm min} \\
    - 7.698P_{\rm max}P_{\rm min} + 0.9015P_{\rm max}^2P_{\rm min} - 0.04207 P_{\rm max}^3P_{\rm min} \\
    - 9.511P_{\rm min}^2 + 3.939 P_{\rm max}P_{\rm min}^2 - 0.3213P_{\rm max}^2P_{\rm min}^2 \\
    \left.+ 2.822P_{\rm min}^3 - 0.4262P_{\rm max}P_{\rm min}^3 - 0.2448P_{\rm min}^4\right)
     \mathrm{mb},
\end{multline}
rrms$=2.7\%$, max error$=18\%$ at $k_n \approx 1.5\,\text{fm}^{-1}$, $k_p = 0.2\,\text{fm}^{-1}$.
\textbf{\textbf{\textbf{}}}
\begin{subequations}\label{app:S1_p4fit}
\begin{equation}\label{app:S1_p4fit-1}
    -p_{\mathrm{F}n}^5 S'_{np4} = p_{\mathrm{F}n}^5 S_{np4}-\frac{p_{\mathrm{F}n}^2+p_{\mathrm{F}p}^2}{2} p_{\mathrm{F}n}^3 S_{np2} + U'_{np4},
\end{equation}
where
\begin{multline}\label{app:S1_p4fit-2}
    U'_{np4} = 0.0451 \frac{\left(P_{\max }-P_{\min }\right)^{3.27}}{P_{\max }^{0.448}} \left( 0.403 P_{\max }^4
   P_{\min }\right. \\
    - 0.918 P_{\max }^3 P_{\min }^2-5.54 P_{\max }^3 P_{\min }+1.78 P_{\max }^2 P_{\min }^3 \\
    + 8.28 P_{\max }^2 P_{\min }^2+28.7 P_{\max }^2 P_{\min }-0.331P_{\max } P_{\min }^4 \\
    - 16.2 P_{\max } P_{\min }^3-15.7 P_{\max } P_{\min }^2-74.6P_{\max } P_{\min } \\
    - 0.00111 P_{\max }^5-0.118 P_{\max }^4+1.77P_{\max }^3 \\
    - 8.17 P_{\max }^2+17.2 P_{\max }-0.118P_{\min }^5+1.77P_{\min }^4 \\
    + \left. 41.1P_{\min }^3-21.0 P_{\min }^2 + 94.6 P_{\min }+1\right)\; \mathrm{mb}.
\end{multline}
\end{subequations}
For  $U'_{np4}$,
 rrms=$0.5\%$ and max error $=1.9\%$. Notice that in the region where $S'_{np4}$ changes sign and is therefore small, this fit can give large relative error for $S'_{np4}$. In this case, however, precise calculation of $S'_{np4}$ is not necessary as being small it drops  out of the equations.
\end{enumerate}

{
\section{Electrical conductivity of the NS core matter}\label{sec:econd}
Electrical conductivity in a multicomponent plasma is an aspect of the general diffusion process. In this appendix we discuss the relation between the electrical conductivity and momentum transfer rates $J_{ab}$ and give practical expressions for the $npe\mu$ matter composition. Below we do not consider thermal diffusion.

In the discussion in the main part of the manuscript, the diffusion in NS cores was described via the generalized Ohm law (\ref{eq:Ohm}) which we repeat here for convenience
\begin{eqnarray}
    &n_a&\left(\bm{d}_{(a)}+\frac{h_a}{hn}\left[\bm{J}\times\bm{B} \right]\right)=\sum_bJ_{ab}(\bm{w}_b-\bm{w}_a)\nonumber\\
    &&+n_aq_a[\bm{w}_a\times\bm{B}].\label{eq:Ohm_app}
\end{eqnarray}

According to eqs.~(\ref{eq:entrprod_lin}), (\ref{eq:phivar_diffusion}), and (\ref{eq:Ohm}), the entropy production rate in the diffusion process is given by multiplication of eq.~(\ref{eq:Ohm_app}) by $\bm{w}_a$ and summing over species
\begin{eqnarray}
    T\varsigma&=&-\sum_a n_a\bm{w_a}\left(\bm{d}_a+\frac{h_a}{hn} \left[\bm{J}\times \bm{B}\right]\right)\nonumber\\
    &=&
    \frac{1}{2}\sum_{ab}J_{ab} (\bm{w}_b-\bm{w}_a)^2.\label{eq:ent_gen_diff_1}
\end{eqnarray}
Notice that the eq.~(\ref{eq:Ohm_app}) is valid in general frame (without imposing conditions of fit) and the second line in eq.~(\ref{eq:ent_gen_diff_1}) shows that the entropy generation rate is frame-independent.
Assume now that the only source for diffusion is the electric field, so that eq.~(\ref{eq:da_three}) becomes
\begin{equation}\label{eq:da_E}
    \bm{d}_{(a)}=-q_a\bm{E}.
\end{equation}
Then the first line of eq.~(\ref{eq:ent_gen_diff_1}) reduces to 
\begin{eqnarray}\label{eq:electric_heat}
    T\varsigma&=&\bm{J}\bm{E}-\bm{u}_g\left[\bm{J}\times \bm{B}\right] \equiv \bm{J}\bm{E}',
\end{eqnarray}
where 
\begin{equation}\label{eq:u_g}
    \bm{u}_g=\sum_a \mathfrak{X}_a \bm{w}_a,\quad  \mathfrak{X}_a=\frac{n_a h_a}{nh},
\end{equation}
may be called the relativistic `center of mass' velocity of a liquid, and
\begin{equation}
    \bm{E}'=\bm{E}+\left[\bm{u}_g\times \bm{B}\right]
\end{equation}
is the electric field in the center of mass frame, in which $\bm{u}_g=0$. 
The terms $\mathfrak{X}_a$ in eq.~(\ref{eq:u_g}) are the  `mass' fractions which in strongly degenerate matter ($h_a\approx \mu_a$) are
\begin{equation}
    \mathfrak{X}_a\approx \frac{\mu_a n_a}{\sum_b \mu_b n_b}.
\end{equation}
The center-of-mass frame is defined by the condition
\begin{equation}\label{eq:cm_frame}
    \bm{u}_g=\sum_a \mathfrak{X}_a\bm{w}_a=0.
\end{equation}
Since the center-of-mass frame and hence $\bm{E}'$ is unique, the eq.~(\ref{eq:electric_heat}) is frame-independent in accordance with eq.~(\ref{eq:ent_gen_diff_1}).
Notice that only\footnote{
{Of course, this is strictly true only if  $\bm{d}_{(a)}=-q_a\bm{E}$. More general case is considered, e.g., in~\cite{Gusakov2017PhRvD}, where $\bm{d}_{(a)}$ includes additional diffusion terms $\propto \bm{\nabla}\mu_a$. 
}
}  in the center-of-mass frame, the heat release is given by the standard $\bm{J}\bm{E}$ expression.

When the diffusion driving forces are given by eq.~(\ref{eq:da_E}), all diffusion velocities and any their linear combination (i.e. $\bm{J}$ and $\bm{u}_g$) will be linear in components of $E$. Namely,
\begin{equation}
    \bm{J}=\hat{\sigma}\bm{E},
\end{equation}
where $\hat{\sigma}$ is the electrical conductivity tensor. This tensor, however, is not frame-independent since the electric field (at $\bm{B}\neq 0$) depends on the frame. Indeed, let us denote the electrical conductivity in the center-of-mass frame as $\hat{\sigma}'$ and let
\begin{equation}
    \bm{u}_g=\hat{\mathfrak{U}}\bm{E}.
\end{equation}
Then
\begin{equation}
    \bm{J}=\hat{\sigma}'\bm{E}'=\hat{\sigma}'(\bm{E}+[\hat{\mathfrak{U}}\bm{E}\times \bm{B}])=\hat{\sigma}\bm{E},
\end{equation}
where
\begin{equation}
    \sigma_{ij}=\sigma'_{im}(\delta_{mj}+\epsilon_{mkl}  B_k\mathfrak{U}_{lj}).
\end{equation}

The electrical conductivity tensor can be found by inversion of the eq.~(\ref{eq:Ohm_app}) in any frame, however according to eq.~(\ref{eq:electric_heat}) it physically motivated to do this in the center-of-mass frame characterized by eq.~(\ref{eq:cm_frame}).

Let us, however, start from the case $B=0$ in general frame, characterized by
\begin{equation}\label{eq:frame_gen}
    \sum_a \mathfrak{Z}_a \bm{w}_a=0,\quad \sum_a\mathfrak{Z}_a=1,
\end{equation}
where $\{\mathfrak{Z}_a\}$ is a constant set.
The general Ohm law eq.~(\ref{eq:Ohm_app}) takes the Stephan-Maxwell form
\begin{equation}\label{eq:SM_B0}
    n_a\bm{d}_a=\sum_b J_{ab}(\bm{w}_b-\bm{w}_a).
\end{equation}
Remind that the vectors in right hand side, given by eq.~(\ref{eq:da_three}) [not necessary by eq.~(\ref{eq:da_E})], obey
\begin{equation}
    \sum_a n_a \bm{d}_a=0.
\end{equation}
The system (\ref{eq:SM_B0}) is singular and can not be inverted directly. 
Assume that the inverse relationship has the form
\begin{equation}\label{eq:Fick_law}
    \bm{w}_a=-\sum_b\mathfrak{D}_{ab} n_b\bm{d}_b
\end{equation}
with $\mathfrak{D}_{ab}$ being the matrix of diffusion coefficients. Due to singularity of this equation, diffusion coefficients $\mathfrak{D}_{ab}$ are not uniquely defined. From a general point of view it is convenient to require this matrix to be symmetric 
\cite{Curtiss1968JChPh,Condiff1969JChPh} that satisfy the Onsager relations manifestly. Since eq.~(\ref{eq:frame_gen}) should hold for any set $\bm{d}_a$, the symmetric diffusion matrix must obey \cite{CurtissBird99}
\begin{equation}
    \sum_a \mathfrak{Z}_a\mathfrak{D}_{ab}=
    \sum_b \mathfrak{D}_{ab}\mathfrak{Z}_b=0.
\end{equation}

It was shown in 
\cite{CurtissBird99} that the following linear system relates symmetric matrix $\mathfrak{D}_{ab}$ and the symmetric matrix $J_{ab}$
\begin{equation}\label{eq:CurtissSystem}
    \sum_c \left(J_{ac}-\frac{\mathfrak{Z}_c}{\mathfrak{Z}_a} J_{aa}\right)\mathfrak{D}_{cb}=-\delta_{ab}+\mathfrak{Z}_a,
\end{equation}
where the diagonal elements $J_{aa}$ are defined in such a way to satisfy
\begin{equation}
    \sum_a J_{ab}=0.
\end{equation}
The advantage of eq.~(\ref{eq:CurtissSystem}) is that it automatically leads to the symmetric diffusion coefficient matrix $\mathfrak{D}_{ab}$, as proved in \cite{CurtissBird99}. However, no simple method of the analytical solution of this equation is known. The explicit expressions  $\mathfrak{D}_{ab}$ in therms of $J_{ab}$ are given in \cite{CurtissBird99} for two- three- and four- component system.

Notice that if $\mathfrak{D}_{ab}$ is found in some frame, then the matrix $\overline{\mathfrak{D}}_{ab}$ in a different frame, described by a different set of coefficients $\{\overline{\mathfrak{Z}}_a\}$, is given by
\begin{equation}\label{eq:D_frame_change}
    \overline{\mathfrak{D}}_{ab}=\mathfrak{D}_{ab} -\sum_c\overline{\mathfrak{Z}}_c \mathfrak{D}_{cb}- \sum_c\overline{\mathfrak{Z}}_c \mathfrak{D}_{ac}+\sum_{cd}\overline{\mathfrak{Z}}_c \overline{\mathfrak{Z}}_d \mathfrak{D}_{cd}.
\end{equation}

For the electric current from eq.~(\ref{eq:Fick_law}) using eq.~(\ref{eq:da_E})
one obtains
\begin{equation}
    \bm{J}=\sigma \bm{E},
\end{equation}
where
\begin{equation}\label{eq:sigma0}
\sigma=\sum_{ab} q_aq_b n_an_b \mathfrak{D}_{ab}.
\end{equation}
In the absence of magnetic field, $\sigma$ is a scalar quantity, moreover from (\ref{eq:sigma0}) and (\ref{eq:D_frame_change}) it is clear that it does not depend on the choice of frame.

Now let us turn to the case of non-zero magnetic field. Following the formalism of Sec.~\ref{sec:tensor} we direct the $Z$ axis along the direction of $\bm{B}$ and introduce cyclic components of vector. We define in this section $\widetilde{w}_a=w_{ax}+iw_{ay}$ and similarly for $\bm{J}$, $\bm{d}$ and $\bm{E}$ vectors (in this definition we omit normalization constants  for brevity). Equation~(\ref{eq:Ohm_app}) takes the form \cite{Iakovlev1991Ap&SS}
\begin{equation}\label{eq:Ohm_irred}
    n_a\widetilde{d}_a-i \mathfrak{X}_a\widetilde{J} B = \sum_bJ_{ab}(\widetilde{w}_b-\widetilde{w}_a)-iq_an_a \widetilde{w}_a B.
\end{equation}
Formally, if $\mathfrak{D}_{ab}$ is known from the non-magnetized problem, one can first transform eq.~(\ref{eq:Ohm_irred}) as
\begin{equation}\label{eq:wa_1}
\widetilde{w}_a=-\sum \mathfrak{D}_{ab}\left(n_b\widetilde{d}_b+iq_bn_b \widetilde{w}_b B -i\mathfrak{X}_b B \widetilde{J} \right)    
\end{equation}
Exclusion of $\bm{w}_a$ from eq.~(\ref{eq:wa_1}) requires another linear system solution which can be written as 
\begin{equation}
\widetilde{w}_a=-\sum_{bc} \mathfrak{R}_{ac} \mathfrak{D}_{cb}\left(n_b\bm{d}_b -i\mathfrak{X}_b B \widetilde{J} \right),    
\end{equation}
where $\mathfrak{R}_{ab}$ is the following matrix inverse 
\begin{equation}\label{eq:R_ab}
    \mathfrak{R}_{ab}=\left(\delta_{ab}- i 
    \mathfrak{D}_{ab}q_bB \right)^{-1}.
\end{equation}
Electric current can be now calculated as %
{$\widetilde{J}=\sum_a n_a q_a \widetilde{w}_a = \widetilde{\sigma}\widetilde{E}$, where $\widetilde{d}_{(a)} = - q_a \widetilde{E}$, cf. eq.~(\ref{eq:da_E}). Then the complex electrical conductivity is
\begin{equation}\label{eq:sigma_arb_frame}
    \widetilde{\sigma}=\frac{\sum_{abc} q_a n_a \mathfrak{R}_{ac} \mathfrak{D}_{cb} n_bq_b}
    {1-iB\sum_{abc} q_a n_a \mathfrak{R}_{ac} \mathfrak{D}_{cb}\mathfrak{X}_b}.
\end{equation}
If the calculations are performed in the center-of-mass frame, i.e. $\mathfrak{Z}_a=\mathfrak{X}_a$ and $\widetilde{J}=\widetilde{\sigma}'\widetilde{E}'$, 
the sum in the denominator in eq.~(\ref{eq:sigma_arb_frame}) vanishes
and 
one obtains
\begin{equation}\label{eq:sigma_formal}
    \widetilde{\sigma}' = \sum_{abc} q_aq_bn_an_b\mathfrak{R}_{ac}\mathfrak{D}_{cb}.
\end{equation}%
}

The complex electrical conductivity $\widetilde{\sigma}'$ in eq.~(\ref{eq:sigma_formal}) is related to the Cartesian components of the electrical conductivity tensor as [cf. eq.~(\ref{eq:kappa_Cartesian})]
$\widetilde{\sigma}'=\sigma_\perp-i\sigma_\wedge$.

The analytical calculations along these lines are cumbersome and not easily tractable since they require additional matrix inversion/linear system solution in order to find $\mathfrak{R}_{ab}$. Therefore, one usually directly inverts eq.~(\ref{eq:Ohm_irred}) replacing one of the equations (i.e. for $a=1$) with eq.~(\ref{eq:cm_frame}) (see, e.g., \cite{Dommes2020PhRvD}). In any case, tractable analytical expressions can be obtained only for small number of species $r$.   For two- three- and four-component matter, the resulting general expressions for $\widetilde{\sigma}'$ can be found in \cite{Iakovlev1991Ap&SS}.

Finally we show below the solution of Ref.~\cite{Iakovlev1991Ap&SS} adapted for the $npe\mu$ NS core composition. We take into account that $q_n=0$ and neglect lepton-neutron interactions. The resulting expression reads \cite{Iakovlev1991Ap&SS,Shternin2008JETP}
\begin{eqnarray}\label{eq:ysh4}
\widetilde{\sigma'}^{-1}&=&\frac{d_0+iBd_1-B^2d_2}{a_0+iBa_1}+\frac{\mathfrak{X}_{n}^2}{J_{ pn}}B^2, \\
a_0&=&e^2(J_{ep}n_\mu^2+J_{\mu p}n_{ e}^2+J_{ e\mu}n_{ p}^2), \nonumber\\
a_1&=&-e^3n_{e}n_\mu n_{ p}, \nonumber\\
d_0&=&J_{ ep}J_{\mu p}+J_{ e\mu}J_{\mu p}+J_{ e\mu}J_{ e p}, \nonumber\\
d_1&=&-en_{ e}(1-2\mathfrak{X}_{e})J_{\mu p}-e n_\mu
(1-2\mathfrak{X}_\mu)J_{e p}\nonumber\\
&&-e
n_{ p}(1-2\mathfrak{X}_{e}-2\mathfrak{X}_\mu)J_{ e \mu}, \nonumber\\
d_2&=&e^2\left[n_{e}n_\mu(1-\mathfrak{X}_{ e}-\mathfrak{X}_\mu)^2-n_{
p}(\mathfrak{X}_{e}^2n_\mu+\mathfrak{X}_\mu^2 n_{ e})\right], \nonumber
\end{eqnarray}
where $e=|e|$ is an elementary charge.

The last term in eq.~(\ref{eq:ysh4}) which  contains $J_{pn}$ is responsible for increase of the resistivity at large $B$ \cite{Haensel1990A&A,Shternin2008JETP} observed in plasma containing neutral species \cite{BykovToptygin2007PhyU}.
}

\bibliographystyle{spphys}       
\def\apj{Astrophys. J.}
\def\apjl{Astrophys. J. Lett.}
\def\apjs{Astroph. J. Suppl. Ser.}
\def\mnras{Mon. Not. R. Astron. Soc.}
\def\aap{Astron. Astrophys.}
\def\prc{Phys. Rev. {\rm C}}
\def\prb{Phys. Rev. {\rm B}}
\def\pre{Phys. Rev. {\rm E}}
\def\prd{Phys. Rev. {\rm D}}
\def\prl{Phys. Rev. Lett.}
\def\plb{Physics Letters {\rm B}}
\def\apss{Astroph. Space Sci.}
\def\pla{Phys. Lett.  A}
\def\ssr{Space Sci. Rev.}
\def\araa{Ann. Rev. Astron. Astrophys.}
\def\aj{Astron. J.}
\def\jphys{J. Phys.}
\def\npa{Nucl. Phys. A}
\def\nphysa{Nucl. Phys. A}
\def\npb{Nucl. Phys.  B}
\def\ijmpe{Int. J. Mod. Phys. E}
\def\ijmpd{Int. J. Mod. Phys. D}
\def\ijmpa{Int. J. Mod. Phys. A}
\def\sovast{Soviet Astronomy}
\def\physrep{Physics Reports}
\def\jcp{Journal of Chemical Physics}
\def\pasa{Publications of the Astronomical Society of Australia}
\bibliography{bib-comp}   

\begin{thebibliography}{100}
\providecommand{\url}[1]{{#1}}
\providecommand{\urlprefix}{URL }
\expandafter\ifx\csname urlstyle\endcsname\relax
  \providecommand{\doi}[1]{DOI \discretionary{}{}{}#1}\else
  \providecommand{\doi}{DOI \discretionary{}{}{}\begingroup
  \urlstyle{rm}\Url}\fi

\bibitem{Schmitt2018}
A.~Schmitt, P.~Shternin, in \emph{The Physics and Astrophysics of Neutron
  Stars}, ed. by L.~Rezzolla, P.~Pizzochero, D.I. Jones, N.~Rea,
  I.~Vida{\~{n}}a (Springer International Publishing, Cham, 2018), pp. 455--574

\bibitem{Kaspi2017ARA&A}
V.M. {Kaspi}, A.M. {Beloborodov}, \araa \textbf{55}(1), 261 (2017).
\newblock \doi{10.1146/annurev-astro-081915-023329}

\bibitem{Potekhin1999AA}
A.Y. {Potekhin}, \aap \textbf{351}, 787 (1999)

\bibitem{Potekhin2015SSRv}
A.Y. {Potekhin}, J.A. {Pons}, D.~{Page}, Space Science Reviews \textbf{191},
  239 (2015).
\newblock \doi{10.1007/s11214-015-0180-9}

\bibitem{Ofengeim2015EL}
D.D. {Ofengeim}, D.G. {Yakovlev}, EPL (Europhysics Letters) \textbf{112}, 59001
  (2015).
\newblock \doi{10.1209/0295-5075/112/59001}

\bibitem{Baiko2016MNRAS}
D.A. {Baiko}, \mnras \textbf{458}, 2840 (2016).
\newblock \doi{10.1093/mnras/stw504}

\bibitem{Harutyunyan2016PhRvC}
A.~{Harutyunyan}, A.~{Sedrakian}, \prc \textbf{94}(2), 025805 (2016).
\newblock \doi{10.1103/PhysRevC.94.025805}

\bibitem{Iakovlev1991Ap&SS}
D.G. {Iakovlev}, D.A. {Shalybkov}, \apss \textbf{176}(2), 171 (1991).
\newblock \doi{10.1007/BF00646697}

\bibitem{Yakovlev1991Ap&SS}
D.G. {Yakovlev}, D.A. {Shalybkov}, \apss \textbf{176}, 171 (1991).
\newblock \doi{10.1007/BF00646697}

\bibitem{Rezzolla2013book}
L.~{Rezzolla}, O.~{Zanotti}, \emph{{Relativistic Hydrodynamics}} (Oxford
  University Press, Oxford, 2013)

\bibitem{RomatschkeRomatschke2017book}
P.~Romatschke, U.~Romatschke, \emph{Relativistic Fluid Dynamics In and Out of
  Equilibrium: And Applications to Relativistic Nuclear Collisions}.
\newblock Cambridge Monographs on Mathematical Physics (Cambridge University
  Press, Cambridge, 2019).
\newblock \doi{10.1017/9781108651998}

\bibitem{Gusakov2016PhRvD}
M.E. {Gusakov}, V.A. {Dommes}, \prd \textbf{94}(8), 083006 (2016).
\newblock \doi{10.1103/PhysRevD.94.083006}

\bibitem{Rau2020PhRvD}
P.B. {Rau}, I.~{Wasserman}, \prd \textbf{102}(6), 063011 (2020).
\newblock \doi{10.1103/PhysRevD.102.063011}

\bibitem{AnderssonComer2021LRR}
N.~{Andersson}, G.L. {Comer}, Living Reviews in Relativity \textbf{24}(1), 3
  (2021).
\newblock \doi{10.1007/s41114-021-00031-6}

\bibitem{Dommes2021arXiv}
V.A. {Dommes}, M.E. {Gusakov}, \prd \textbf{104}(12), 123008 (2021).
\newblock \doi{10.1103/PhysRevD.104.123008}

\bibitem{Huang2011AnPhy}
X.G. {Huang}, A.~{Sedrakian}, D.H. {Rischke}, Annals of Physics
  \textbf{326}(12), 3075 (2011).
\newblock \doi{10.1016/j.aop.2011.08.001}

\bibitem{Finazzo2016PhRvD}
S.I. {Finazzo}, R.~{Critelli}, R.~{Rougemont}, J.~{Noronha}, \prd
  \textbf{94}(5), 054020 (2016).
\newblock \doi{10.1103/PhysRevD.94.054020}

\bibitem{Hernandez2017JHE}
J.~{Hernandez}, P.~{Kovtun}, Journal of High Energy Physics \textbf{2017}(5), 1
  (2017).
\newblock \doi{10.1007/JHEP05(2017)001}

\bibitem{Landau1987Fluid}
L.D. Landau, E.M. Lifshitz, \emph{Fluid Mechanics, Second Edition: Volume 6
  (Course of Theoretical Physics)}, 2nd edn. (Butterworth-Heinemann, 1987)

\bibitem{Kovtun2019JHEP}
P.~{Kovtun}, Journal of High Energy Physics \textbf{2019}(10), 34 (2019).
\newblock \doi{10.1007/JHEP10(2019)034}

\bibitem{EckartPhysRev58}
C.~Eckart, Phys. Rev. \textbf{58}, 919 (1940).
\newblock \doi{10.1103/PhysRev.58.919}.
\newblock \urlprefix\url{https://link.aps.org/doi/10.1103/PhysRev.58.919}

\bibitem{HiscockPhysRevD85}
W.A. Hiscock, L.~Lindblom, Phys. Rev. D \textbf{31}, 725 (1985).
\newblock \doi{10.1103/PhysRevD.31.725}.
\newblock \urlprefix\url{https://link.aps.org/doi/10.1103/PhysRevD.31.725}

\bibitem{Mueller1967ZPhy}
I.~{M{\"u}ller}, Zeitschrift fur Physik \textbf{198}(4), 329 (1967).
\newblock \doi{10.1007/BF01326412}

\bibitem{Israel1976AnPhy}
W.~{Israel}, Annals of Physics \textbf{100}(1), 310 (1976).
\newblock \doi{10.1016/0003-4916(76)90064-6}

\bibitem{Israel1979AnPhy}
W.~{Israel}, J.M. {Stewart}, Annals of Physics \textbf{118}(2), 341 (1979).
\newblock \doi{10.1016/0003-4916(79)90130-1}

\bibitem{Gavassino2021FrASS}
L.~{Gavassino}, M.~{Antonelli}, Frontiers in Astronomy and Space Sciences
  \textbf{8}, 92 (2021).
\newblock \doi{10.3389/fspas.2021.686344}

\bibitem{Tsumura2007PhLB}
T.~{Tsumura}, T.~{Kunihiro}, K.~{Ohnishi}, Physics Letters B \textbf{646}(2-3),
  134 (2007).
\newblock \doi{10.1016/j.physletb.2006.12.074}

\bibitem{Tsumura2008PhLB}
K.~{Tsumura}, T.~{Kunihiro}, Physics Letters B \textbf{668}(5), 425 (2008).
\newblock \doi{10.1016/j.physletb.2008.07.109}

\bibitem{Van2012PhLB}
P.~{V{\'a}n}, T.S. {Bir{\'o}}, Physics Letters B \textbf{709}(1-2), 106 (2012).
\newblock \doi{10.1016/j.physletb.2012.02.006}

\bibitem{Freistuehler2014RSPSA}
H.~{Freistuhler}, B.~{Temple}, Proceedings of the Royal Society of London
  Series A \textbf{470}(2166), 20140055 (2014).
\newblock \doi{10.1098/rspa.2014.0055}

\bibitem{Freistuehler2017RSPSA}
H.~{Freist{\"u}hler}, B.~{Temple}, Proceedings of the Royal Society of London
  Series A \textbf{473}(2201), 20160729 (2017).
\newblock \doi{10.1098/rspa.2016.0729}

\bibitem{Freistuehler2018JMP}
H.~{Freist{\"u}hler}, B.~{Temple}, Journal of Mathematical Physics
  \textbf{59}(6), 063101 (2018).
\newblock \doi{10.1063/1.5007831}

\bibitem{Bemfica2018PhRvD}
F.S. {Bemfica}, M.M. {Disconzi}, J.~{Noronha}, \prd \textbf{98}(10), 104064
  (2018).
\newblock \doi{10.1103/PhysRevD.98.104064}

\bibitem{Bemfica2019arXiv}
F.S. {Bemfica}, M.M. {Disconzi}, C.~{Rodriguez}, Y.~{Shao}, arXiv e-prints
  arXiv:1911.02504 (2019)

\bibitem{Bemfica2019PhRvD}
F.S. {Bemfica}, M.M. {Disconzi}, J.~{Noronha}, \prd \textbf{100}(10), 104020
  (2019).
\newblock \doi{10.1103/PhysRevD.100.104020}

\bibitem{Bemfica020arXiv}
F.S. {Bemfica}, M.M. {Disconzi}, J.~{Noronha}, arXiv e-prints arXiv:2009.11388
  (2020)

\bibitem{Hoult2020JHEP}
R.E. {Hoult}, P.~{Kovtun}, Journal of High Energy Physics \textbf{2020}(6), 67
  (2020).
\newblock \doi{10.1007/JHEP06(2020)067}

\bibitem{Pandya2021PhRvD}
A.~{Pandya}, F.~{Pretorius}, \prd \textbf{104}(2), 023015 (2021).
\newblock \doi{10.1103/PhysRevD.104.023015}

\bibitem{Noronha2021arXiv}
J.~{Noronha}, M.~{Spali{\'n}ski}, E.~{Speranza}, arXiv e-prints
  arXiv:2105.01034 (2021)

\bibitem{Becattini2015EPJC}
F.~{Becattini}, L.~{Bucciantini}, E.~{Grossi}, L.~{Tinti}, European Physical
  Journal C \textbf{75}, 191 (2015).
\newblock \doi{10.1140/epjc/s10052-015-3384-y}

\bibitem{Zubarev1974}
D.N. {Zubarev}, \emph{Nonequilibrium statistical thermodynamics} (Consultants
  Bureau, New York, 1974)

\bibitem{Gavassino2020PhRvD}
L.~{Gavassino}, M.~{Antonelli}, B.~{Haskell}, \prd \textbf{102}(4), 043018
  (2020).
\newblock \doi{10.1103/PhysRevD.102.043018}

\bibitem{DeGroot1980Book}
S.R. De~Groot, \emph{Relativistic Kinetic Theory. Principles and Applications}
  (North-Holland, Netherlands, 1980)

\bibitem{Bhattacharyy2014JHEP}
S.~{Bhattacharyya}, Journal of High Energy Physics \textbf{2014}, 165 (2014).
\newblock \doi{10.1007/JHEP08(2014)165}

\bibitem{Bhattacharyy2014JHEPa}
S.~{Bhattacharyya}, Journal of High Energy Physics \textbf{2014}, 139 (2014).
\newblock \doi{10.1007/JHEP07(2014)139}

\bibitem{Becattini2019PhRvD}
F.~{Becattini}, D.~{Rindori}, \prd \textbf{99}(12), 125011 (2019).
\newblock \doi{10.1103/PhysRevD.99.125011}

\bibitem{Dowling2020PhRvD}
N.~{Dowling}, S.~{Floerchinger}, T.~{Haas}, \prd \textbf{102}(10), 105002
  (2020).
\newblock \doi{10.1103/PhysRevD.102.105002}

\bibitem{Metens1990PhFlB}
T.~{Metens}, R.~{Balescu}, Physics of Fluids B \textbf{2}(9), 2076 (1990).
\newblock \doi{10.1063/1.859428}

\bibitem{vanErkelens1977PhyA}
H.~{van Erkelens}, W.A. {van Leeuwen}, Physica A Statistical Mechanics and its
  Applications \textbf{89}(2), 225 (1977).
\newblock \doi{10.1016/0378-4371(77)90103-0}

\bibitem{Landau5eng}
L.D. Landau, E.M. Lifshitz, \emph{Statistical Physics, Part 1}, \emph{Course of
  Theoretical Physics}, vol.~5, 3rd edn. (Elsevier, Oxford, 1980)

\bibitem{deGrootMazur1984}
S.R. {de Groot}, P.~{Mazur}, \emph{Non-equilibrium thermodynammics} (Dover
  Publications, New York, 1984)

\bibitem{Landau10eng}
L.~Pitaevskii, E.~Lifshitz, \emph{Physical Kinetics}.
\newblock Course of theoretical physics by L. D. Landau and E. M. Lifshitz,
  Vol. 10 (Butterworth-Heinemann, 2008)

\bibitem{zhdanov2002transport}
V.~Zhdanov, \emph{Transport Processes in Multicomponent Plasma} (Taylor \&
  Francis, 2002)

\bibitem{Mahan93book}
G.D. {Mahan}, \emph{{Many-particle physics}} (Plenum Press, New York, 1993)

\bibitem{HarutyunyanParticles2018}
A.~Harutyunyan, A.~Sedrakian, D.H. Rischke, Particles \textbf{1}(1), 155
  (2018).
\newblock \doi{10.3390/particles1010011}.
\newblock \urlprefix\url{https://www.mdpi.com/2571-712X/1/1/11}

\bibitem{BaymPethick}
G.~Baym, C.~Pethick, \emph{Landau Fermi-Liquid Theory: Concepts and
  Applications} (John Wiley \& Sons, inc., New York, Chichester, Brisbane,
  Toronto, Singapore, 1991).
\newblock \doi{10.1002/9783527617159}

\bibitem{BaymChin1976NuPhA}
G.~{Baym}, S.A. {Chin}, \nphysa \textbf{262}(3), 527 (1976).
\newblock \doi{10.1016/0375-9474(76)90513-3}

\bibitem{Gusakov2009PhRvC}
M.E. {Gusakov}, E.M. {Kantor}, P.~{Haensel}, \prc \textbf{79}(5), 055806
  (2009).
\newblock \doi{10.1103/PhysRevC.79.055806}

\bibitem{vanWeert1984PhLA}
C.G. {van Weert}, M.C.J. {Leermakers}, Physics Letters A \textbf{101}(4), 195
  (1984).
\newblock \doi{10.1016/0375-9601(84)90377-3}

\bibitem{vanWeert1985PhyA}
C.G. {Van Weert}, M.C.J. {Leermakers}, Physica A Statistical Mechanics and its
  Applications \textbf{131}(3), 465 (1985).
\newblock \doi{10.1016/0378-4371(85)90126-8}

\bibitem{vanWeert1986PhyA}
C.G. {van Weert}, M.C.J. {Leermakers}, A.M.J. {Schakel}, Physica A Statistical
  Mechanics and its Applications \textbf{138}(3), 404 (1986).
\newblock \doi{10.1016/0378-4371(86)90024-5}

\bibitem{Hayata2015PhRvD}
T.~{Hayata}, Y.~{Hidaka}, T.~{Noumi}, M.~{Hongo}, \prd \textbf{92}(6), 065008
  (2015).
\newblock \doi{10.1103/PhysRevD.92.065008}

\bibitem{CercignaniKremer2002book}
C.~{Cercignani}, G.M. {Kremer}, \emph{The relativistic Boltzmann equation:
  theory and applications} (Birkh\"{a}user, Boston; Basel; Berlin, 2002)

\bibitem{Grad1949}
H.~Grad, Communications on Pure and Applied Mathematics \textbf{2}(4), 331
  (1949).
\newblock \doi{10.1002/cpa.3160020403}

\bibitem{ChapmanCowling1999}
S.~{Chapman}, T.G. {Cowling}, D.~{Burnett}, \emph{The Mathematical Theory of
  Non-uniform Gases: An Account of the Kinetic Theory of Viscosity, Thermal
  Conduction and Diffusion in Gases}, 3rd edn. (Cambridge University Press,
  Cambridge, 1999)

\bibitem{Denicol2014}
G.S. {Denicol}, Journal of Physics G Nuclear Physics \textbf{41}(12), 124004
  (2014).
\newblock \doi{10.1088/0954-3899/41/12/124004}

\bibitem{Gabbana2020PhR}
A.~{Gabbana}, D.~{Simeoni}, S.~{Succi}, R.~{Tripiccione}, \physrep
  \textbf{863}, 1 (2020).
\newblock \doi{10.1016/j.physrep.2020.03.004}

\bibitem{GarciaPerciante2020JSP}
A.L. {Garc{\'\i}a-Perciante}, M.E. {Rubio}, O.A. {Reula}, Journal of
  Statistical Physics \textbf{181}(1), 246 (2020).
\newblock \doi{10.1007/s10955-020-02578-0}

\bibitem{Denicol2016arXiv}
G.S. {Denicol}, J.~{Noronha}, arXiv e-prints arXiv:1608.07869 (2016)

\bibitem{Varshalovich}
D.A. {Varshalovich}, A.N. {Moskalev}, V.K. {Khersonskii}, \emph{{Quantum Theory
  of Angular Momentum}} (World Scientific Publishing Co, 1988).
\newblock \doi{10.1142/0270}

\bibitem{Anderson1987}
R.H. {Anderson}, C.J. {Pethick}, K.F. {Quader}, \prb \textbf{35}, 1620 (1987).
\newblock \doi{10.1103/PhysRevB.35.1620}

\bibitem{Viehland1974JChPh}
L.A. {Viehland}, C.F. {Curtiss}, \jcp \textbf{60}(2), 492 (1974).
\newblock \doi{10.1063/1.1681066}

\bibitem{vanErkelens1978PhyA}
H.~{van Erkelens}, W.A. {van Leeuwen}, Physica A Statistical Mechanics and its
  Applications \textbf{91}(1), 88 (1978).
\newblock \doi{10.1016/0378-4371(78)90059-6}

\bibitem{Dommes2020PhRvD}
V.A. {Dommes}, M.E. {Gusakov}, P.S. {Shternin}, \prd \textbf{101}(10), 103020
  (2020).
\newblock \doi{10.1103/PhysRevD.101.103020}

\bibitem{Anderson1974Phy}
J.L. {Anderson}, H.R. {Witting}, Physica \textbf{74}(3), 466 (1974).
\newblock \doi{10.1016/0031-8914(74)90355-3}

\bibitem{Bhatnagar1954PhRv}
P.L. {Bhatnagar}, E.P. {Gross}, M.~{Krook}, Physical Review \textbf{94}(3), 511
  (1954).
\newblock \doi{10.1103/PhysRev.94.511}

\bibitem{Formanek2021arXiv}
M.~{Formanek}, C.~{Grayson}, J.~{Rafelski}, B.~{M{\"u}ller}, Annals of Physics
  \textbf{434}, 168605 (2021).
\newblock \doi{10.1016/j.aop.2021.168605}

\bibitem{Rocha2021PhRvL}
G.S. {Rocha}, G.S. {Denicol}, J.~{Noronha}, \prl \textbf{127}(4), 042301
  (2021).
\newblock \doi{10.1103/PhysRevLett.127.042301}

\bibitem{ZimanBook}
J.M. Ziman, \emph{Electrons and Phonons}.
\newblock Oxford Classical Texts in the Physical Sciences (Clarendon
  Press/Oxford University Press, Oxford, UK; New York, USA, 2001)

\bibitem{JensenSmith1968PhLA}
H.~{H{\o}jg{\aa}rd Jensen}, H.~{Smith}, J.W. {Wilkins}, \pla \textbf{27}, 532
  (1968).
\newblock \doi{10.1016/0375-9601(68)90904-3}

\bibitem{BrookerSykes1968PhRvL}
G.A. {Brooker}, J.~{Sykes}, \prl \textbf{21}, 279 (1968).
\newblock \doi{10.1103/PhysRevLett.21.279}

\bibitem{SykesBrooker1970AnPhy}
J.~{Sykes}, G.A. {Brooker}, Annals of Physics \textbf{56}, 1 (1970).
\newblock \doi{10.1016/0003-4916(70)90002-3}

\bibitem{FlowersItoh1979ApJ}
E.~{Flowers}, N.~{Itoh}, \apj \textbf{230}, 847 (1979).
\newblock \doi{10.1086/157145}

\bibitem{PethickSchwenk2009PhRvC}
C.J. {Pethick}, A.~{Schwenk}, \prc \textbf{80}(5), 055805 (2009).
\newblock \doi{10.1103/PhysRevC.80.055805}

\bibitem{PolyaninBook}
A.D. {Polyanin}, A.V. {Manzhirov}, \emph{{Handbook of integral equations}}, 2nd
  edn. ({Chapman \& Hall/CRC}, Boca Raton; London; New York, 2008)

\bibitem{Ichiyanagi1994PhR}
M.~{Ichiyanagi}, \physrep \textbf{243}(3), 125 (1994).
\newblock \doi{10.1016/0370-1573(94)90052-3}

\bibitem{Abrikosov1959RPPh}
A.A. {Abrikosov}, I.M. {Khalatnikov}, Reports on Progress in Physics
  \textbf{22}(1), 329 (1959).
\newblock \doi{10.1088/0034-4885/22/1/310}

\bibitem{Heiselberg1992NuPhA}
H.~{Heiselberg}, G.~{Baym}, C.J. {Pethick}, J.~{Popp}, \npa \textbf{544}, 569
  (1992).
\newblock \doi{10.1016/0375-9474(92)90620-Y}

\bibitem{Shternin2020PhRvD}
P.S. {Shternin}, M.~{Baldo}, \prd \textbf{102}(6), 063010 (2020).
\newblock \doi{10.1103/PhysRevD.102.063010}

\bibitem{Blaizot1996NuPhA}
J.P. {Blaizot}, \nphysa \textbf{606}(1), 347 (1996).
\newblock \doi{10.1016/0375-9474(96)00209-6}

\bibitem{Ferziger1972Book}
J.H. {Ferziger}, H.G. {Kaper}, \emph{{Mathematical theory of transport
  processes in gases}} ({North - Holland}, Amsterdam, 1972)

\bibitem{Goldreich1992ApJ}
P.~{Goldreich}, A.~{Reisenegger}, \apj \textbf{395}, 250 (1992).
\newblock \doi{10.1086/171646}

\bibitem{Shalybkov1995MNRAS}
D.A. {Shalybkov}, V.A. {Urpin}, \mnras \textbf{273}(3), 643 (1995).
\newblock \doi{10.1093/mnras/273.3.643}

\bibitem{Beloborodov2016ApJ}
A.M. {Beloborodov}, X.~{Li}, \apj \textbf{833}(2), 261 (2016).
\newblock \doi{10.3847/1538-4357/833/2/261}

\bibitem{Passamonti2017MNRAS}
A.~{Passamonti}, T.~{Akg{\"u}n}, J.A. {Pons}, J.A. {Miralles}, \mnras
  \textbf{465}(3), 3416 (2017).
\newblock \doi{10.1093/mnras/stw2936}

\bibitem{Gusakov2017PhRvD}
M.E. {Gusakov}, E.M. {Kantor}, D.D. {Ofengeim}, \prd \textbf{96}(10), 103012
  (2017).
\newblock \doi{10.1103/PhysRevD.96.103012}

\bibitem{Castillo2020MNRAS}
F.~{Castillo}, A.~{Reisenegger}, J.A. {Valdivia}, \mnras \textbf{498}(2), 3000
  (2020).
\newblock \doi{10.1093/mnras/staa2543}

\bibitem{Machleidt2011PhR}
R.~{Machleidt}, D.R. {Entem}, \physrep \textbf{503}(1), 1 (2011).
\newblock \doi{10.1016/j.physrep.2011.02.001}

\bibitem{Shternin2013PhRvC}
P.S. {Shternin}, M.~{Baldo}, P.~{Haensel}, \prc \textbf{88}(6), 065803 (2013).
\newblock \doi{10.1103/PhysRevC.88.065803}

\bibitem{Shternin2021Univ}
P.~{Shternin}, I.~{Vida{\~n}a}, Universe \textbf{7}(6), 203 (2021).
\newblock \doi{10.3390/universe7060203}

\bibitem{Baym1990PhRvL}
G.~{Baym}, H.~{Monien}, C.J. {Pethick}, D.G. {Ravenhall}, \prl \textbf{64}(16),
  1867 (1990).
\newblock \doi{10.1103/PhysRevLett.64.1867}

\bibitem{Heiselberg:1993cr}
H.~Heiselberg, C.J. Pethick, \prd \textbf{48}, 2916 (1993).
\newblock \doi{10.1103/PhysRevD.48.2916}

\bibitem{ShterninYakovlev2007}
P.S. {Shternin}, D.G. {Yakovlev}, \prd \textbf{75}(10), 103004 (2007).
\newblock \doi{10.1103/PhysRevD.75.103004}

\bibitem{ShterninYakovlev2008}
P.S. {Shternin}, D.G. {Yakovlev}, \prd \textbf{78}(6), 063006 (2008).
\newblock \doi{10.1103/PhysRevD.78.063006}

\bibitem{Shternin2008JETP}
P.S. {Shternin}, JETP \textbf{107}, 212 (2008).
\newblock \doi{10.1134/S1063776108080050}

\bibitem{Stetina2019arXiv}
S.~{Stetina}, arXiv e-prints arXiv:1902.10745 (2019)

\bibitem{Stetina2018PhRvC}
S.~{Stetina}, E.~{Rrapaj}, S.~{Reddy}, \prc \textbf{97}(4), 045801 (2018).
\newblock \doi{10.1103/PhysRevC.97.045801}

\bibitem{Baiko2001AA}
D.A. {Baiko}, P.~{Haensel}, D.G. {Yakovlev}, \aap \textbf{374}, 151 (2001).
\newblock \doi{10.1051/0004-6361:20010621}

\bibitem{Sedrakian1994PhLB}
A.D. {Sedrakian}, D.~{Blaschke}, G.~{R{\"o}pke}, H.~{Schulz}, \plb
  \textbf{338}, 111 (1994).
\newblock \doi{10.1016/0370-2693(94)91352-8}

\bibitem{Wambach1993NuPhA}
J.~{Wambach}, T.L. {Ainsworth}, D.~{Pines}, \npa \textbf{555}, 128 (1993).
\newblock \doi{10.1016/0375-9474(93)90317-Q}

\bibitem{Benhar2010PhRvC}
O.~{Benhar}, A.~{Polls}, M.~{Valli}, I.~{Vida{\~n}a}, \prc \textbf{81}(2),
  024305 (2010).
\newblock \doi{10.1103/PhysRevC.81.024305}

\bibitem{BenharValli2007PhRvL}
O.~{Benhar}, M.~{Valli}, \prl \textbf{99}(23), 232501 (2007).
\newblock \doi{10.1103/PhysRevLett.99.232501}

\bibitem{CarboneBenhar2011JPhCS}
A.~{Carbone}, O.~{Benhar}, Journal of Physics Conference Series \textbf{336},
  012015 (2011).
\newblock \doi{10.1088/1742-6596/336/1/012015}

\bibitem{Baldo1999Book}
M.~Baldo (ed.), \emph{Nuclear Methods and the Nuclear Equation of State},
  \emph{International Review of Nuclear Physics}, vol.~8 (World Scientific,
  Singapore, 1999).
\newblock \doi{10.1142/2657}

\bibitem{Zhang2010PhRvC}
H.F. {Zhang}, U.~{Lombardo}, W.~{Zuo}, \prc \textbf{82}(1), 015805 (2010).
\newblock \doi{10.1103/PhysRevC.82.015805}

\bibitem{Shternin2017JPhCS}
P.~{Shternin}, M.~{Baldo}, H.~{Schulze}, Journal of Physics Conference Series
  \textbf{932}, 012042 (2017)

\bibitem{Migdal1990PhR}
A.B. {Migdal}, E.E. {Saperstein}, M.A. {Troitsky}, D.N. {Voskresensky}, Phys.
  Rep. \textbf{192}, 179 (1990).
\newblock \doi{10.1016/0370-1573(90)90132-L}

\bibitem{Blaschke2013PhRvC}
D.~{Blaschke}, H.~{Grigorian}, D.N. {Voskresensky}, \prc \textbf{88}(6), 065805
  (2013).
\newblock \doi{10.1103/PhysRevC.88.065805}

\bibitem{Kolomeitsev2015PhRvC}
E.E. {Kolomeitsev}, D.N. {Voskresensky}, \prc \textbf{91}(2), 025805 (2015).
\newblock \doi{10.1103/PhysRevC.91.025805}

\bibitem{LiMachleidt1993PhRvC}
G.Q. {Li}, R.~{Machleidt}, \prc \textbf{48}(4), 1702 (1993).
\newblock \doi{10.1103/PhysRevC.48.1702}

\bibitem{LiMachleidt1994PhRvC}
G.Q. {Li}, R.~{Machleidt}, \prc \textbf{49}(1), 566 (1994).
\newblock \doi{10.1103/PhysRevC.49.566}

\bibitem{Machleidt1987PhR}
R.~{Machleidt}, K.~{Holinde}, C.~{Elster}, \physrep \textbf{149}(1), 1 (1987).
\newblock \doi{10.1016/S0370-1573(87)80002-9}

\bibitem{Wiringa1995PhRvC}
R.B. {Wiringa}, V.G.J. {Stoks}, R.~{Schiavilla}, \prc \textbf{51}, 38 (1995).
\newblock \doi{10.1103/PhysRevC.51.38}

\bibitem{Potekhin2013A&A}
A.Y. {Potekhin}, A.F. {Fantina}, N.~{Chamel}, J.M. {Pearson}, S.~{Goriely},
  \aap \textbf{560}, A48 (2013).
\newblock \doi{10.1051/0004-6361/201321697}

\bibitem{Machleidt2001PhRvC}
R.~{Machleidt}, \prc \textbf{63}(2), 024001 (2001).
\newblock \doi{10.1103/PhysRevC.63.024001}

\bibitem{Carlson1983NuPhA}
J.~{Carlson}, V.R. {Pandharipande}, R.B. {Wiringa}, \npa \textbf{401}, 59
  (1983).
\newblock \doi{10.1016/0375-9474(83)90336-6}

\bibitem{Li2008PhRvC}
Z.H. {Li}, U.~{Lombardo}, H.J. {Schulze}, W.~{Zuo}, \prc \textbf{77}(3), 034316
  (2008).
\newblock \doi{10.1103/PhysRevC.77.034316}

\bibitem{Li2012PhRvC}
Z.H. {Li}, H.J. {Schulze}, \prc \textbf{85}(6), 064002 (2012).
\newblock \doi{10.1103/PhysRevC.85.064002}

\bibitem{Baldo2014PhRvC}
M.~{Baldo}, G.F. {Burgio}, H.J. {Schulze}, G.~{Taranto}, \prc \textbf{89}(4),
  048801 (2014).
\newblock \doi{10.1103/PhysRevC.89.048801}

\bibitem{Gulminelli2015PhRvC}
F.~{Gulminelli}, A.R. {Raduta}, \prc \textbf{92}(5), 055803 (2015).
\newblock \doi{10.1103/PhysRevC.92.055803}

\bibitem{Agrawal2005PhRvC}
B.K. {Agrawal}, S.~{Shlomo}, V.K. {Au}, \prc \textbf{72}(1), 014310 (2005).
\newblock \doi{10.1103/PhysRevC.72.014310}

\bibitem{Friedrich1986PhRvC}
J.~{Friedrich}, P.G. {Reinhard}, \prc \textbf{33}(1), 335 (1986).
\newblock \doi{10.1103/PhysRevC.33.335}

\bibitem{Koehler1976NuPhA}
H.S. {K{\"o}hler}, \nphysa \textbf{258}(2), 301 (1976).
\newblock \doi{10.1016/0375-9474(76)90008-7}

\bibitem{Reinhard1995NuPhA.584}
P.G. {Reinhard}, H.~{Flocard}, \nphysa \textbf{584}(3), 467 (1995).
\newblock \doi{10.1016/0375-9474(94)00770-N}

\bibitem{Nazarewicz1996PhRvC}
W.~{Nazarewicz}, J.~{Dobaczewski}, T.R. {Werner}, J.A. {Maruhn}, P.G.
  {Reinhard}, K.~{Rutz}, C.R. {Chinn}, A.S. {Umar}, M.R. {Strayer}, \prc
  \textbf{53}(2), 740 (1996).
\newblock \doi{10.1103/PhysRevC.53.740}

\bibitem{Bennour1989PhRvC}
L.~{Bennour}, P.H. {Heenen}, P.~{Bonche}, J.~{Dobaczewski}, H.~{Flocard}, \prc
  \textbf{40}(6), 2834 (1989).
\newblock \doi{10.1103/PhysRevC.40.2834}

\bibitem{Reinhard1999PhRvC}
P.G. {Reinhard}, D.J. {Dean}, W.~{Nazarewicz}, J.~{Dobaczewski}, J.A. {Maruhn},
  M.R. {Strayer}, \prc \textbf{60}(1), 014316 (1999).
\newblock \doi{10.1103/PhysRevC.60.014316}

\bibitem{ChabantPhd1995}
E.~{Chabanat}, Interactions effectives pour des conditions extremes d'isospin.
\newblock Ph.D. thesis, University Claude Bernard Lyon-1, Lyon, France (1995)

\bibitem{Dexheimer2008ApJ}
V.~{Dexheimer}, S.~{Schramm}, \apj \textbf{683}(2), 943 (2008).
\newblock \doi{10.1086/589735}

\bibitem{Dexheimer2017PASA}
V.~{Dexheimer}, \pasa \textbf{34}, e066 (2017).
\newblock \doi{10.1017/pasa.2017.61}

\bibitem{Bombaci2018A&A}
I.~{Bombaci}, D.~{Logoteta}, \aap \textbf{609}, A128 (2018).
\newblock \doi{10.1051/0004-6361/201731604}

\bibitem{Fortin2016PhRvC}
M.~{Fortin}, C.~{Provid{\^e}ncia}, A.R. {Raduta}, F.~{Gulminelli}, J.L.
  {Zdunik}, P.~{Haensel}, M.~{Bejger}, \prc \textbf{94}(3), 035804 (2016).
\newblock \doi{10.1103/PhysRevC.94.035804}

\bibitem{Douchin2001A&A}
F.~{Douchin}, P.~{Haensel}, \aap \textbf{380}, 151 (2001).
\newblock \doi{10.1051/0004-6361:20011402}

\bibitem{Sharma2015A&A}
B.K. {Sharma}, M.~{Centelles}, X.~{Vi{\~n}as}, M.~{Baldo}, G.F. {Burgio}, \aap
  \textbf{584}, A103 (2015).
\newblock \doi{10.1051/0004-6361/201526642}

\bibitem{Pearson2018MNRAS}
J.M. {Pearson}, N.~{Chamel}, A.Y. {Potekhin}, A.F. {Fantina}, C.~{Ducoin}, A.K.
  {Dutta}, S.~{Goriely}, \mnras \textbf{481}(3), 2994 (2018).
\newblock \doi{10.1093/mnras/sty2413}

\bibitem{Akmal1998PhRv}
A.~{Akmal}, V.R. {Pandharipande}, D.G. {Ravenhall}, \prc \textbf{58}(3), 1804
  (1998).
\newblock \doi{10.1103/PhysRevC.58.1804}

\bibitem{Heiselberg1999Ap}
H.~{Heiselberg}, M.~{Hjorth-Jensen}, \apjl \textbf{525}(1), L45 (1999).
\newblock \doi{10.1086/312321}

\bibitem{Heiselberg2000PhR}
H.~{Heiselberg}, M.~{Hjorth-Jensen}, \physrep \textbf{328}(5-6), 237 (2000).
\newblock \doi{10.1016/S0370-1573(99)00110-6}

\bibitem{Prakash1988PhRvL}
M.~{Prakash}, T.L. {Ainsworth}, J.M. {Lattimer}, \prl \textbf{61}(22), 2518
  (1988).
\newblock \doi{10.1103/PhysRevLett.61.2518}

\bibitem{HPY2007Book}
P.~{Haensel}, A.Y. {Potekhin}, D.G. {Yakovlev}, \emph{{Neutron Stars 1:
  Equation of State and Structure}} (Springer Science+Buisness Media, 2007)

\bibitem{Page1992ApJ}
D.~{Page}, J.H. {Applegate}, \apjl \textbf{394}, L17 (1992).
\newblock \doi{10.1086/186462}

\bibitem{Gusakov2005MNRAS}
M.E. {Gusakov}, A.D. {Kaminker}, D.G. {Yakovlev}, O.Y. {Gnedin}, \mnras
  \textbf{363}(2), 555 (2005).
\newblock \doi{10.1111/j.1365-2966.2005.09459.x}

\bibitem{Ofengeim2017PhRv}
D.D. {Ofengeim}, M.~{Fortin}, P.~{Haensel}, D.G. {Yakovlev}, J.L. {Zdunik},
  \prd \textbf{96}(4), 043002 (2017).
\newblock \doi{10.1103/PhysRevD.96.043002}

\bibitem{Logoteta2021Univ}
D.~Logoteta, Universe \textbf{7}(11), 408 (2021).
\newblock \doi{10.3390/universe7110408}

\bibitem{SedrakianClark2019EPJA}
A.~{Sedrakian}, J.W. {Clark}, European Physical Journal A \textbf{55}(9), 167
  (2019).
\newblock \doi{10.1140/epja/i2019-12863-6}

\bibitem{Andersson2021Univ}
N.~{Andersson}, Universe \textbf{7}(1), 17 (2021).
\newblock \doi{10.3390/universe7010017}

\bibitem{Vollhardt1990}
D.~{Vollhardt}, P.~{W\"{o}lfle}, \emph{The superfluid phases of Helium 3}
  (Taylor \& Francis, Bristol, 1990).
\newblock \doi{10.1201/b12808}

\bibitem{Manuel2021Univ}
C.~{Manuel}, L.~{Tolos}, Universe \textbf{7}(3), 59 (2021).
\newblock \doi{10.3390/universe7030059}

\bibitem{Curtiss1968JChPh}
C.F. {Curtiss}, \jcp \textbf{49}(7), 2917 (1968).
\newblock \doi{10.1063/1.1670528}

\bibitem{Condiff1969JChPh}
D.W. {Condiff}, \jcp \textbf{51}(10), 4209 (1969).
\newblock \doi{10.1063/1.1671780}

\bibitem{CurtissBird99}
C.F. Curtiss, R.B. Bird, Industrial \& Engineering Chemistry Research
  \textbf{38}(7), 2515 (1999).
\newblock \doi{10.1021/ie9901123}.
\newblock \urlprefix\url{https://doi.org/10.1021/ie9901123}

\bibitem{Haensel1990A&A}
P.~{Haensel}, V.A. {Urpin}, D.G. {Iakovlev}, \aap \textbf{229}(1), 133 (1990)

\bibitem{BykovToptygin2007PhyU}
A.M. {Bykov}, I.~{Toptygin}, Physics Uspekhi \textbf{50}(2), 141 (2007).
\newblock \doi{10.1070/PU2007v050n02ABEH006177}

\end{thebibliography}

\end{document}